# UMA PROPOSTA METODOLÓGICA PARA A APRENDIZAGEM:

## REFLEXÃO SOBRE AS PRÁTICAS PEDAGÓGICAS DA ESTATÍSTICA AO ELABORAR OS INSTRUMENTOS DE PESQUISA SOCIAIS

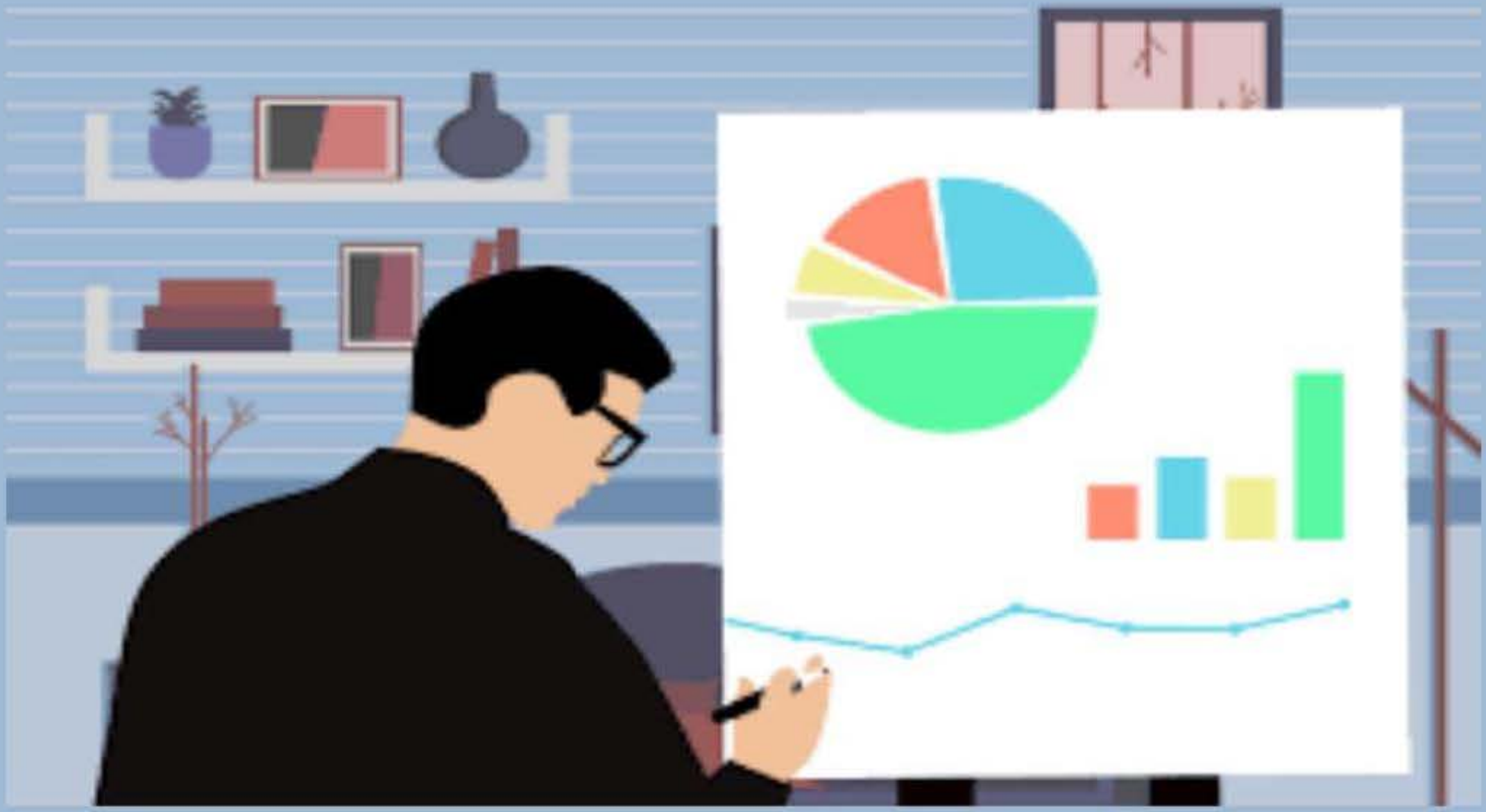

MANOEL BENEDITO NIRDO DA SILVA CAMPOS

MANOEL BENEDITO NIRDO DA SILVA CAMPOS

# UMA PROPOSTA METODOLÓGICA PARA A APRENDIZAGEM:

## REFLEXÃO SOBRE AS PRÁTICAS PEDAGÓGICAS DA ESTATÍSTICA AO ELABORAR OS INSTRUMENTOS DE PESQUISA SOCIAIS

**FOLHA DE APROVAÇÃO**

**UMA PROPOSTA METODOLÓGICA PARA A APRENDIZAGEM: REFLEXÃO SOBRE AS PRÁTICAS PEDAGÓGICAS DA ESTATÍSTICA AO ELABORAR OS INSTRUMENTOS DE PESQUISA SOCIAIS.**

MANOEL BENEDITO NIRDO DA SILVA CAMPOS

ORIENTADOR PROFESSOR DR. ARISTIDES ESCOBAR ARGAÑA

BANCA EXAMINADORA

Prof. Dr. Domingo Borja_Ortiz___

PRESIDENTE

Profª Drª  Catarina Costa Fernandes_____

MEMBRO

Profª Drª Margarita Prieto Yegros

MEMBRO

Aprovado: 26/07/2008

**Assunción-Paraguay**

**2008**

# AGRADECIMENTOS

A Deus, sabedoria superior do universo, do qual sou criatura e filho.

Ao meu orientador e grande amigo, professor Dr. Aristides Escobar Argaña Pelos grandes ensinamentos e pelo incentivo no desenvolvimento desta pesquisa.

A UTIC pela oportunidade para realização desta pesquisa.

Aos meus pais, Manoel (em memória) e Antônia, que sempre se preocuparam com a educação de seus filhos e souberam mostrar os bons procedimentos e o caminho seguro do trabalho e da luta.

A Carmen, minha esposa, que com sua maneira especial de ser, trouxe muita inspiração e sabedoria em momentos de dificuldade que não imaginava estar me ajudando.

Aos meus filhos, em especial ao Emanuel e Matheus, pela compreensão da minha ausência e pelo amoroso convívio, aumentando minha fé e perseverança na busca deste sonho.

Aos amigos, que fiz na pós-graduação pelas discussões e contribuições ao trabalho.

Aos alunos do Curso de Matemática, Informática, Ciências Contábeis, Enfermagem que se dispuseram, com boa vontade, a participar desta pesquisa.

A Estér, Secretária do Instituto de Ciências Exatas e Naturais do *Campus* Universitário de Rondonópolis, pelas suas atitudes positivas e pela compreensão frente aos problemas e pelas valiosas contribuições espirituais, acadêmicas e administrativas.

# RESUMO

Esta pesquisa apresenta uma proposta diferenciada de ensino que faculta, ao aluno, ser o agente na construção do conhecimento, superando as dificuldades que a Matemática apresenta. Objetivando compreender como a utilização das ferramentas estatísticas pode contribuir para a melhoria do processo ensino-aprendizagem e a construção do conhecimento estatístico, estudados junto aos alunos do *Campus* Universitário de Rondonópolis/UFMT. Para alcançar o objetivo proposto, realizou-se uma análise acerca das atividades didáticas no ensino de estatística básica no qual se optou por uma abordagem quali-quantitativa, enfocando o cotidiano, com uso de software, criação e simulação de modelos, bem como procurando estabelecer a frequência da atitude dos alunos, através de questionários e Escala Likerts de quatro pontos, buscando dados que traçassem o perfil dos atores envolvidos que auxiliassem na compreensão das atividades didáticas planejadas. O estudo justifica-se pela inexistência de referenciais teóricos metodológicos da temática em questão. Foram organizadas atividades didáticas de Estatística, cujas aplicações ocorreram em aulas alternadas, sendo aulas expositivas tradicionais e aulas práticas em um Laboratório de Informática. Constituindo-se no principal foco de caracterização e reflexão desta pesquisa científica durante o processo ensino-aprendizagem da Estatística. Os resultados apresentaram uma correta atitude dos pesquisados sobre as estratégias metodológicas usadas, assumidas com referência na concepção da pesquisa/ação. Este estudo contribuiu para motivar, despertar e responder indagações e dar significado e entendimentos aos trabalhos com Modelagem Matemática, pois pode promover a melhoria do processo de ensino-aprendizagem e se configura como uma ferramenta indispensável para a educação.

## RESUMEN

Esta investigación presenta una propuesta diferente que proporciona la educación, el estudiante, ya sea el agente en la construcción del conocimiento, la superación de las dificultades que tiene la matemática. El objetivo es comprender cómo el uso de herramientas estadísticas puede contribuir a mejorar el proceso de enseñanza-aprendizaje y la construcción de conocimiento estadístico, estudiado con los estudiantes de la Universidad Campus Rondonópolis / UFMT. Para lograr el objetivo propuesto, se realizó un análisis de las actividades didácticas en las estadísticas básicas de la educación en la que optó por un enfoque cualitativo y cuantitativo, se centra en la vida cotidiana, utilizando modelos de software, creación y simulación, así como tratar de establecer la frecuencia de la actitud de los estudiantes, a través de cuestionarios y escalas Likerts de cuatro puntos, ir a buscar los datos que dibujen el perfil de los actores involucrados que ayudaron en la comprensión de las actividades educativas planificadas. El estudio se justifica por la falta de un marco teórico metodológico del tema en cuestión. Se organizaron actividades educativas de Estadística cuyas aplicaciones ocurrido en clases alternas, con las clases magistrales y clases prácticas en un laboratorio de computación. Convirtiéndose en el foco principal de la caracterización y la reflexión de esta investigación científica en los procesos de enseñanza-aprendizaje de Estadística. Los resultados mostraron la actitud correcta de los encuestados sobre las estrategias metodológicas utilizadas, tomadas en relación con el diseño de la investigación / acción. Este estudio ha contribuido a motivar, despertar y responder preguntas y dar sentido y entendimiento para trabajar con los modelos matemáticos, por lo tanto, puede promover la mejora de la enseñanza y el aprendizaje y se configura como una herramienta indispensable para la educación.

## LISTA DE FIGURAS

# LISTA DE TABELAS

# LISTA DE QUADROS

# LISTA DE GRÁFICOS

## LISTA DE ABREVIATURAS E SIGLAS

| | |
|---|---|
| UFMT | Universidade Federal de Mato Grosso |
| CUR | Campus Universitário de Rondonópolis |
| IES | Instituição de Ensino Superior |
| ICEN | Instituto de Ciências Exatas e Naturais |
| ICHS | Instituto de Ciências Humanas e Sociais |
| PDI | Plano de Desenvolvimento Institucional |
| SESU | Secretaria de Ensino Superior |
| CNE | Conselho Nacional de Educação |
| INEP | Instituto Nacional de Estudos e Pesquisas Educacionais |
| MEC | Ministério de Educação e Desportos |
| UFSC | Universidade Federal de Santa Catarina |
| PCNs | Parâmetros Curriculares Nacionais |
| CAPES | Coordenação de Aperfeiçoamento de Pessoal de Nível Superior |
| UFRGS | Universidade Federal do Rio Grande do Sul |
| UNICAMP | Universidade Estadual de Campinas |
| UFMG | Universidade Federal de Minas Gerais |
| USP | Universidade de São Paulo |
| PAD | Programa de Apoio Didático |
| ANPED | Associação Nacional de Pós-Graduação e Pesquisa em Educação |
| SINAPE | Simpósio Nacional de Probabilidade e Estatística |
| PED | Programa de Estágio Docente |
| PAE | Programa de Aperfeiçoamento de Estatística |
| ABE | Associação Brasileira de Estatística |
| UESC | Universidade Estadual de Santa Catarina |
| UNESP | Universidade Estadual do Oeste Paulista |
| UNISANTA | Universidade de Santa Cecília |
| MT | Mato Grosso |
| PRAPEM | Prática Pedagógica em Matemática |
| PROEM | Programa de Estudos e Pesquisa no Ensino de Matemática |
| SBEM | Sociedade Brasileira de Educação Matemática |
| ENEM | Encontro Nacional de Educação Matemática |
| COLE | Conselho de Leitura do Brasil |
| EAE | Escala de Atitude de Estatística |

## SUMÁRIO

# 1 INTRODUÇÃO

Esta pesquisa apresenta resultados de experiências de ensino, abordando o ensino de Estatística no contexto dos cursos de graduação que ofertam a disciplina de Estatística no *Campus* Universitário de Rondonópolis/UFMT, bem como reflexão sobre a aplicação dos conteúdos em seu cotidiano profissional. Por essa razão, estratégias didáticas foram propostas com a finalidade de um melhor entendimento dos conteúdos estatísticos.

A estatística possibilita ao aluno abertura de novos horizontes, pois traz subsídios para a pesquisa e a tomada de decisões tanto no ambiente acadêmico como no empresarial. Além do uso da estatística para simples acompanhamento de resultados e descrição de cenários, a estatística pode ser utilizada como uma ferramenta de projeção e perspectivas aliada a um planejamento estratégico auxiliando seu desenvolvimento dentro do ambiente em que se insere.

A *American Statistical Association* (ASA) estabelece os seguintes objetivos com relação à Estatística e à Probabilidade: (1) possibilitar a compreensão dos dados, o que significam e qual a sua origem; (2) usar técnicas gráficas e numéricas simples para sintetizá-los; (3) criar um modelo matemático para descrevê-los e (4) fazer inferências sobre uma população usando amostras oriundas da mesma (GONÇALEZ, 2002, p. 24).

Embora os Parâmetros Curriculares Nacionais – PCN's – façam referência direta à inclusão de elementos de Estatística e Probabilidades nos Ensinos Fundamentais e Médios, os estudantes estão chegando aos cursos superiores com pouco ou nenhum contato anterior com a Estatística. Os alunos ingressantes no curso superior vêm com conhecimentos insuficientes (ou até inexistentes) de termos como, por exemplo, probabilidade, variabilidade, raciocínio aleatório, como esclarece:

> Grande parte desses alunos está utilizando as ferramentas estatísticas de forma mecânica, usando as fórmulas e algoritmos sem a real compreensão do(s) objeto(s) matemático(s) que as justifica e, consequentemente, sem a percepção da aplicabilidade destas ferramentas na área de atuação. Se o processo de construção dos saberes estatísticos necessários não for vivenciando de forma adequada pelo aluno, este não terá as condições necessárias para reaplicá-los na sua vida profissional (CORDANI, 2001 *apud* NOVAES, 2004, p. 10).

A Estatística é uma ferramenta indispensável para qualquer profissional que necessita analisar informações em suas tomadas de decisões diárias, na sua vida profissional, no seu trabalho. Nesse sentido, se o processo de aprendizagem estatística não for vivenciado de forma adequada pelo aluno, este não terá as condições necessárias para reaplicá-la na sua vida profissional.

Neste sentido, a Estatística é um ramo da Matemática e, desde sua origem, sempre houve preocupação em registrar diferentes dados e realizar estimativas. Segundo Crespo (2002), na antiguidade, vários povos já registravam o número de habitantes, de nascimento, de óbitos, faziam estimativas de riquezas individuais e sociais, distribuíam equitativamente terras ao povo, cobravam impostos e realizavam inquéritos quantitativos por processos que, hoje chamaríamos de "estatísticas".

Este autor explica que, na Idade Média, colhiam-se informações, geralmente com finalidade tributárias ou bélicas. A partir do século XVI, começaram a surgir as primeiras análises sistemáticas de fatos sociais, como batizados, casamentos, funerais, originando as primeiras tábuas e tabelas e os primeiros números relativos. No século XVIII, o estudo de tais fatos foi adquirindo, aos poucos, feição, verdadeiramente, científica. Godofredo Achenwall batizou a nova ciência (ou método) com o nome de Estatística, determinando o seu objetivo e suas relações com as ciências.

Especificamente no Brasil, em 1800, surgiu o primeiro senso geral da população brasileira, realizado por José Maria da Silva Paranhos, conhecido como visconde do Rio Branco. Em 1930, criou-se o IBGE (Instituto Brasileiro de Geografia e Estatística) e em 1972, também surgiu o primeiro computador que ajudou a dar um grande salto na Estatística, por fim a inclusão nos Ensinos Fundamental e Médio (PCN's, 1997).

As relações que se estabelecem com outros tempos e espaços, além de se tornar o estudo de como chegar a conclusões sobre o todo, partindo da observação de partes desse todo,

> [...] atualmente, o público leigo (leitor de jornais e revistas) posiciona-se em dois extremos divergentes e igualmente errôneos quanto à validade das conclusões estatísticas: ou crê em sua infalibilidade ou afirma que elas nada provam. Os que assim pensam ignoram os objetivos, o campo e o rigor do método estatístico; ignoram a Estatística, quer teórica quer prática, ou a conhecem muito superficialmente (CRESPO, 2002, p. 11-12).

Na era da energia nuclear, os estudos estatísticos têm avançado rapidamente e, com seus processos e técnicas, têm contribuído para a organização dos negócios e recursos do mundo moderno.

Do exposto, é aconselhável que, nos dias de hoje, tenha-se uma visão bem mais ampla da Estatística considerando-a como uma ciência fundamental e básica do método científico experimental (GONÇALVES, 2002).

Dessa forma, Wada (1996), em seu trabalho de doutorado, coletou, junto aos professores que lecionavam a Estatística na Universidade Estadual de Campinas (UNICAMP), opiniões com relação aos objetivos dessa disciplina em diversos cursos descritos a seguir:

> O objetivo da Estatística como disciplina é a de apresentar a importância da sua metodologia, principalmente na área de experimentação e também na interpretação de resultados obtidos através de amostragem. As técnicas básicas devem ser ensinadas com o objetivo de análise para casos simples e a importância de um profissional qualificado para situações mais elaboradas. O aluno deverá apresentar um bom conhecimento de matemática em nível de segundo grau e também de cálculo.
> Nota-se, nesse objetivo, uma valorização do método estatístico e a necessidade de um bom conhecimento de Matemática por parte do aluno. Os alunos deveriam ter conhecimentos sobre os conteúdos de Matemática básica, tais como: saber contar, somar, frações, etc.. O objetivo da disciplina de Estatística deveria ser o de transmitir conceitos tais como: aleatoriedade, organização de dados, exploração de dados, comparações.
> Observa-se uma grande preocupação em realizar atividades com muitos exercícios com a finalidade da fixação dos conteúdos.
> O objetivo da Estatística deve ser o de resumir as informações fundamentais, devendo o aluno ter uma boa formação sobre as ciências básicas, ter conhecimentos sobre conceitos gerais de análise de dados e construção de modelos probabilísticos.
> Nota-se uma valorização aos pré-requisitos e uma boa formação geral.
> O conhecimento sobre a disciplina de Estatística é fundamental nos dias de hoje, tendo em vista que a aplicação da estatística encontra-se em todas as áreas da ciência. O aluno deve ter noções bem fundamentadas de Estatística Introdutória, que possibilitem ao mesmo, quando necessário, poder se aprofundar em determinados tópicos, ou pelo menos, saber diagnosticar a necessidade da Estatística (WADA, 1996, *apud* GONÇALEZ, 2002, p. 3 ).

As dificuldades pelas quais passam os professores no ato do ensino, e os alunos, em termos de aprendizagem no campo da Estatística, vêm inquietando muitos pesquisadores na área da didática da Matemática.

Etimologicamente, tem-se que a palavra Estatística é uma parte da Matemática Aplicada que fornece métodos para coleta, organização, descrição, análise e interpretação de dados e para utilização deles na tomada de decisões. A coleta, a organização e a descrição dos dados estão a cargo da Estatística Descritiva, enquanto a análise e a interpretação desses dados são realizadas pela Estatística Indutiva ou Inferencial (CRESPO, 2000).

Nesta mesma linha de compreensão, a Estatística é conceituada por Mandim (1995, p.11) "como uma metodologia ou conjunto de técnicas que utiliza a coleta de dados, sua classificação, sua apresentação ou representação, sua análise e sua interpretação visando a sua utilização dentro de um processo decisório".

Ainda de acordo com o autor, a Estatística é um conjunto ou coleção de dados numéricos que fornecem informações acerca de um fenômeno observado: estatísticas econômicas (que fornecem dados sobre preços, vendas, produção etc.), estatísticas demográficas (que fornecem dados sobre nascimentos, mortes etc.).

Compreende-se que:

> Estatística Descritiva – é a parte da Estatística que procura **descrever** e **analisar** certo grupo de observações, normalmente denominado de amostra, procurando expressar estas observações através de medidas e formas de representação (tabelas, gráficos, curvas, etc.). É também denominada **Estatística Dedutiva**.
> Estatística Inferencial – é a parte da Estatística que compreende um processo de generalização, a partir da análise e interpretação de dados amostrais. É também denominada **Estatística Indutiva**.
> Teoria da Probabilidade – como todo processo de generalização envolve uma margem de risco ou incerteza, temos então a parte da Estatística que se ocupa em estudá-la utilizando métodos e técnicas apropriadas, o que constitui os fundamentos da **Teoria da Probabilidade** (MANDIM, 1995, p. 11, grifo nosso).

Em geral, as pessoas, quando se referem ao termo estatístico, o fazem no sentido da organização e descrição dos dados, desconhecendo que o aspecto essencial da Estatística é o de proporcionar métodos inferenciais, que permitem conclusões que transcendam os dados obtidos.

Modelos matemáticos utilizam informações descritas em termos matemáticos, usando representações numéricas e geométricas. A construção de modelos e a elaboração sobre eles é o que se chama modelagem. Esta forma de modelagem consiste na representação do real.

Modelagem é a operação de modelar, e modelar, por sua vez, significa fazer ou representar por meio de modelo. Dentre os vários significados atribuídos à palavra “modelo”, aqueles que parecem mais próximos de um contexto matemático estão relacionados à informática, à física e à economia. Por exemplo, em um contexto de informática, modelo seria a representação simplificada e abstrata de fenômeno ou situação concreta e serve de referência para a observação, estudo ou análise (FERREIRA, 2000).

Também, segundo Ferreira (2000), um modelo econômico é a representação simplificada de relações entre variáveis econômicas, em geral, sob a forma de um sistema de equações, e que, com o uso de técnicas econométricas, pode fornecer simulações ou previsões.

Trabalhar com Modelagem pode motivar os alunos e o próprio professor, facilitando a aprendizagem. O conteúdo passa a ter significado deixando de ser abstrato e passa a ser concreto. Prepara para futuras profissões nas mais diversas áreas do conhecimento, devido à interatividade do conteúdo com outras disciplinas. Também desenvolve o raciocínio lógico e dedutivo, desenvolvendo o aluno como cidadão crítico e transformador de sua realidade.

A Modelagem, processo que envolve a realidade, proporciona ao aluno uma análise global da realidade em que ele envolve. Na concepção de Machado, “Como totalidade, a realidade só pode ser apreendida nesta perspectiva globalizante. O conhecimento matemático nasce do real e a ele se dirige como ocorre em todos os outros campos” (MACHADO, 1987, p. 15).

Portanto, nos dias de hoje, um dos grandes desafios é fazer o aluno compreender o seu papel na sociedade, de agente ativo e transformador da sua realidade.

De acordo com D’Ambrosio (1986), a Modelagem por meio da dinâmica de realidade – reflexão sobre a realidade – resulta em uma ação planejada. Nesse sentido, a construção de modelos sobre os quais o indivíduo opera, depende de sua experiência e conhecimento dos recursos naturais, incrementando as opções de atividades ciclo realidade – reflexão – ação – realidade como o ponto importante da questão, desvendando o comportamento individual, social e cultural.

## 1.1 Ambiente da pesquisa

A presente pesquisa realizou-se em uma instituição de ensino superior, localizada na cidade de Rondonópolis – MT. O *Campus* Universitário de Rondonópolis (CUR) foi criado e homologado em 31 de março de 1976, mediante a Resolução nº. 01/76 do Conselho Universitário da então Universidade Estadual de Mato Grosso, muito embora a Lei Estadual nº 3575 de 2 de dezembro de 1974 já autorizasse a sua criação como Centro Pedagógico de Rondonópolis (CPR). Oferecendo, simultaneamente, os Cursos de Ciências e Estudos Sociais, na forma de Licenciatura Curta, o Centro Pedagógico de Rondonópolis iniciou suas atividades em 05 de maio de 1976.

Com a divisão do estado em 1977, deu-se início ao processo de federalização do Centro, integrando-o à Universidade Federal de Mato Grosso, uma vez que o município de Rondonópolis passava a pertencer ao estado de Mato Grosso, agora, dividido em duas unidades federativas. De fato, em 5 de julho de 1979, foi instituída a Fundação da Universidade Federal de Mato Grosso do Sul, mediante a Lei Federal nº. 6.674, que em seu artigo 13º transferia para a Universidade Federal do Mato Grosso a responsabilidade pelo Centro Pedagógico de Rondonópolis [...] o Centro Pedagógico de Rondonópolis, atualmente vinculado à Universidade Estadual de Mato Grosso, passa a integrar com todos os seus bens e direitos, a Universidade Federal de Mato Grosso, com sede em Cuiabá.

Por meio de ato do Conselho Diretor de nº. 05/80, datado de 9 de janeiro de 1980, e com a lotação no quadro de pessoal administrativo (Portaria GR 016/80) e docente (Portaria GR 015/80), dos servidores, o Centro Pedagógico de Rondonópolis integrou-se como *Campus* à estrutura da Universidade Federal de Mato Grosso. Esta integração evidenciou a necessidade de uma nova adequação à estrutura organizacional da UFMT. Neste sentido, a administração do Centro coube a um coordenador, coadjuvado por seu vice e um Conselho de Departamentos. Assumindo de forma *pro tempore*, o Prof. Etewaldo de Oliveira Borges esteve na Coordenação do Centro, no período de 1979 a 1984.

Os dois cursos que compunham o Centro permaneceram e foram criados dois Departamentos, coordenados por chefes e subchefes, designados pelo Reitor, com base em lista tríplice. De acordo com Maria das Graças Kida (1985), essa estrutura existia apenas no aspecto formal, pois, na realidade, só em agosto de 1983, após cinco anos, com o processo de abertura, discussões e reivindicações, ocorreram eleições, e o Centro passou a contar com Vice-coordenador e subchefes de Departamentos. Cada curso organizava-se, a partir dos Colegiados de Departamento e de Curso. As necessidades administrativas passaram a contar com uma Secretaria Geral e uma Biblioteca Regional, cuja coordenação, inicialmente, coube ao bibliotecário e ao professor Javert Melo Vieira.

As demandas da comunidade local e a necessidade de expansão da própria Universidade aceleraram a política de interiorização. Com base nas diretrizes prescritas pelas normas da Universidade e ratificadas pela Resolução nº. CD 04/80, de 8 de maio de 1980, que aprovava a estrutura organizacional do Centro e definia normas sobre os cursos, procedeu-se aos estudos para a elaboração do projeto de criação de novos cursos, já no segundo semestre do mesmo ano.

Tais estudos permitiram a opção por três cursos de graduação a serem oferecidos já no primeiro semestre do ano subsequente, a saber: Ciências Contábeis; Letras, com habilitação em Português e Literatura Portuguesa; e Pedagogia, com habilitações em Supervisão Escolar e Magistério das Matérias Pedagógicas do 2º Grau.

Aprovados em 27 de janeiro de 1981, por meio da Resolução nº. CD 019/81, esses cursos abrem seus vestibulares em fevereiro do mesmo ano, tendo como limite o número de 30 vagas por curso. As aprovações, diante das instalações disponíveis, revelaram uma questão importante a ser resolvida – o espaço físico. Desde a sua criação, os dois primeiros cursos funcionavam, inicialmente, em algumas salas de aula da Escola Adolfo Augusto de Moraes e no salão paroquial da Igreja Santa Cruz e, posteriormente, na Escola Estadual de 1º e 2º Graus Joaquim Nunes Rocha. O curso de Ciências Contábeis funcionava no prédio da APAE. Os antigos cursos já demandavam espaços maiores, e a criação dos cursos novos, por sua vez, exigiu ainda mais a construção de uma sede própria do *Campus*, fazendo com que, em abril de 1983, fosse inaugurada a primeira etapa do prédio e feita a transferência dos cursos existentes para as novas instalações, com exceção dos cursos de Ciências Contábeis e Ciências, que continuaram funcionando no prédio da APAE.

O crescimento do município de Rondonópolis e da região sul do Estado exigia a oferta de novos cursos. Tal demanda resultou, primeiramente, na criação dos referidos cursos de Pedagogia e Letras, em 1981, posteriormente, na implantação dos cursos de História e

Geografia, extinguindo-se, assim, o curso de Estudos Sociais, em 1988, quando também os cursos de Matemática e Biologia substituíram o de Ciências.

Com a Resolução CD nº. 027 de 12 de fevereiro de 1992, que dispôs sobre a reorganização administrativa da UFMT, foi criado o Conselho Administrativo dos Institutos de Rondonópolis (CADIR). Dessa forma, passaram, a funcionar, neste Campus, os seguintes Institutos: o Instituto de Ciências Humanas e Sociais (ICHS), que abrange os Departamentos de Educação, Letras, História e Ciências Contábeis; e o Instituto de Ciências Exatas e Naturais (ICEN), compreendendo os Departamentos de Ciências Biológicas, Geografia e Matemática.

Num processo crescente de expansão, o então chamado Centro Pedagógico de Rondonópolis desenvolveu o projeto "Unestado", dando sequência à interiorização, iniciada pela UFMT em 1979, mas apenas iniciada neste *Campus* a partir de 1989. Tratava-se de projeto extensionista, com a realização de cursos de atualização em fundamentos didático-pedagógicos para professores da rede pública de ensino das cidades de Pedra Preta, Jaciara, Juscimeira, Poxoréo e Guiratinga.

Essa interiorização teve continuidade com a instalação da Licenciatura Parcelada em Pedagogia no município de Guiratinga a partir de 1995, atendendo a uma clientela específica, composta de cinquenta professores da rede pública de ensino, todos atuantes em municípios circunvizinhos a Rondonópolis. No mesmo ano, foi dado inicio ao curso de Bacharelado em Ciências Contábeis, no município de Primavera do Leste, atendendo a oitenta alunos daquela região.

As Licenciaturas Parceladas em Pedagogia e Letras, iniciadas em 1996, ano em que ingressaram cento e oitenta alunos, atendiam às demandas dos municípios de Alto Taquari, Campo Verde, Guiratinga, Jaciara, Juscimeira, Paranatinga, Pedra Preta, Poxoréo, Primavera do Leste, São José do Povo e Tesouro, e aulas eram ministradas no próprio *Câmpus*. A instituição conta, hoje, com mais de 17 cursos de ensino superior e dois mestrados.

O curso de Informática desta instituição apresenta uma estrutura curricular formal. O projeto contempla algumas práticas e abordagens de caráter integrado, multi e, sobretudo, transdisciplinar. Não é visível a separação clássica entre o conhecimento estatístico, matemático, químico, físico, sociológico, antropológico, filosófico, jurídico ou artístico.

Quatro ciências alimentaram a construção do Projeto Pedagógico do curso de Informática por meio das suas disciplinas em uma subdivisão de quatro (4) eixos, de acordo com a natureza e enfoque de cada uma.

O primeiro eixo de disciplinas do Curso de Informática contempla aquelas que focalizam a dimensão da formação básica com foco nas Ciências Exatas.

As disciplinas que focalizam a dimensão e as relações operacionais, redes de computadores e sistemas distribuídos na organização do banco de dados encontram-se alocadas no segundo eixo formação tecnológica. Finalmente, o terceiro e o quarto eixo se encarregam das disciplinas de formação complementar e formação humanística.

O pesquisador lecionava, há alguns anos, esta disciplina Probabilidade e Estatística, o discente estuda a Estatística Descritiva e Inferencial com o objetivo de orientá-lo na coleta, no resumo, na apresentação, na análise e na interpretação dos dados estatísticos. Os métodos de estatística inferencial são ferramentas imprescindíveis no teste das hipóteses científicas. Esta disciplina proporciona ao discente uma ampla visão das possibilidades de aplicação da Análise Multivariada à pesquisa do curso de informática.

## 1.2 Problema de pesquisa

O tema da pesquisa foi delimitado de forma a apresentar processo ensino-aprendizagem das disciplinas de Estatística de maneira significativa, uma vez que muitos alunos não entendem o porquê de se estudar Estatística e qual seria sua aplicação prática em sua formação profissional (BRASIL, 1998).

Nessa perspectiva, essa proposta metodológica para aprendizagem é uma alternativa de pesquisa em sala de aula e permite a participação do professor e do aluno. Utilizar-se das práticas pedagógicas da Estatística e da modelagem Matemática é permitir o envolvimento e participação eficaz dos alunos e do professor. Segundo Melo (2007), novas metodologias de ensino não mudam a relação pedagógica entre o aluno e o professor, porém modificam algumas das funções do professor e do aluno.

De acordo com Becker (1997, p.117), "é fato notório que poucos são os alunos que, realmente, aprendem matemática, para não falar do pouco que resta desta aprendizagem nos antigos alunos".

Diante desses fatos, da dificuldade de ensino da Matemática e Estatística e dos problemas em torno da questão da aprendizagem, torna-se fundamental o estudo das possibilidades de compreensão para estudar conteúdos matemáticos e estatísticos e uma interação entre a modelagem Matemática como proposta metodológica. Assim, observa-se a necessidade e o interesse em conhecer as atividades didáticas e a compreensão sociocultural da Matemática (STIELER, 2007). O autor sugere que esta investigação o professor necessita repensar as metodologias do processo de ensino-aprendizagem visando trabalhar a formação continuada, possibilitando ao aluno o aprofundamento na disciplina de acordo com a própria motivação e reflexão dos modelos de concepção e percepção das práticas pedagógicas adotadas em sala de aula, contribuindo com a discussão e lançando novos horizontes sobre a educação.

Percebe-se, então, a importância da implementação da proposta metodológica para a aprendizagem de forma diferenciada, seguindo critérios metodológicos baseados nas teorias e nos métodos vigentes e em critérios técnicos, visando à inter-relação da universidade com a vivência dos alunos e a construção da aprendizagem. Conforme Zabalza (2004), o aluno aprende, quando ele interage com os colegas e o professor, pois "[...] O aprendizado dos alunos depende não apenas deles [...] mas também das condições em que se dá o processo de aprendizagem e da capacidade dos professores para ajudá-los" (ZABALZA, 2004, p. 197 *apud* STIELER, 2007, p. 22).

O autor argumenta que as tendências didáticas da modernidade insistem em orientar o processo de ensino-aprendizagem para a autonomia do sujeito. Propor alternativas e sugestões para que a aprendizagem possa se efetivar, desenvolvendo o espírito crítico, o raciocínio lógico e o mundo de pensar estatístico, acompanhado pelo professor.

Nesta investigação, apresenta-se um estudo que foi desenvolvido com a utilização da metodologia da modelagem Matemática nos cursos que ofertam a disciplina de Estatística durante o 2º semestre letivo de 2007. O ementário estabelecia estudar a natureza da Estatística; Distribuição de frequências; Medidas de tendência central e de dispersão; Noções de probabilidade; Distribuições Teóricas de Probabilidades; Técnicas de amostragem; Estatísticas não Paramétricas.

De acordo, ainda, com o Plano de Ensino, o objetivo desta disciplina é capacitar o aluno a ter uma leitura prática de informações e dados estatísticos, organizar e manipular estes dados, analisar teorias e métodos quantitativos de padrões em Informática, em diferentes níveis de organização, com ênfase em sua aplicação, e, por fim, apresentar resultados.

A metodologia da disciplina é a sistematização do saber dos alunos a respeito dos temas e ampliação dos referencias teóricos por meio de aulas expositivas dialogadas, aulas práticas laboratoriais, grupos de discussão norteados por roteiros e exercícios seguidos de aula expositiva dialogada para esclarecimentos e conclusão final do tema, apresentação de seminário e debates.

Como todas as aulas foram planejadas para esta pesquisa, as avaliações formais previstas pela Instituição foram, também, instrumentos de análise e de total importância no processo de ensino-aprendizagem.

No intuito de analisar a modelagem como proposta metodológica diferenciada para o processo de ensino aprendizagem de Estatística nos cursos de Informática, Matemática, Ciências Contábeis e Enfermagem do Campus Universitário de Rondonópolis/UFMT por meio do diagnóstico de estratégias didáticas, foram enviadas questionários para os alunos e, principalmente, para os professores que lecionam a disciplina de Estatística, buscando caracterizar o processo ensino-aprendizagem.

## 1.3 Pergunta de pesquisa

Para alcançar os objetivos propostos, o presente estudo parte do seguinte questionamento: A modelagem Matemática utilizada como uma metodologia de aprendizagem apresenta estratégia didática que podem desenvolver práticas de atividades para a construção do conhecimento estatístico?

## 1.4 Justificativa

A escolha dessa metodologia permite o contato dos alunos com problemas do cotidiano, como também permite analisar e interpretar as soluções das atividades estatísticas e, dessa forma, compreender o mundo em que estão inseridos. Essa metodologia tem apresentado resultados animadores quanto à motivação, à participação e ao envolvimento dos atores no processo ensino-aprendizagem (STIELER, 2007).

O processo ensino-aprendizagem e a construção do conhecimento estatístico de modo a utilizar estratégias apontam a modelagem Matemática como proposta diferenciada para o ensino de Estatística, com viés interdisciplinar, uma vez que desde sua origem, ela teve esse movimento de dialogar com outras áreas (OLIVEIRA, 1999).

O ensino de Estatística tem sido verdadeiramente interdisciplinar. Alternativa de ensino, com desenvolvimento de atividades que levam o aluno a construir o seu próprio conhecimento por meio de relações concretas e por procedimentos que o valorizam como pessoa.

De acordo ainda com Polya (1978), fazer com que o aluno passe a pensar, é fundamental para a aprendizagem, em qualquer área do conhecimento. Isso é inquestionável, principalmente, quando se trata do conteúdo de Estatística, ramo da Matemática aplicada.

O desenvolvimento do processo de ensino-aprendizagem da Estatística, nas universidades, tem sido alvo de pesquisas em algumas partes do mundo, e muitos pesquisadores publicaram trabalhos a respeito, procurando justificar a relevância do assunto.

Conforme Fonseca (1982), da mesma maneira que:

> [...] ocorre com a maior parte das disciplinas, o estudo da estatística pressupõe uma sequenciação que pauta de uma introdução geral à compreensão de suas finalidades e de seus instrumentos básicos, chegando até a avaliação crítica de suas limitações. Entre esses extremos há todo um conjunto sistematizado de proporções conceituadas e de modelos de aplicações, que traduzem as aquisições teóricas e os resultados das pesquisas que constantemente enriquecem o próprio conteúdo da disciplina.

A utilidade da disciplina, apontada pelos pesquisadores, refere-se à necessidade de que todos os indivíduos têm de dominar alguns conhecimentos de Estatística para atuarem na sociedade. São conhecimentos fundamentais para analisar índices do custo de vida, para realizar levantamentos de preços, escolherem amostras e outras situações do cotidiano.

Do ponto de vista acadêmico, espera-se que o estudo contribua para melhorar as discussões relacionadas ao processo ensino-aprendizagem, há que se ressaltar ainda que Estatística seja uma dos temas transversais no ensino fundamental e médio a informar noções de raciocínio probabilístico e estatístico, nos Parâmetros Curriculares Nacionais (BRASIL, 1999), no entanto, verifica-se uma escassez de estudos relacionados ao ensino-aprendizagem com modelagem Matemática.

Debater a modelagem Matemática é uma iniciativa positiva, pois precisa encontrar alternativa para estabelecer uma relação harmoniosa com as estratégias didáticas e as práticas pedagógicas, mas, sobretudo, faz-se necessária uma relação harmoniosa com a Estatística (VENDRAMINI, 2004).

## 1.5 Minha história

Sou professor do Departamento de Matemática/ICEN/CUR da Universidade Federal de Mato Grosso (UFMT). Neste Departamento, decidiu-se que todo docente vinculado a ele pode ministrar aula de qualquer uma das disciplinas que está sob sua responsabilidade.

Assim, além de responsáveis pela maior parte do curso de Matemática, nas modalidades Bacharelado e Licenciatura, nossa maior demanda vem dos cursos, que têm, em suas grades curriculares, a disciplina de Estatística, a saber: Ciências Contábeis, Biologia, Engenharia Agrícola e Ambiental, Engenharia Mecânica, Pedagogia, Biologia, Informática, Biblioteconomia, Psicologia e Enfermagem.

A minha primeira experiência com o ensino de Estatística ocorreu em 1988, numa turma de secretariado, na Escola Estadual de 1º e 2º graus Raimundo Pinheiro em Cuiabá-MT. As aulas de Estatística tinham como base o estudo da estatística descritiva com fórmulas e cálculos, seguindo o modelo de aulas que recebi na graduação.

Trabalhei com a disciplina Estatística, desde quando assumi o cargo de professor do Departamento de Matemática. Desde essa época, minha preocupação com a disciplina foi aumentando, pois era fácil perceber a existência de um grande número de problemas envolvendo os alunos, os professores, o conteúdo trabalhado, a metodologia de ensino usada etc. Os problemas se manifestavam de várias formas, como por exemplo, no grande número de reprovações e desistências na disciplina, no desinteresse dos alunos e na falta de ânimo dos professores.

As aulas expositivas sempre me acompanharam, mas, mesmo com pouco embasamento teórico, procurei algumas alternativas metodológicas, como História da Matemática, jogos, materiais concretos, trabalhos em grupos, entre outras atividades, com intuito de melhorar a aprendizagem.

Pude constatar, portanto, que tratar os problemas simplesmente como aplicações de conteúdo, com os objetivos citados acima, não implicava fazer com que os alunos vissem sentido em tudo àquilo que estavam estudando. Eles apenas tinham a oportunidade de ver algumas ilustrações daquele conteúdo em problemas que raramente estavam relacionados com a futura profissão. Além disso, perguntava-me se os alunos estavam “aprendendo”, realmente, o conteúdo da disciplina, qualquer que fosse o significado dessas “aprendizagens”.

Minha tarefa voltou-se para a educação de alunos nas mais diferentes situações, quer seja com os desmotivados com o aprender, com aqueles prestes a engrossar a ala dos excluídos da escola, aqueles com dificuldades de aprendizagem ou com tantas outras situações de exclusão que permeiam o cotidiano do professor. Destaco que sempre procurei transformar minhas aulas num ambiente agradável e propício para aprendê-lo. Mesmo assim, tenho clareza do quanto ainda devo acrescentar à minha caminhada como professor.

Igualmente, minha caminhada no ensino superior foi permeada de dificuldades, de barreiras criadas por profissionais conservadores e por outros motivos. No entanto, reconheço que, cada vez mais, aumentava o desejo de estudar e de participar de cursos, seminários, encontros e palestras, no intuito de melhorar minha prática pedagógica, de diversificar metodologias e de ampliar horizontes. O contato com alguns pensadores e escritores trouxe novo (re) dimensionamento ao meu trabalho, além de incentivo e apoio a alguns projetos em nível de universidade e, também, a buscar especialização capaz de acrescentar conhecimentos e outras percepções à minha vida docente.

Por termos vivenciado algumas situações de orientação de ensino sobre a disciplina, principalmente quando éramos Coordenador de Ensino e Graduação dos Cursos de Matemática e Informática do Departamento de Matemática do *Campus* Universitário de Rondonópolis-MT, percebemos as contribuições do estudo da Estatística à formação do aluno. Pudemos verificar, também, que não é suficiente apenas o estudo desse assunto. Ainda assim, ele proporciona possibilidades para esse desenvolvimento, que se efetivará, caso haja uma prática pedagógica coerente com o objeto de estudo. Acredita-se que tais assuntos sejam tão importantes quanto o estudo da Estatística da Modelagem Matemática, que, trabalhado significativamente, também contribui com essa formação.

Não se pode negar que a Estatística representa um entrave imenso para um número significativo de alunos. É um mito que continua vivo no senso comum; basta observar as referências feitas à disciplina: "odeio Matemática", "tenho verdadeiro pavor da Matemática", "a ciência dos números é assustadora", "a Matemática é o terror dos jovens, adultos e crianças" e outros tantos pronunciamentos.

Após essas reflexões, reafirmo os motivos que determinaram a intenção de pesquisar, como tema desta tese em Ciências da Educação, a aplicação de uma proposta alternativa para o ensino da Estatística: uma proposta diferenciada de ensino, que faculta ao aluno ser agente na construção do conhecimento, superando, com motivação e descontração, as dificuldades que se apresentam.

Assim, com várias inquietações sobre o ensino e a aprendizagem de Estatística, em especial com aquelas que diziam respeito aos, tão comuns, problemas de aplicações. A expectativa foi à busca de alternativas para trabalhar com a matemática e a estatística de forma contextualizada e significativa de proposta metodológica para aprendizagem, por isso fiz uma revisão bibliográfica sobre o assunto, que passo a descrever no capítulo 2.

## 1.6 Objetivos da pesquisa

### 1.6.1 Geral

Compreender como a utilização da Modelagem Matemática pode contribuir para a melhoria do processo de ensino-aprendizagem e a construção do conhecimento estatístico, estudados nos curso que ofertam a disciplina de Estatística, do *Campus* de Rondonópolis, da Universidade Federal de Mato Grosso.

### 1.6.2 Específicos

a) Caracterizar as estratégias didáticas por meio de uma metodologia proposta para o ensino de estatística e conceitos matemáticos básicos necessários;

b) Identificar os tipos de estratégia didática nas resoluções de problemas de Estatística, estabelecendo um perfil de aprendizagem dos alunos e professores, de modo a motivar e dar significado e entendimento aos assuntos estudados;

c) Analisar as soluções das situações-problema propostas, com vistas à formação do profissional, habilitado para tratar as questões estatísticas com as demais áreas do conhecimento.

Para alcançar os objetivos definidos, de acordo com a metodologia elaborada para o estudo, a pesquisa está dividida em cinco capítulos: na primeira parte, enfatizou-se a temática, considerando sua visão introdutória, o problema e os objetivos da pesquisa e sua importância na investigação. Tratou-se da apresentação de conceitos e significados importantes, os quais permitiram a redefinição do objeto e do papel da Estatística em relação à Modelagem Matemática.

O capítulo segundo traz diferentes abordagens da construção teórica, bem como as estratégias didáticas, buscando a interdisciplinaridade com as diversas ciências que abordam a temática em questão.

O capítulo terceiro apresenta tanto os procedimentos metodológicos, como a estrutura conceitual para abordagem sistêmica e multidisciplinar aplicada ao estudo das disciplinas de Estatística dos cursos que ofertam a referida disciplina no Campus Universitário de Rondonópolis/UFMT.

O capítulo quarto tece uma discussão sobre as atividades realizadas e a análise dos resultados observados em relação à aprendizagem dos alunos no decorrer dos trabalhos com atividades de Estatística e Modelagem Matemática. A sinergia entre esses elementos proporciona condições favoráveis para o desenvolvimento da estatística.

Os capítulos que são apresentados nesta tese contemplam estudos significativos da temática teórica e empírica. Este estudo é um convite essencial na atual conjuntura para a tomada de decisões, alerta a respeito da importância da Modelagem Matemática e sinaliza quais preocupações devem direcionar seu processo de implementação.

# 2 FUNDAMENTAÇÃO TEÓRICA

## 2.1 Ensino de estatística

O ensino de estatística, praticado de forma diferenciada em um curso do ensino superior, deve ser desenvolvido de maneira diferenciada, procurando articular a teoria com a prática.

Essa compreensão é reafirmada por Silva (2000), quando ela coloca que ensinar sem articulação da prática investigativa pode resultar atitude negativa de desempenho nas atividades de aprendizagem. A autora investigou 643 estudantes das áreas de ciências humanas, biológicas e exatas de uma universidade particular, cujo resultado indicou que os alunos apresentavam atitudes negativas em relação à Estatística e apontou necessidades em relação à disciplina. Aqui se revela uma das causas do baixo desempenho nesse componente curricular, pois a autora mostra que existe correlação positiva e significativa entre as atitudes dos alunos em relação à Matemática e a nota final da disciplina Estatística, ou seja, os estudos negativos com relação à Matemática são transferidos para a Estatística.

Os autores Mcleod e Adams (1989, *apud* NOVAES, 2004, p. 24-28) definem a ansiedade como um componente emocional, sentido na presença do objeto, no momento da experiência com este, durante alguns segundos, minutos ou no máximo horas. As experiências educacionais que vão se acumulando em relação a um objeto, da mesma forma, podem desenvolver atitudes em relação ao mesmo. As atitudes são menos intensas que as emoções, porém mais duradouras. Segundo Ragazzi (1976) e Silva (2000), também conforme citado por Novaes (2004), a atitude é a prontidão de uma pessoa para responder a determinados objetos de uma maneira favorável ou desfavorável; ela consiste em.

> [...] uma variável que pode influenciar a aprendizagem de Estatística, levando o aluno a ter interesse, querer aprender mais e estudar quando apresenta atitudes positivas em relação à disciplina, mas também pode tornar os alunos nervosos, ansiosos, com medo e sem interesse de aprendê-la, quando esse aluno apresenta atitudes negativas em relação a ela (SILVA, 2000, p.15).

De acordo com os referidoS estudos, os alunos associam as dificuldades que tiveram na sua vida escolar anterior, com a Matemática, às atuais, com a disciplina Estatística.

Em seu trabalho, Silva (2000) sugere que os alunos precisam desenvolver atitudes positivas em relação à Estatística, como condição para obter melhores resultados. Pode-se interpretar essa afirmação como a necessidade de uma construção de situações didático-pedagógicas adequadas para a construção dos processos de aprendizagem de Estatística pelos alunos. Moore (1997), citado por Silva (2000), sugere que o ensino dessa disciplina deveria apresentar problemas com dados reais, concentrando-se em aspectos que não necessitem de memorização, mas sim de interpretação, em estratégias para uma exploração efetiva de dados, com um diagnóstico básico preliminar para a interferência.

Os resultados apontados por essas pesquisas e as sugestões de organização das atividades com alunos no processo de aprendizagem de Estatística, apresentadas nesses trabalhos, despertaram este investigador para a organização de uma situação-problema que o possa auxiliar a detectar erros, compreender a origem desses erros, bem como fazer um estudo de novas propostas de formas de abordagem do conteúdo da disciplina.

Outra pesquisadora muito importante da didática da estatística é Carmem Batanero, da Universidade de Granada, Espanha. Investigações realizadas por diversos autores sobre erros e dificuldades na compreensão do processo ensino-aprendizagem de Estatística são analisadas por essa estudiosa. Batanero (2000a) afirma que grande parte da investigação teórica e experimental, realizada atualmente em didática da Matemática, mostra que os alunos produzem respostas erradas ou simplesmente não são capazes de dar nenhuma resposta quando se pede para que realizem outras tarefas. Nos casos em que não se trata de distorção, os professores acreditam que a tarefa é muito difícil para o aluno. Porém esses erros não acontecem de forma aleatória, imprevisível: com frequência, é possível encontrar regularidades, associações com variáveis próprias das atividades propostas, dos sujeitos ou de circunstâncias presentes ou passadas.

A formação de professores neste âmbito específico é quase inexistente. Só recentemente, segundo Batanero (2001a), o ensino de Estatística foi inserido em alguns cursos de licenciatura das universidades na Espanha e muito ainda não o contemplam. No Brasil, quando o contemplam, nem sempre é no enfoque crítico.

Outra causa de erro identificado por Batanero (2001a) refere-se ao obstáculo cognitivo descrito por Braussen (1983), o qual é classificado em obstáculos antagônicos: quando o aluno ainda não tem maturidade para entender determinado conceito; quando a forma de ensinar, ao invés de ajudar, complica a compreensão (didático-metodológico); ou quando se referem às dificuldades históricas encontradas no desenvolvimento do próprio conceito (epistemologias).

Batanero (1994) destacou alguns problemas com relação ao ensino da Estatística: a) estrutura curricular: o que ensinar e quando ensinar; b) material didático (livros de textos e softwares educacionais); c) avaliação; d) formação de professores e e) crenças e atitudes dos professores. Em relação à aprendizagem dos alunos, a autora apresenta três dados: a) o significado dos conceitos e procedimentos estatísticos, incluindo suas propriedades, problemas relacionados, representações e instrumentos, ou seja, a epistemologia dos conceitos (o que são, como surgiram, que problemas permitem solucionar, que dificuldades são previsíveis em sua aprendizagem); b) capacidades cognitivas dos alunos (nível de desenvolvimento do pensamento estatístico dos alunos); e c) aspecto afetivo (atitudes e sentimentos em relação à Estatística).

No que se refere, especificamente, ao ensino superior, Batanero (1994) apresenta três argumentos: a) com relação à natureza da Estatística, verificou-se que ela lida com conceitos abstratos, usa notações e terminologias complexas e seus problemas são abertos[1], são problemas do mundo real que obrigam o estatístico a tomar decisões em situações de incerteza; b) Quanto à linguagem, a Estatística tem uma fundamentação teórica na Matemática, mais notadamente na Teoria das Probabilidades, embora nem sempre exista a necessidade de uma Matemática avançada; para os cursos de ciências humanas, por exemplo, a dificuldade de seu aprendizado é maior; c) Com relação à estrutura curricular, não é dada à Estatística, no conjunto das disciplinas que compõem os cursos de ciências humanas, a carga horária necessária para o seu bom desenvolvimento.

Outras dificuldades encontradas, descritas por Batanero, referem-se à falta de conhecimentos básicos necessários à compreensão correta de um conceito ou procedimento.

---

1 Problemas em aberto são aqueles que podem não ter solução ou uma única solução.

Os resultados de um desses estudos sobre o raciocínio estatístico de 325 universitários indicaram índices de dificuldade elevados para a maioria dos itens de uma prova que se referia à interpretação de dados apresentados em tabela (VENDRAMINI, SILVA; CANALE, 2003). Os itens referentes à interpretação de dados expostos em gráficos apresentaram índices de dificuldade mais baixos daqueles apresentados em tabelas. Uma das explicações para esse resultado é o fato de a tabela apresentada na prova ter sido de dupla entrada e exigir a comparação das informações nela contidas para se chegar às respostas corretas das questões propostas.

Em outro estudo realizado com 447 estudantes, 27,5% do ensino fundamental (5ª a 8ª); 51,5% do ensino médio; 21,0% do superior foram detectadas dificuldades associadas a frequências relativas. Os resultados revelaram que os itens mais difíceis da prova (total de 30 itens), com menos de 15% de acertos, referiam-se direta ou indiretamente a operações com números decimais ou fracionários (BUENO et al*., 2003).*

De acordo com Batanero (1994), a importância de uma Didática da Estatística, que seria o estudo do ensino-aprendizagem da Estatística envolvendo e relacionando não só os conhecimentos da estrutura da Estatística, como também os da Matemática, Psicologia, está na formação do educador. Para a autora, várias dificuldades encontradas no ensino de Estatística dizem respeito, por exemplo, a professores de Matemática que ensinam Estatística no ensino básico e em outros cursos em geral. Neste sentido, o presente trabalho priorizou o processo ensino-aprendizagem da Estatística no curso de Informática, considerando quais as práticas pedagógicas do ensino de Estatística com os alunos.

Acredita que iniciar o ensino de estatística com problemas do dia a dia pode facilitar o entendimento dos conceitos, familiarizarem o aluno com a situação e prepará-los então para a introdução dos modelos estatísticos. o aprendizado pela experiência aplicada em sua área é a forma mais correta a ser adotada para obter melhores resultados com relação à atitude dos alunos em relação à disciplina estatística. A atitude é a prontidão de uma pessoa para responder a determinado objeto de maneira favorável ou desfavorável (CARZOLA, 1999).

Na busca pelo desenvolvimento dos alunos com novas habilidades com o uso da estatística para suas áreas de interesse através do projeto e a atitude positiva em relação à estatística, aplica-se essa metodologia para se trabalhar na sala de aula com os conceitos que pretendem ser usados ao longo de suas carreiras.

### 2.1.1 O Ensino da Estatística e da Probabilidade

Com o objetivo de sabermos as recomendações do MEC para o Ensino da Estatística e Probabilidade, tanto no Ensino Fundamental quanto no Ensino Médio e Superior, fizemos uma busca pelos problemas pedagógicos de pesquisas sobre o ensino-aprendizagem da Estatística e também por professores que lecionam essa disciplina. De acordo com os Parâmetros Curriculares Nacionais[2] (PCN), podem-se destacar os seguintes objetivos:

> Coleta organização de dados e utilização de recursos visuais adequados (fluxogramas, tabelas e gráficos) para sintetizá-los, comunicá-los e permitir a elaboração de conclusões.
> Leitura e interpretação de dados expressos em tabelas e gráficos.
> Compreensão do significado da média aritmética como um indicador da tendência de uma pesquisa.
> Representação e contagem dos casos possíveis em situações combinatórias. (BRASIL, 1997, p. 61)

Podemos complementar as afirmações anteriores com algumas pesquisas na área de Estatística. Segundo Vendramini, Silva e Canale (2003), os resultados de estudos sobre o raciocínio estatístico de 325 universitários indicam índices de dificuldade elevados para a maioria dos itens de uma prova que se refere à interpretação de dados apresentados em tabela.

Os itens referentes à interpretação de dados apresentados em gráficos apresentam índices de dificuldade mais baixos daqueles apresentados em tabelas. Uma das explicações para esse resultado é o fato de a tabela presente na prova ter sido de dupla entrada e exigir a comparação das informações nela contida para se chegar às respostas corretas das questões propostas.

Segundo, ainda, Bueno et al. (2003), em outro estudo realizado com 447 estudantes, 27,5% do Ensino Fundamental (5ª a 8ª); 51,5% do Ensino Médio; 21,0% do Ensino Superior, foram detectados dificuldades associadas às frequências relativas. Os resultados revelaram que os itens com maior dificuldade na prova (total de 30 itens), com menos de 15% de acertos, referiam-se direta ou indiretamente a operações com números decimais ou fracionários.

---

2 PCN’s – Documentos que explicitam as habilidades básicas das competências específicas de cada área, em atendimento ao estabelecido na LDB nº 9394/96, lei de diretrizes e bases da Educação Nacional.

Procuramos relatar, também, pontos importantes encontrados por vários pesquisadores na área de Estatística e Probabilidade, assim como indicar algumas instituições e grupos que têm se dedicado a esta, com o intuito de facilitar o caminho para futuros pesquisadores, descritos a seguir.

Um desses grupos se encontra na Espanha, na Faculdade de Ciências e da Educação na Universidade de Granada, com o qual se pode entrar em contato através da Internet pelo este endereço eletrônico (http://www.urg.es/local/batanero/). Como um dos seus membros efetivos, a Dra. Carmem Batanero Bernabeu, já esteve no Brasil, ministrando palestras na Universidade de São Paulo (USP) e cursos na Universidade de Campinas (UNICAMP) em setembro de 1999, é importante falarmos sobre seu trabalho.

O grupo de Batanero destaca a necessidade de se refletir sobre os principais objetivos do ensino da estatística, ao dizer:

> que os alunos cheguem a compreender e apreciar o papel da estatística e da probabilidade na sociedade, incluindo campos de aplicação e o modo como a estatística contribui para o seu desenvolvimento; que os alunos cheguem a compreender e a valorizar o método estatístico, isto é, a classe de perguntas que o uso inteligente da estatística pode responder as formas básicas de raciocínio probabilístico e estatístico, seu potencial e limitações (BATANERO, 2001b, p. 50).

Outro grupo que destacamos, encontra-se no Brasil, na Faculdade de Educação da Universidade Estadual de Campinas (UNICAMP). Trata-se do Círculo de Estudos Memória e Pesquisa em Educação Matemática (CEMPEM). Essa Faculdade é responsável pelo ensino de pós-graduação para estudantes, principalmente de Matemática e Pedagogia.

Encontra-se, no CEMPEM, um subgrupo que trabalha com a Estatística, a Probabilidade e a Análise Combinatória. Destacamos alguns dos seus membros: Professores Dario Fiorentine, Dione Luchessi de Carvalho, Anna Regina Lanner de Moura, da Faculdade de Educação, e Vera Lúcia Figueiredo e João Frederico Meyer, do Instituto de Estatística e Ciência da Computação.

Wada (1996), em seu trabalho de doutorado, coletou, junto aos professores que lecionam a Estatística na Universidade Estadual de Campinas (UNICAMP), opiniões com relação aos objetivos da disciplina de Probabilidade e Estatística em diversos cursos, os quais são descritos a seguir:

> O grupo de quatorze professores da UNICAMP, que respondeu o questionário espera que os alunos ao saírem da Universidade atuem profissionalmente prestando consultoria a profissionais e empresas de outras áreas. Devem exercer atividades ligadas à área de Qualidade, atuando em indústrias, e podem também atuar em serviço como bancos, secretarias etc. (WADA, 1996, p.101).

Nesse objetivo, destaca-se a necessidade de dominar cálculos com as operações matemáticas fundamentais e de trabalhar com dados de organização, apresentação, exploração e comparações.

Na opinião de um dos docentes, o aluno deve ser

> Exposto aos fundamentos básicos da metodologia estatística, iniciando o seu aprendizado. Segundo o docente, a meta de se formar um estatístico pela universidade, ainda que oficialmente, é difícil, dada a duração de quatro anos do curso de graduação (WADA, 1996, p. 102).

Observa-se uma grande preocupação em realizar atividades com muitos exercícios com a finalidade da fixação dos conteúdos. Ainda sobre os conteúdos e habilidades,

> Para os professores, os alunos, de um modo geral, devem saber usar a metodologia estatística na prática, para diagnosticar e resolver problemas, o que significa saber analisar e inferir resultados em distintas áreas do conhecimento, atuando interdisciplinar mente na profissão. Consideram também a necessidade de conhecimentos teóricos a respeito dos princípios e técnicas da metodologia estatística (WADA, 1996, p.103).

Destaca-se também a valorização da Estatística como disciplina a serviço de outras áreas da ciência. De acordo com Wada (1996), o curso de Estatística deve desenvolver noções sobre população, amostra, análise e inferência estatística; tentar transmitir para esses alunos a importância da utilidade da disciplina para o desenvolvimento de seus futuros trabalhos; fazer uma espécie de "marketing" da profissão do Estatístico, transmitindo à importância de quem deve fazer as análises mais complexas do que é o Estatístico.

Os alunos deveriam ter uma visão bem clara de sua própria área de interesse, para compreender a utilidade da ferramenta estatística na sua atividade, deveriam, dessa forma, estar nos últimos anos de seu curso de graduação para poder realizar um curso de Estatística. Sobre os conhecimentos a serem adquiridos depende do tipo de curso, isto é, na área de Ciências Exatas, poderia ser oferecido um curso que incluísse maior rigor formal nos conceitos e metodologia a serem discutidos. Em outros casos, como na área de Ciências Humanas, há maior dificuldade de oferecer um curso com maior rigor formal, mas os conceitos e as principais ideias metodológicas poderiam ser transmitidos de forma mais intuitiva e baseados em exemplos concretos (WADA, 1996).

A ênfase é, sobretudo, em relação às funções da Estatística as que poderiam despertar o interesse do aluno, pois

> Os conhecimentos a serem adquiridos pelos alunos são: 'Probabilidade, Inferência, Amostragem, Análise de Dados Categóricos, Análise de Regressão, Análise de Sobrevivência, Análise Multivariada entre as principais'[4]. Na opinião do mesmo docente, os alunos devem familiarizar-se com 'alguma linguagem de programação como Turbo-Pascal, uso de planilhas como Lotus-123 e pacotes estatísticos de ampla difusão como SAS ou SPSS e outros pacotes de uso mais específicos como GLIM, STATA, Harvard-Graphics, etc.' [4] (WADA, 1996, p. 105).

A importância da Estatística em outras áreas de conhecimento deve ser relacionada não apenas saber efetuar cálculos simples, mas também saber fazer uso do computador.

No tocante aos objetivos da disciplina, há dois imprescindíveis: (1) Informar o aluno a respeito das possibilidades de que a Estatística oferece em quase todo tipo de problemas e quais são suas ferramentas básicas; (2) Capacitar o aluno de tal forma que ele consiga resolver problemas simples e, em caso de problemas complexos, mostrarem a necessidade de recorrer a um assessoramento de um estatístico, oferecendo assim um diálogo de bom nível com um mínimo de linguagem em comum. (WADA, 1996).

Destaca-se a importância da Estatística em outras áreas de atuação, permitindo ao aprendiz uma boa comunicação com o Estatístico. O objetivo da Estatística deveria ser o de mostrar a existência de um método científico no planejamento, execução e análise de qualquer experimento. Os alunos deveriam aprender a conviver com as incertezas, saber interpretá-las e, sempre que possível, mantê-las sob controle. Para aprender é primordial que os professores dos cursos de origem tenham conhecimento da Estatística e que os professores de Estatística aprendam um pouco a linguagem e as necessidades dos outros cursos. (WADA, 1996, p.107).

Faz-se relevante o método estatístico e a adequação dos conteúdos de acordo com a natureza de cada curso.

Segundo *o National Council of Teachers of Mathematics* (NCTM) (1989, 1995), tem-se enfatizado a necessidade atual de desenvolver nos estudantes habilidades[3] que os auxiliem não só na interpretação e crítica de informações, como também na solução de problemas matemáticos. A Estatística e Probabilidade são tópicos apropriados para serem trabalhados na Matemática do currículo escolar porque: (1) mostram aplicações importantes da Matemática em todos os níveis; (2) suprem métodos para tratar com a incerteza; (3) ensinam a lidar com bons e maus argumentos estatísticos, com os quais o sujeito se defronta com frequência; (4) auxiliam os consumidores a distinguirem o uso de procedimentos estatísticos e consistentes e inconsistentes; e (5) são tópicos altamente motivadores e de grande interesse para a maioria dos estudantes.

A *American Statistical Association* (ASA) estabelece os seguintes objetivos com relação à Estatística e Probabilidade: (1) possibilitar a compreensão dos dados, o que significam e qual a sua origem; (2) usar técnicas gráficas e numéricas simples para sintetizá-los; (3) criar um modelo matemático para descrevê-los e (4) fazer inferências sobre uma população, usando amostras oriundas da mesma.

E, apesar da expectativa em torno dos novos recursos técnicos das tecnologias da comunicação e da informação, as práticas nas salas de aula não superam o modelo tradicional de ensino traduzido, tanto no “apego” ao livro didático como norteador principal dos

3 O conceito de habilidade é amplamente discutido no trabalho de Vendramini (2000).

conteúdos curriculares, como na ciência empirista[4] e positivista[5] como método. Tais modelos são calcados em mecanismos e estratégias.

Segundo a LDB Nº. 9394/96 e a Resolução CNE/C nº. 2 de 27/08/2004, as propostas pedagógicas de aprendizagem dos cursos deverão contemplar, então, nos seus currículos atividades de ensino-aprendizagem que possibilitem não apenas, sólida formação nas áreas específicas, mas também a participação crítica e comprometida com a construção. Assim, a implantação e a validação de propostas pedagógicas que visem garantir a formação dos estudantes e a qualidade do ensino.

Segundo ainda os PCN's, técnicas e raciocínios estatísticos e probabilísticos são, sem dúvida, instrumentos tanto das Ciências da Natureza, quanto das Ciências Humanas, sendo destacada, portanto, a importância de uma cuidadosa abordagem dos conteúdos de estatística e probabilidade no Ensino Fundamental, Médio e Superior.

Os tópicos relativos à Estatística se encontram no "Bloco Tratamento da Informação" (BRASIL, 1997a e 1998) http://www.mec.gov.br/sef/estrut2/pcn/pdf/livro03.pdf e http://www.mec.gov.br/sef/estrut2/pcn/pdf/matematica.pdf), que é um dos cinco blocos de conteúdos conceituais e procedimentais para o ensino da Matemática e tem um destaque especial nos Temas Transversais (BRASIL, 1997b) (http://www.mec.gov.br/sef/sef/pcn1a4.shtm). Os tópicos incluem a leitura e interpretação de informações estatísticas, coleta, organização, resumo e apresentação de informações,

4 Os empiristas afirmam que a razão, a verdade e as ideias racionais são adquiridas através da experiência. Segundo a concepção empirista, antes da experiência, a razão é como uma "folha em branco", uma "tabula rasa". Para o empirismo, os conhecimentos começam com a experiência dos sentidos, ou seja, com as sensações. As percepções se dão por combinação ou associação. A causa da associação das percepções é a repetição, as ideias trazidas pela experiência – sensação, percepção e hábito – são levadas à memória e de lá a razão as apanha para formar os pensamentos. Os conhecidos e famosos filósofos empiristas são os filósofos ingleses do século XVI ao XVII, Francis Bacon, John Locke, George Bekley e David Hume. (Cf. Chauí, 1998).

5 O positivismo, fundado por Augusto Comte no século XIX, distingue três etapas progressivas na humanidade, as quais são denominadas de Lei dos três estados. Essas leis marcaram as fases da evolução do pensar humano, indo do Teológico, ao Metafísico e à ciência Positiva, ponto final do progresso humano. Comte enfatiza a idéia do homem como ser social e propõe o estudo científico da sociedade. Elabora uma disciplina para estudar os fatos sociais, a Sociologia, que num primeiro momento denominou de Física Social. Os fatos humanos devem ser estudados através dos procedimentos métodos e técnicas empregadas pela ciência da Natureza. A concepção positivista é uma das correntes mais influentes nas Ciências Humanas em todo o século XX. Dessa forma, a Psicologia positiva afirma que o objeto de estudo é o psiquismo enquanto comportamento observável que pode ser tratado a partir do método experimental das Ciências Naturais. (os Pensadores, 1996 – Augusto Comte).

construção de tabelas e gráficos, cálculo e interpretação de medidas de tendência central e de dispersão, bem como os rudimentos da teoria de probabilidades. Os PCN's sugerem que se inicie esse trabalho desde a Educação Infantil. Segundo esses parâmetros, ao ensinar matemática nas séries iniciais do Ensino Fundamental (1ª a 4ª séries), deve-se despertar na criança o espírito de investigação e organização de dados, buscando desenvolver habilidades de leitura e interpretação de informação já organizadas em gráficos e tabelas e de produção de textos para a sua interpretação.

### 2.1.2 Material Didático "Oficina de Estatística"

Apresentamos algumas oficinas já cadastradas no Núcleo de Apoio à Pesquisa do Instituto de Ciências Exatas e Naturais do *Campus* Universitário de Rondonópolis – NAP/ICEN/CUR, a saber: (a) Projeto da Cesta Básica dos principais supermercados de Rondonópolis; (b) Criação e Desenvolvimento de um Laboratório Virtual de Estatística; (c) Oficina de formação à distância para o ensino da Estatística, etc. O núcleo foi criado com o objetivo de democratizar o acesso das atividades de Pesquisa do instituto, desenvolvendo estudos nas diversas áreas abrangidas por este.

A organização de pesquisa em Núcleo apresenta-se como uma possível solução para o trabalho individual e isolado que tem caracterizado a produção da pesquisa. O trabalho solitário tem gerado duplicidade de esforços, desconhecimento de outros trabalhos semelhantes e desinformação entre os pesquisadores de uma unidade ou comunidade científica. Por outro lado, servirá de intermediário na identificação de problemas práticos que possam vir a ser solucionados através de ferramentas quantitativas para apoio à decisão, notadamente, mas não apenas, ferramentas estatísticas. Será oferecida assessoria técnica tanto para a comunidade do Campus quanto externamente, o que propiciará convênios entre a Universidade e Empresas, e/ou outras instituições.

O núcleo de pesquisa será um órgão complementar, com propostas interdisciplinares, destinados a coordenar, apoiar, planejar, elaborar e executar programas e projetos relativos à pesquisa em áreas determinadas, sempre com características multi e interdisciplinares.

Com o intuito de facilitar o caminho para futuros pesquisadores, incorporamos também ao núcleo de pesquisa um artigo publicado pela Sociedade Brasileira de Educação Matemática intitulado "O Ensino de Estatística no Brasil" (CARZOLA, 2009), de Irene Mauricio Cazorla, da Universidade Estadual de Santa Cruz-UESC, Brasil, conforme relato a seguir.

A Educação Estatística no Brasil tem seu marco histórico na Conferência Internacional "Experiências e Expectativas do Ensino de Estatística - Desafios para o Século XXI", realizado na Universidade Federal de Santa Catarina (http://www.inf.ufsc.br/cee/) (UFSC, 1999) e começa a tomar forma, enquanto área de pesquisa, com tendência crescente e perspectivas de consolidação.

A história da Estatística está associada à história do Instituto Brasileiro de Geografia e Estatística – IBGE, cujas raízes foram fincadas ainda durante o Império. Seu ensino remonta ao final do século XVIII, ligado ao cálculo de probabilidades, destinado à formação de engenheiros militares, conforme a Associação Brasileira de Estatística - ABE (ABE, 2005).

Aos poucos, as disciplinas de Estatística foram sendo incorporadas nos diversos cursos de Agronomia, Medicina, Ciências Sociais dentre outros. Essas disciplinas, conhecidas como disciplinas de serviço (WADA, 1996), têm como objetivo instrumentalizar diversos profissionais (usuários) para o uso adequado das ferramentas estatísticas.

O primeiro curso de graduação, Bacharelado em Estatística, foi criado na UFRJ, em 1946 e, hoje, segundo o MEC (Brasil, 2005b), o Brasil conta com 27 cursos de bacharelado credenciados (http://www.educacaosuperior.inep.gov.br/), sete cursos de mestrado (USP, UNICAMP, UFSCar, UFMG, UFRJ, UFPE e UFRN) e cinco de doutorado (USP, UNICAMP, UFMG, UFRJ e UFSCar), (http://www1.capes.gov.br/Scripts/Avaliacao/MeDoReconhecidos/Area/GArea.asp).

Observa-se que existem dois programas de Mestrado e Doutorado em Agronomia, um com ênfase em Estatística e Experimentação Agronômica, da Escola Superior de Agronomia Luiz de Queiroz - ESALQ, da USP (http://www.esalq.usp.br/pg/11134.htm) e, o outro, em Estatística e Experimentação Agropecuária da Universidade Federal de Lavras – UFLA (http://www.prpg.ufla.br/estatistica/est_index.htm).

Na Educação Básica, os tópicos de Estatística fazem parte da disciplina de Matemática e, antes dos PCN's, era um dos últimos tópicos do livro-texto, ou seja, quase nunca ensinados (PANAINO, 1998). Hoje essa situação mudou substancialmente. Segundo Cazorla (2002), os PCN's enfatizam a necessidade dos sujeitos serem capazes de comunicar-se, solucionar problemas, tomar decisões, fazer inferências, para agir como consumidores prudentes ou para tomar decisões em suas vidas pessoais e profissionais. Devem-se desenvolver atitudes positivas em relação à Estatística, para que os sujeitos possam "compreender a importância da Estatística na atividade humana e de que ela pode induzir a erros de julgamento, pela manipulação de dados e pela apresentação incorreta das informações (ausência da frequência relativa, gráficos, escalas inadequadas)".

A necessidade de dar respostas aos problemas enfrentados no ensino de conceitos e procedimentos estatísticos deve-se, principalmente, à oficialização de seu ensino na Educação Básica, através dos Parâmetros Curriculares Nacionais (http://www.mec.gov.br/sef/relat/gestao.shtm) (BRASIL, 2005a). Nos eventos nacionais e regionais, ligados à Educação Matemática ou Estatística, observa-se um número crescente de professores de Matemática da Educação Básica que procuram minicursos, oficinas, relatos de experiências, a fim de encontrar material e metodologias que lhes permitam trabalhar esses conceitos e procedimentos, tendo em vista a lacuna na formação inicial. Contudo, os resultados das pesquisas relativas ao ensino da Estatística se encontram espalhados na comunidade acadêmica.

Os dados coletados na internet foram sobre os cursos de bacharelado em Estatística no portal do Ministério de Educação e Desporto - MEC (http://www.educacaosuperior.inep.gov.br/curso.stm) (Brasil, 2005b); sobre os cursos de mestrado e doutorado do portal da Coordenação de Aperfeiçoamento de Pessoal de Nível Superior – CAPES (http://www1.capes.gov.br/Scripts/Avaliacao/MeDoReconhecidos/Area/GArea.asp) (CAPES, 2005); sobre as universidades, dissertações e teses, dos portais dessas instituições; dos pesquisadores e grupos de pesquisa na Plataforma Lattes (http://lattes.cnpq.br/), através do Sistema Eletrônico de Currículos (http://buscatextual.cnpq.br/buscatextual/index.jsp), que coleta informações curriculares dos pesquisadores num formato padrão e do Diretório dos Grupos de Pesquisa no Brasil (http://dgp.cnpq.br/buscaoperacional/), que contém os recursos humanos engajados no grupo, as linhas de pesquisa em andamento, as especialidades do conhecimento, os cursos de mestrado e doutorado com os quais os grupos interagem e a produção científica e tecnológica do grupo; das sociedades científicas relacionadas e eventos, dos portais ou anais.

Serão apresentados, ainda, exemplos de tabelas e gráficos que podem representar, de maneira sintética, as informações sobre o comportamento de variáveis numéricas levantadas por meio de processos de pesquisa baseado em proposta metodológica, bem como destacadas as medidas estatísticas que podem apresentar uma população e uma amostra: medida de posição, dispersão, assimetria e curtose.

Para muitos, a Estatística não passa de um conjunto de tabelas de dados numéricos, e os estatísticos são as pessoas que coletam esses dados. A Estatística originou-se com a coleta e construção de tabelas de dados para o governo, porém a situação evoluiu e essa coleta de dados representa somente um dos aspectos da Estatística. Hoje em dia, pode-se adotar a seguinte definição: a Estatística é uma ciência (ou método) baseada na teoria das probabilidades, cujo objetivo principal é nos auxiliar a tomar decisões ou tirar conclusões em uma situação de incerteza, a partir de informações numéricas. A (Figura 1) demonstra as etapas metodológicas da linha de pesquisa na área da Estatística.

Em pesquisas de Marketing, segundo Mattar (1992), para apresentação em forma de tabela, os cálculos estatísticos mais empregados na análise e interpretação dos dados são os seguintes:

**Figura 1 - Fluxograma das etapas metodológica da Estatística**

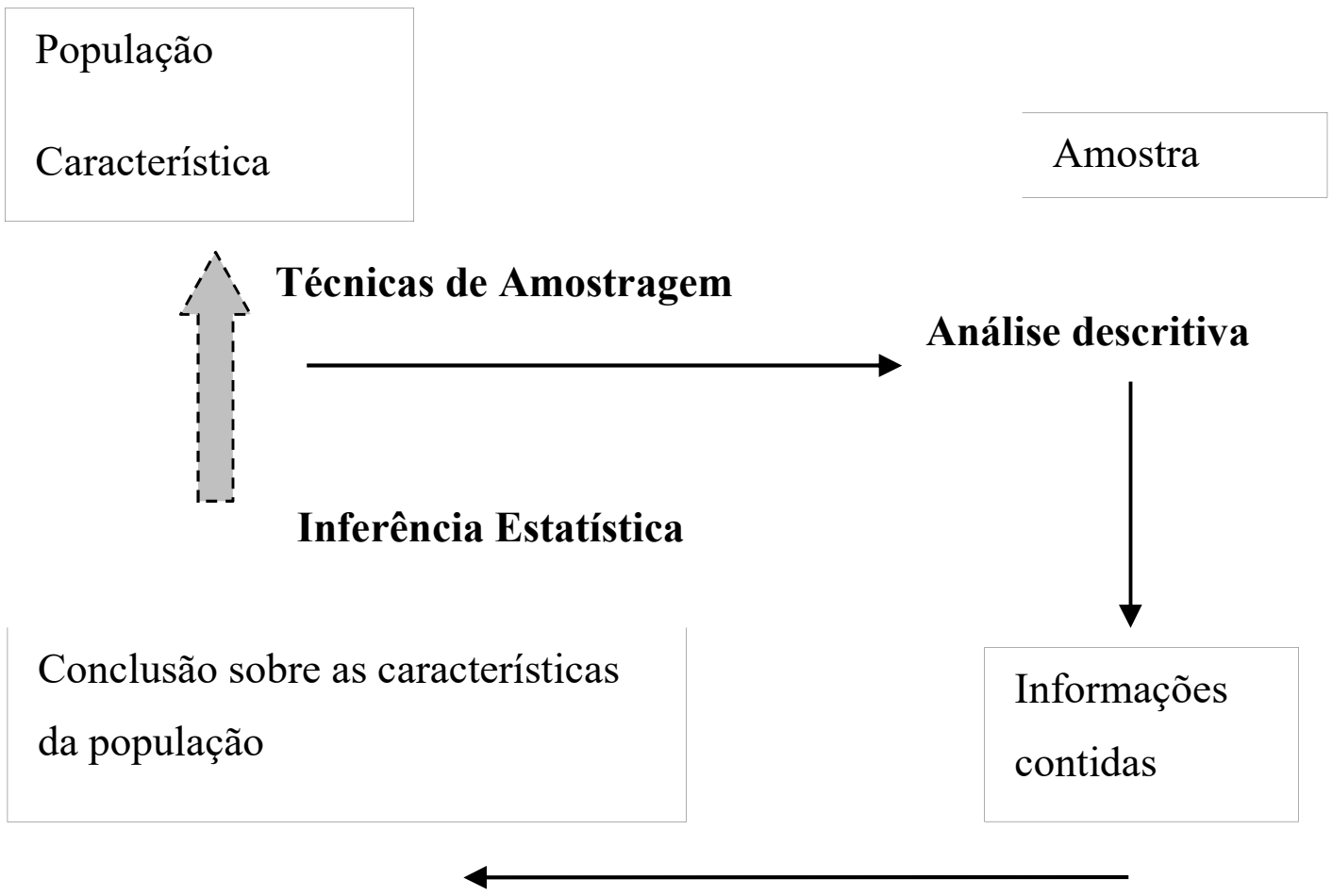

Fonte.: CAMPOS, Manoel B. N. S. (2008)

**Tabela 1- Cálculos Estatísticos mais usados para análise e interpretação de dados**

| Escala da | Medidas de | Variabilidade |
|---|---|---|
| Variável | Posição | Dispersão |
| Nominal | Moda | Distribuição de frequência (absoluta e relativa) |
| | Mediana | Ordenamento |
| Ordinal | Quartis, decis e percentis | |
| | | Amplitude<br>Distribuição de frequência acumulada (absoluta e relativa) |
| Intervalar | Média aritmética | Desvio-médio<br>Desvio-padrão<br>Coeficiente de variação |

Fonte: Pesquisa de Marketing Mattar (1992)

Uma área importante em muitas aplicações da Estatística e que pode ser aproveitada no ensino desta disciplina é a da tecnologia de amostragem. Pode-se citar como exemplos de utilização: pesquisa de mercado, pesquisa de opinião pública, ensaios de medicamentos e em praticamente todo experimento.

### 2.1.3 Aprendizagem e a Estatística

As concepções de ensino aprendizagem de Estatística do professor têm forte influência na natureza da Matemática presente em sala de aula. Além disso, como nos alerta Valente (1998, p.91), "a aprendizagem pode ocorrer basicamente de duas maneiras: a informação é memorizada ou é processada pelos esquemas mentais e agregada a esses esquemas. Neste último caso o conhecimento é construído".

No processo ensino aprendizagem naturalmente, o homem tem a capacidade de aprender:

> Tomemos como ponto de partida o fato de que a aprendizagem da criança começa muito antes da aprendizagem escolar. A aprendizagem escolar nunca parte do zero. Toda aprendizagem da criança na escola tem uma pré-história. Por exemplo, a criança começa a estudar aritmética, mas já antes de ir à escola adquiriu determinada experiência referente à quantidade, encontrou já várias operações de divisão e adição, complexas e simples; portanto, a criança teve uma pré-escola de aritmética, e o psicólogo que ignora este fato está cego (HUPPES, 2002, p. 45 *apud* VYGOTSKY, 1998, p.109).

Observando que, em todo plano de ensino, consta, na bibliografia básica, uma relação de livros que serão utilizados pelo professor e pelos alunos, pode-se dizer que o livro-texto é parte do material empregado no processo de ensino-aprendizagem e pode contribuir para a evolução do desempenho na solução de problemas dos alunos.

Escolhemos alguns exercícios dos livros que foram utilizados pelos sujeitos desta pesquisa durante o desenvolvimento do componente curricular Estatística, para fazer uma ilustração da abordagem dos processos ensino-aprendizagem, dos tipos de exercícios que estão sendo propostos e se esses exercícios facilitam a possibilidade de os alunos saírem do nível técnico e atingirem o nível de aprendizagem.

Tentamos verificar, ainda, se os exercícios remetem à possibilidade de uma análise mais ampla observando, se já é possível detectar tendências de a população a partir dos dados coletados na amostra, com as ferramentas da Estatística Descritiva, abrir caminho para a análise inferencial a ser feita logo adiante. Discutimos também o que ficaria, nesse caso, a cargo de professor, para fazer evoluir o processo de ensino- aprendizagem dos alunos.

Os livros utilizados são estes:

Livro 1 – TEIXEIRA, Daniel Mandim. *Estatística descomplicada.* Brasília: VEST-CON Editora Ltda., 1995.

Livro 2 – LAPPONI, Juan Carlos. *Estatística usando Excel.* São Paulo: Lapponi Treinamento e Editora Ltda., 2000.

Livro 3 – FONSECA, Jairo Simon da. *Curso de Estatística.* São Paulo: Editora Atlas, 1996.

Livro 4 – CRESPO, Antônio Arnot. *Estatística Fácil.* São Paulo: Editora Saraiva, 2000.

Livro 5 – OLIVEIRA, Francisco Estevam Martins de. *Estatística e Probabilidade.* 2.ed. São Paulo: Atlas, 1999.

Foram classificados os exercícios e atividades escolhidas nesses livros, segundo quatro (4) categorias, estabelecidas conforme interesse desta pesquisa:

● Categoria 1: Exercícios não contextualizados, do tipo "Calcule a média aritmética das seguintes séries":

a) 3,4,1,3,6,5,6

b) 7,8,8,10,12

c) 3,2 ; 4; 0,75 ; 5 ; 2,13 ; 4,75

d) 70,75,76,80,82,83,90

● Categoria 2: Exercícios contextualizados, que sugerem o caminho a ser seguido na resolução, do tipo “Considere a distribuição das estaturas de 140 alunos. Calcule a média, a mediana e a moda”.

● Categoria 3: Exercícios contextualizados como os anteriores, porém que solicitam a interpretação dos resultados obtidos.

● Categoria 4**:** Situação-problema colocada, não é solicitado ao aluno, por exemplo, “Calcule isso ou aquilo”. Exige-se o funcionamento autônomo do aluno para escolher entre os conhecimentos aquele que deverá ser usado para solucionar o problema.

Destacamos que a situação da linha do processo ensino-aprendizagem que foi proposta aos alunos nesta pesquisa está classificada na quarta categoria. Por isso, optamos por fazer uma pesquisa qualitativa e construtivista. Segundo Ludke e André (1986, p.13), tal pesquisa “(...) envolve a obtenção de dados descritivos, obtidos no controle direto do pesquisador com a situação estudada, enfatiza mais o processo que o produto e se preocupa em retratar a perspectiva dos participantes”.

Exercícios colocados na categoria 1, à luz da teoria que estamos analisando, não deveriam ser predominantes, pois servem, unicamente, para fixar o uso da fórmula e, portanto, devem ser uma parte ínfima do todo.

Os resultados da aprendizagem alcançados têm que se embasar em pontos fortes do aluno, pois isto facilita a apreensão do processo ensino-aprendizagem. Devem-se privilegiar, também, os métodos de ensino utilizados, tendo em vista o grau de preocupação por uma inovação didática, a adequação das estratégias didáticas utilizadas em distintas disciplinas para o tipo de formação de que se precisa. A existência de atividades complementares dedicadas à formação do aluno com técnicas de estudo, de aprendizagem de atitudes com as quais estaria apto a resolver sozinho o problema de sua área de atuação, conectando as informações das outras áreas envolvidas na questão. Assim, o trabalho com situações próximas daquelas que o aluno terá que enfrentar na profissão para a qual está se preparando deve ser o eixo condutor da atividade educacional, embora no desenvolvimento do conteúdo o professor possa mostrar outras situações e fazer generalizações.

**Exercícios na Categoria 1 :**

Exercícios deste tipo são descontextualizados, pois cobram apenas um algoritmo.

A) Complete o esquema para o cálculo da mediana na distribuição.

Temos:

(Livro 4 – nº1, p.96)

Como:

$$\frac{\sum fi}{2} = \frac{...}{2} = \ldots.$$

Vem: Md = . . . .

Nos exercícios (A) acima e (B) abaixo, não contextualizados, o aluno pode ir diretamente à descrição do texto para responder. Não precisa fazer nenhuma adaptação para encontrar a resposta, não tem instrumentos de validação, a não ser a resposta no final do livro.

B) A série estatística é chamada cronológica quando:

a) O elemento variável é o tempo;

b) O elemento variável é o local;

c) Não tem elemento variável.

Resposta: alternativa a.

(Livro 1 – nº 2, p.21)

Com os exemplos demonstrados, podemos trabalhar apenas o nível técnico de conhecimento. Nesse sentido, os alunos não são instigados a extrair todas as informações contidas no problema e discutir o significado dos resultados obtidos.

**Exercícios na Categoria 2:**

Nesta categoria, incluímos os exercícios que são contextualizados e sugerem o caminho a ser seguido, solicitam o que deve ser feito, mas não pedem o que se pode concluir com os dados obtidos.

C) Sejam os seguinte dados:

Pede-se:

a) Qual é a classe modal?

b) Qual é o valor da moda bruta da distribuição?

c) Determinar a moda pelo método de Czuber.

Esse tipo de exercício trabalha o processo ensino-aprendizagem apenas no nível técnico: capacita o aluno a usar as fórmulas. Mas levar o aluno a pensar é fundamental para a aprendizagem, em qualquer área do conhecimento.

**Exercícios na Categoria 3:**

Nesta categoria, incluímos os exercícios, que além de contextualizados, também exigem que o aluno interprete os resultados obtidos.

Dado o (Gráfico 1), supondo as exportações brasileiras – março: 1995, determinem as exportações correspondentes aos milhões de dólares na figura e construa uma tabela reproduzindo os dados obtidos.

Responda qual é o Estado brasileiro que mais exportou?

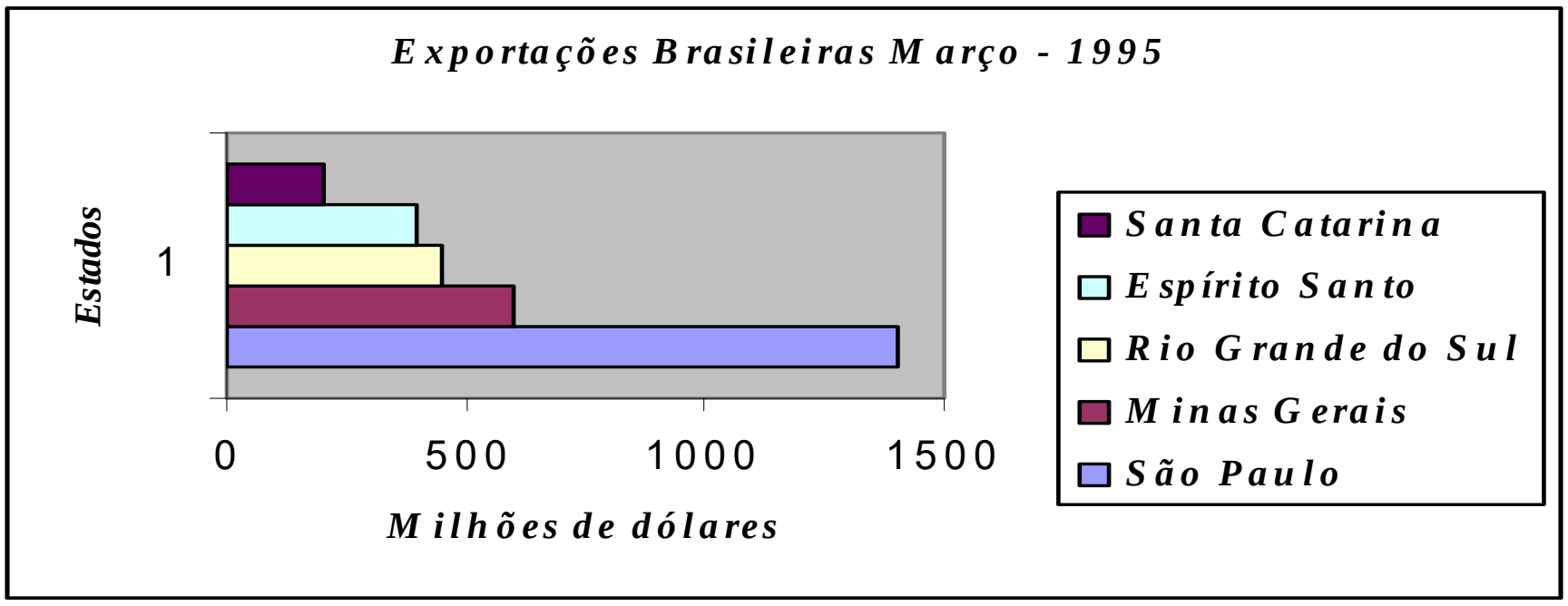

**Gráfico 1- Exportações Brasileiras Março – 1995**
Fonte: (Livro 1-nº 20, p.68)

E) A distribuição abaixo indica o número de acidentes ocorridos com 70 motoristas de uma empresa de ônibus:

Determine:

a) O número de motoristas que não sofreram nenhum acidente;

b) O número de motoristas que sofreram pelo menos 4 acidentes;

c) O número de motoristas que sofreram menos de 3 acidentes;

d) O número de motoristas que sofreram no mínimo 3 e no máximo 5 acidentes;

e) A porcentagem dos motoristas que sofreram no máximo 2 acidentes.

(Livro 4 – nº 8, p. 68)

**Análise da Solução:**

a) Para o aluno que conhece a variável discreta de variação relativamente pequena, cada valor pode ser tomado como um intervalo de classe e, nesse caso, a distribuição é chamada sem intervalo de classe; ele é capaz de encontrar que o número de motorista que não sofreu nenhum acidente foi o total de 20 motoristas.

b) O conhecimento dos vários tipos de frequência ajuda-nos a responder a muitas questões com relativa facilidade, como a do número de motoristas que sofreram pelo menos 4 acidentes: são 15 motoristas.

c) É evidente que o número de motoristas que sofreram menos de 3 acidentes é o daqueles que formam as classes de ordem 0, 1 e 2 . Assim, o número de motoristas é dado por:

$f1 + f2 + f3 = \Sigma fi = F3 = 20 + 10 + 16 = 46$ motoristas.

d) O que foi solicitado nesse item permite ao aluno fazer pequenas inferências apenas com as ferramentas da Estatística Descritiva dos que sofreram no mínimo 3 e no máximo 5, sendo igual a 20 motoristas.

e) Observando-se que o número de acidentes ocorridos com 70 motoristas é a frequência total: aplicando uma regra de três simples. Logo, a frequência relativa dos motoristas que sofreram no máximo 2 acidentes é $fr = \underline{46} = 0{,}657 \times 100 = 65{,}7$ % dos motoristas.

**Exercícios na Categoria 4:**

São poucos os exercícios encontrados com as características da categoria ora citada, pois todos eles sugerem alguma interpretação pré-estabelecida. De acordo com a teoria do processo ensino-aprendizagem, o aluno, aos poucos, vai adquirindo autonomia para resolver problemas. Outra possibilidade para esse exercício, num segundo momento, é propô-lo sem nenhuma “dica” de resolução. Neste sentido, o aluno é instigado a procurar entre os conhecimentos aqueles que possam ser utilizados para a solução dos problemas.

f) Um produto é vendido por apenas três empresas, em um determinado mercado. Em determinado ano, para um total de 18.000 unidades vendidas, tivemos a seguinte distribuição das vendas:

Determinar a distribuição percentual das vendas.

(Livro 1, exercício nº 11, p. 35)

Neste exercício, o aluno teve de observar que a variável é ordinal e que não faz sentido calcular a média desses dados. Já foi dada a "dica" de calcular a percentagem das vendas, logo tirou-se a possibilidade de autonomia na estratégia de ensino-aprendizagem, bem como na resolução de problemas.

Nos livros que estão sendo utilizados pelos sujeitos desta pesquisa, não foram encontrados exercícios na categoria quatro. Observamos que os exercícios estão concentrados apenas nas categorias dois e três.

### 2.1.4 Técnicas de ensino

São instrumentos que devem expressar comportamentos face às explicações técnicas, à compreensão como um todo. Os métodos e técnicas de ensino são meios que conduzem o discente à reflexão sobre os conteúdos das técnicas e suas aplicações (BASTOS, 1998).

Para Bastos (1998), os conteúdos são parte de uma produção social e histórica. São interdisciplinares. Não devem ser apreendidos visando exclusivamente à manipulação do fazer, mas à compreensão da tecnologia como um todo e das suas tendências, como elementos de realização flexível com vistas a preparar o imprevisível e o adaptável a ser concretizado no mundo do trabalho.

Segundo Oliveira e Gracio (1999), a proposta para os cursos introdutórios de estatística, em geral, é constituída por um modelo já pronto, cabendo ao professor apenas executá-lo, o que tem causado desinteresse dos alunos e, muitas vezes, os mesmos não conseguem visualizar como a metodologia estatística será aplicada na sua futura prática profissional. Afirmam que há necessidade de criar cursos introdutórios que priorizem o ferramental estatístico adequado a cada área do conhecimento, de forma que os conteúdos estatísticos sejam melhores articulados às áreas de atuação dos diferentes cursos.

O professor deve cultivar a atitude de observação e pesquisa permanente, selecionando partes de métodos e técnicas conhecidas, como também selecionar novas formas de ensino mais ajustadas à realidade pedagógica que tenha de enfrentar, para, dinâmica e objetivamente, tornar o ensino mais consequente (NÉRICI, 1989).

Toda técnica é tecida e envolvida por determinados ideais educativos. Não é a técnica que define o ideal educativo, mas o contrário. Assim é possível usar o retroprojetor sem ser tecnicista. É possível a aula expositiva sem ser tradicional (ARAÚJO, 1996).

Segundo Triola (1998), a utilização também de Software estatístico deve ser sempre suportada por um adequado conhecimento das técnicas estatísticas envolvidas, ou orientada por quem detenha esses conhecimentos. De uma forma genérica e simplificada, todos os aplicativos estatísticos, lidam com a análise estatística de dados estruturada em quatro etapas:

a) Introdução dos dados no programa (ou importação do fiche iro de dados gravado noutra aplicação).

b) Seleção do procedimento de análise estatística a aplicar.

c) Seleção da(s) variável (eis) a utilizar nos caixas de diálogo.

d) Apreciação e interpretação dos resultados.

Hoje em dia, o software estatístico adquiriu uma grande importância nos meios acadêmico, empresarial e administrativo, entre outros, quer pela sua facilidade de utilização, quer pela eficácia no tratamento de grandes conjuntos de dados.

## 2.2 Teorias de aprendizagem e a proposta metodológica

As concepções de ensino-aprendizagem abarcam principalmente as teorias de Skinner, de Piaget, de Vygotsky, de Freire e de Gardner, Essas teorias influenciaram e continuam influenciando os educadores responsáveis pelo ensino da educação matemática. E acreditamos que, a partir delas, pode-se fazer proposta metodológica para melhoria do ensino-aprendizagem da Estatística.

O professor vai intuitivamente e empiricamente construindo a sua própria didática calcada nos modelos que conheceu como aluno e no bom senso que o ajuda a filtrar os procedimentos que entendem como adequados e/ou não adequados. Segundo Viecili (2006, p. 20), exige-se do professor uma postura diferente visando possibilitar que o aluno "aprenda a aprender", mas também como: "[...] com o passar do tempo, um "jeito" de organizar e conduzir o ensino que, geralmente não chega a ser tomado com reflexão nem pelo professor individualmente e, menos ainda, pelo conjunto de professores que lecionam um dado curso".

Dessa forma, as teorias de aprendizagem, fundamentam-se em uma visão de mundo, de sociedade e de homem e, consequentemente, têm seus reflexos na educação. Cada uma dessas visões sobre o processo de aprendizagem causa impactos no desenvolvimento educacional, se apóiam no subjetivismo e/ou no objetivismo.

Compreende-se que:

> a) **Comportamentalismo e Neo-Comportamentalismo**
> O comportamentalismo e o neo-comportamentalismo fundamentam-se numa visão objetivista de mundo, de sociedade, de homem. Essas teorias dão ênfase em tudo o que é visível e tocável. O comportamentalismo, que tem origem na psicologia experimental de Watson (início deste século), defende que a aprendizagem é um processo lento e gradativo. O individuo é visto como "tabula rasa" que vai recebendo informações com grau crescente de complexidade, entendendo assim, o meio externo. O processo de aprendizagem se dá em função das situações de ensino, nas quais os indivíduos têm constantes reforços para respostas corretas. A instrução programada, tão utilizada nas décadas de 60 e 70, tem base nessa visão de aprendizagem.
>
> **b) Construtivismo**
> Piaget desenvolveu uma teoria bastante complexa na busca de explicação sobre a gênese do conhecimento. Para Piaget (1987), todos os indivíduos, independentemente da cultura e da condição social podem passar pelo mesmo processo de desenvolvimento, que ocorre em 4 estágios: sensório motor, pré-operatório, operacional concreto e das operações formais. Mas isto não ocorre em ordem linear, pois cada estágio inicia pela reorganização dos anteriores, e esse processo pode ser retardado, ou mesmo bloqueado, conforme as interações estabelecidas. Os estágios de desenvolvimento do conhecimento fornecem indicadores para a definição da complexidade da situação, ou seja, fornecem condições de aprendizagem favorável ao atual estágio de desenvolvimento do aluno. Assim, para Piaget (apud LA TAILLE; OLIVEIRA; DANTAS, 1992), aprender é atuar sobre o objeto da aprendizagem para compreendê-lo e modificá-lo.
> Como aprender é uma contínua adaptação ao meio externo, aprendem-se quando se entra em conflito cognitivo, ou seja, quando há a confrontação de ideias e situações de algo que não se sabe qual a melhor solução ou resposta.
> O individuo procura encontrar o equilíbrio. Para isso, precisa passar por importante processo de adaptação, pelo qual o sujeito adquire o equilíbrio entre assimilação e acomodação. A assimilação refere-se à ação do sujeito sobre o meio físico e/ou social e, portanto, depende da iniciativa pessoal para a construção do conhecimento. Por meio da incorporação, a estrutura de conhecimento existente se modifica de modo a acomodar-se a novos elementos – tal modificação é denominada acomodação. Equilíbrio é o processo de organização das estruturas cognitivas num sistema coerente, interdependente, que possibilita ao indivíduo a adaptação à realidade. É a partir deste entendimento que as situações de aprendizagem baseiam-se em jogos e desafios, nos quais o sujeito é defrontado com um problema novo para resolver.

**c) Sócio-Interacionismo**
Atualmente está em voga a teoria sócio-interacionista de Vygostsky. Os pontos-chave, enumerados por Rabelo e Lorenzato (1994), que causam impacto ao desenvolvimento de situações de aprendizagem são:
• A aprendizagem está fortemente ligada à cultura e à interação social do indivíduo.
• O desenvolvimento cognitivo é condicionado conforme os intervalos de idade (Zona Proximal de Desenvolvimento).
• O desenvolvimento cognitivo completo depende do meio social em que o indivíduo está inserido.
Note-se que as teorias aqui expostas são construções teóricas apresentadas em uma visão muito simples. Piaget é considerado um dos cientistas mais importantes da atualidade. Vygotsky redefiniu as formas de aprendizagem cooperativa – teoria que determina que qualquer discussão sobre educação e cooperação passe necessariamente por ela.
Baseando-se nessas teorias, deve-se ver a Matemática como forma dinâmica de conhecimento. A resolução de problemas – que inclui a maneira de dar significado à linguagem matemática deve ser central na vida escolar, de tal modo que os alunos possam explorar experimentar e organizar, criando um conhecimento novo no decurso das suas vidas (CARRETERO, 1997, *apud* VIECILI, 2006, p. 21-23).

As teorias de aprendizagem podem propiciar modalidades para o desenvolvimento e a transformação do homem.

O "Método Paulo Freire" consiste de três momentos dialéticos e interdisciplinarmente entrelaçados, segundo:

a) A investigação, temática pela qual aluno e professor buscam, no universo vocabular do aluno e da sociedade onde ele vive as palavras e temas centrais de sua biografia. Esta é a etapa da descoberta do universo vocabular, em que são levantados palavras e temas geradores relacionados à vida cotidiana dos alfabetizando e dos grupos sociais aos quais eles pertencem.

b) A tematização, pela qual eles codificam e decodificam esses temas; ambos buscam o seu significado social, tomando assim consciência do mundo vivido. Descobrem-se assim novos temas geradores, relacionados com os que foram inicialmente levantados.

c) A problematização, na qual eles buscam superar uma primeira visão mágica por uma visão crítica, partindo para a transformação do contexto vivido. Nesta ida e vinda do concreto para o abstrato e do abstrato para o concreto, volta-se ao concreto problematizando-o (HUPPES, 2002, *apud* GADOTTI, 2001, p. 37).

Na interpretação do construtivismo freireano, evidenciou não só que todos podem aprender, mas que todos sabem alguma coisa. De acordo com Freire (1986, p.28), "não há ignorantes absolutos" e que o sujeito é responsável pela construção do conhecimento e pela ressignificação do que aprende.

Atualmente, a educação deve estar comprometida com o desenvolvimento integral do educando. Aprender a ser supõe a preparação da pessoa para elaborar pensamentos autônomos e críticos. Formular os seus próprios juízos de valor, para poder decidir por si mesmo, frente às diferentes circunstâncias encontradas na vida.

Importante salientar as seguintes sugestões, para melhorar a prática pedagógica:

a) O professor deve planejar atividades diversificadas, para os alunos que possuem ritmos de desenvolvimento diferente;

b) Aproveitar os alunos mais adiantados para serem monitores, auxiliando os colegas com mais dificuldades;

c) Trabalhar com atividades coletivas;

d) Levar o aluno a desenvolver o espírito de colaboração.

Em relação à concepção de desenvolvimento e aprendizagem, podem-se apontar três ideias básicas de Vygotsky com relevância no ensino:

> 1. O desenvolvimento psicológico deve ser olhado de maneira prospectiva, na trajetória do indivíduo.
> O conceito de zona de desenvolvimento proximal está estreitamente ligado a essa concepção.
> 2. Os processos de aprendizado movimentam os processos de desenvolvimento.
> Onde no pensamento de Vygotsky, a trajetória do desenvolvimento humano se dá de “fora para dentro”, por meio da internalização dos processos interpsicológicos.
> 3. A importância da atuação dos outros membros do grupo social na mediação entre a cultura e o indivíduo e na promoção dos processos interpsicológicos que serão posteriormente internalizados (HUPPES, 2002, *apud* OLIVEIRA, 1997, p. 59-61).

O indivíduo não tem instrumentos endógenos para percorrer, sozinho, o caminho do pleno desenvolvimento.

Nesse contexto, Gardner (1994) diz que o desenvolvimento da inteligência é determinado por condições genéticas e por condições ambientais, podendo alguma dessas inteligências ser mais desenvolvidas, com forma própria de pensamento, ou de processamento das informações. Como a inteligência é estimulável, deve-se utilizar uma diversidade de elementos mediadores para estimular todos os tipos de inteligência.

O que deixa o ato de educar muito pobre é a quase ausência de atividades construtivas. Os alunos fazem muito pouco. Em geral, repetições lhes são cobrados. Não se avalia o que os alunos fazem, mas sua capacidade de imitar os pensamentos que estão nos livros e no dos professores.

### 2.2.1 Modelagem Matemática

No Brasil, constata-se também que a Modelagem Matemática vem impulsionando a Educação Matemática, fazendo com que propicie características capazes de aprimorar o processo de ensino e aprendizagem dos alunos, especialmente nas propostas de educadores que veem a realidade social como um componente do aprendizado.

A trajetória das pesquisas sobre o desenvolvimento do uso da Modelagem Matemática no país, conforme Burak (1987, 2004), Scheffer (1999), Barbosa (2007), Bisognin (2008), Caldeira (1998, 2004, 2007, 2008), Santos e Bisognin (2007), mostra que diferentes possibilidades podem ser oferecidas aos alunos com o objetivo de aprimorar o desenvolvimento do raciocínio lógico, bem como contribuir para a formação de indivíduos mais críticos e comprometidos com a transformação da sociedade a que pertencem.

Em Modelagem Matemática, na Educação Matemática Brasileira, as pesquisas e as práticas educacionais (BARBOSA, CALDEIRA, ARAÚJO, 2007), originadas a partir do Grupo de Trabalho em Modelagem Matemática do II e III SIPEM (Seminário Internacional de Pesquisa em Educação Matemática) discutem questões referentes à Modelagem Matemática e à Formação de Professores, à Modelagem e prática de sala de aula, aos Aspectos Teóricos da Modelagem Matemática, à Modelagem Matemática e às Tecnologias da Informação e Comunicação. Tudo são referências importantes que refletem o panorama atual da comunidade científica que percebe a Modelagem Matemática como uma alternativa capaz de correlacionar a prática de ensino com a realidade dos alunos.

Dessa forma, pode-se concluir que ensinar matemática por meio da Modelagem Matemática possibilita a transformação de uma educação desvinculada da realidade dos alunos em uma educação comprometida com os aspectos sociais, culturais, políticos, econômicos e ambientais. Burak (1987, p. 129) percebe que

> Modelagem Matemática constitui-se em um conjunto de procedimentos cujo objetivo é estabelecer um paralelo para tentar explicar, matematicamente, os fenômenos presentes no cotidiano do ser humano, ajudando-o a fazer predições e a tomar decisões.

A Modelagem Matemática como ambiente de aprendizagem, o foco permanece, portanto, na matemática e sua capacidade de resolver problemas de outras áreas:

> Modelagem pode ser entendida em termos mais específicos. Do nosso ponto de vista, trata-se de uma oportunidade para os alunos indagarem situações por meio da matemática sem procedimentos fixados previamente e com possibilidades diversas de encaminhamento. Os conceitos e ideias matemáticas exploradas dependem do encaminhamento que só se sabe à medida que os alunos desenvolvem a atividade. Porém, alguns casos podem ser mais propícios a alguns conceitos matemáticos – por exemplo, situações que envolvem variação podem levar a ideias do Cálculo ou Pré- cálculo -, mas nada garante que os alunos se inclinem por eles (BASSANEZI, 1994).

Nos trabalhos de Barbosa (1999, 2001), Bisognin (2008), Santos e Bisognin (2007), Almeida (2004), Bisognin, Bisognin e Rays (2004) , constata-se que a Modelagem Matemática propicia ao aluno uma interatividade que o aproxima da sua realidade sociocultural. Alguns autores criam ou adotam esquemas que ilustram o processo de criação de modelos (CROSS; MOSCARDINI, 1985; GALBRAITH; CLATWORTHY, 1990; HANNON; RUTH, 1994; BASSANEZI; FERREIRA, 1988).

Esses esquemas procuram sintetizar os aspectos que são, segundo seus autores, desejáveis em um processo de Modelagem Matemática. Apresentamos, na (Figura 2), o esquema sugerido por Bassanezi e Ferreira (1988):

**Figura 2 - Exemplo de um esquema de processo de criação de modelos matemáticos**

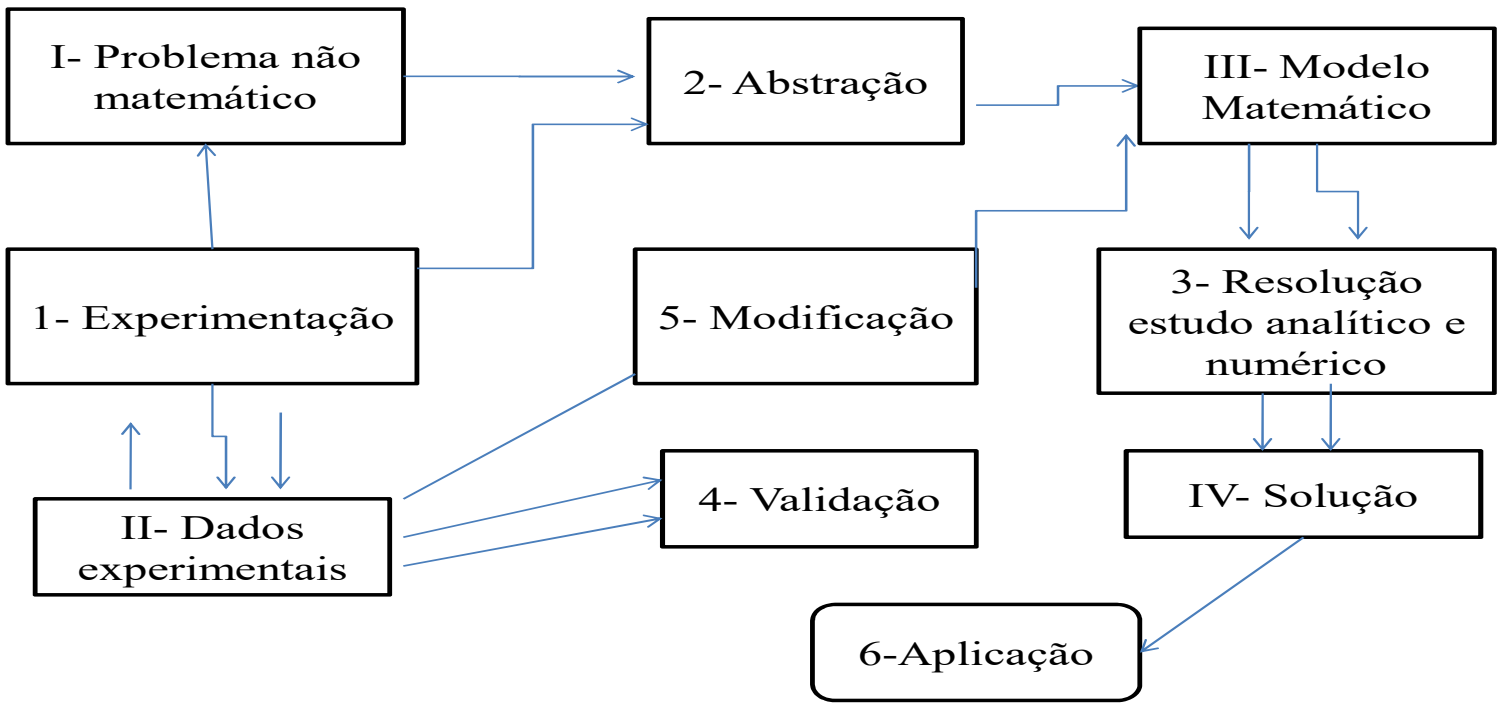

Org.: CAMPOS, Manoel B. N. S. (2008)

Os outros autores citados utilizam esquemas semelhantes ao apresentado acima, alguns com mais, outros com menos detalhes que este.

Ao apresentar sua experiência utilizando a Modelagem Matemática como uma estratégia de ensino e aprendizagem, Bassanezi (1994) começa expondo sua perspectiva (na matemática): o estudo de problemas e situações reais com o uso da matemática como linguagem para sua compreensão, simplificação e resolução, objetivando uma possível revisão ou modificação do objeto em estudo, é parte de um processo que tem sido denominado 'Modelagem Matemática'.

As situações problemas envolvendo Modelagem Matemática de uma situação problema real deve seguir uma sequência de etapas:

> **1. Experimentação**: É uma atividade essencialmente laboratorial onde se processa a obtenção de dados;
> **2. Abstração**: É o procedimento que deve levar à formulação dos Modelos Matemáticos;
> **3. Resolução**: O modelo matemático é obtido quando se substitui a linguagem natural das hipóteses por uma linguagem matemática coerente – é como num

> dicionário, a linguagem matemática admite "sinônimos" que traduzem os diferentes graus de sofisticação da linguagem natural;
> **4. Validação**: É o processo de aceitação ou não do modelo proposto. Nesta etapa, os modelos, juntamente com as hipóteses que lhes são atribuídas, devem ser testados em confronto com os dados empíricos, comparando suas soluções e previsões com os valores obtidos no sistema real. O grau de aproximação desejado destas previsões será o fator preponderante para validação;
> **5. Modificação**: Alguns fatores ligados ao problema original podem provocar a rejeição ou aceitação dos modelos. Quando os modelos são obtidos considerando simplificações e idealizações da realidade, suas soluções geralmente não conduzem às previsões corretas e definitivas, pois o aprofundamento da teoria implica na reformulação dos modelos. Nenhum modelo deve ser considerado definitivo, podendo sempre ser melhorado, poder-se-ia dizer que um bom modelo é aquele que propicia a formulação de novos modelos, sendo esta reformulação dos modelos uma das partes fundamentais do processo de modelagem (BASSANEZI, 2004).

E confirmando que, acerca da ampliação dos limites da Modelagem Matemática em um contexto de ensino e aprendizagem, o mesmo autor acrescenta:

> Nosso trabalho tem sido tentar conectar a experiência de ensino com a perspectiva de modelagem baseados em preocupações teóricas, filosóficas e metodológicas específicas. Nós levamos em conta os recursos humanos, o interesse compartilhado pelo professor, aluno e comunidade; contextos sociais, político e econômico etc. Procuramos, também, resgatar a Etnomatemática, sua interpretação e contribuição no nível da sistematização matemática. (BASSANEZI, 1994, p. 31).

Neste enfoque, há uma mudança no papel do professor. Ele passa a ser orientador do processo de descoberta de possíveis soluções para os problemas encontrados, auxiliando o aluno na elaboração dos modelos e na sua validação.

O modelo de Modelagem Matemática no qual a matemática e realidade são dois conjuntos disjuntos e a modelagem é o meio de fazê-los interagir, conforme a (Figura 3).

**Figura 3- Modelo de Modelagem Matemática**

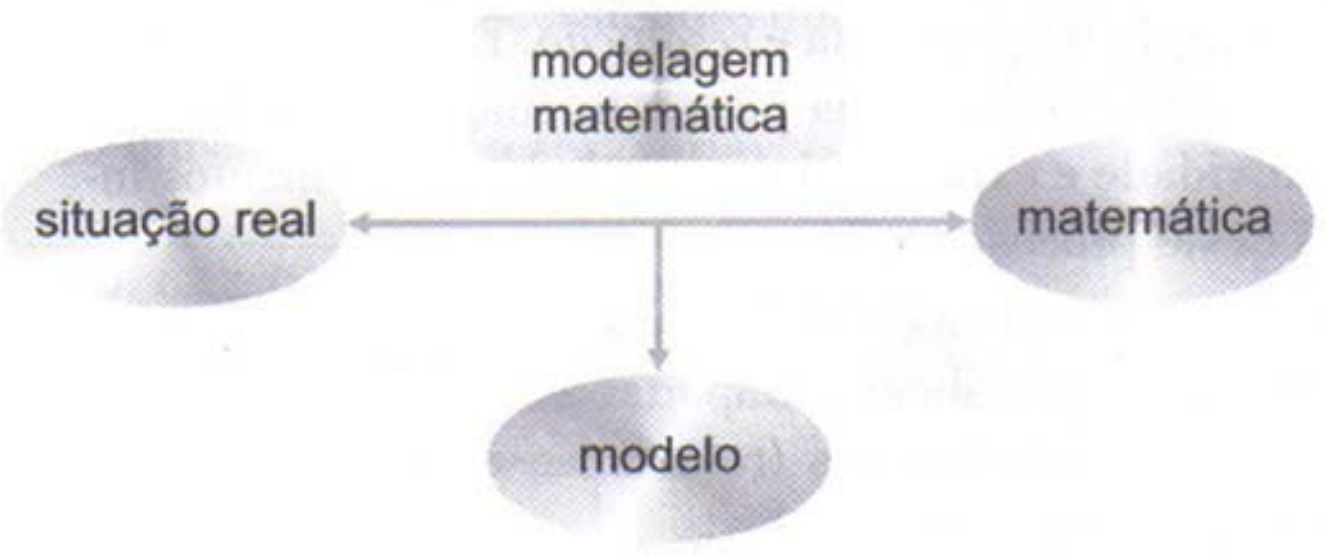

Fonte: BEIMBENGUT; HEIN , 2005, p. 13

Por outro lado, essa interação permite a representação do fenômeno através da linguagem matemática (modelo matemático), a fim de propiciar seu uso de maneira adequada.

### 2.2.2 O Ensino no Laboratório de Matemática

As dificuldades pelas quais passam os professores no ato do ensino, e os alunos, em termos de aprendizagem no campo da Matemática, vêm inquietando muitos pesquisadores na área da didática da Matemática.

Diante de tal inquietação, muito se tem discutido sobre as diferentes concepções metodológicas de Laboratório de Matemática, os objetivos, o papel e a importância deste laboratório na formação de professores, bem como as diferentes propostas de sua utilização nas diversas instituições de Ensino Superior comprometidas com a formação de professores.

Nesse contexto, pois, após a LNDBE de 20.12.1996 e a CNE/CP nº 2 de 19.02.2002 determinarem a obrigatoriedade de 400 horas de estágio supervisionado na matriz curricular dos cursos de Licenciatura, muitas instituições de Ensino Superior passaram a sentir a necessidade de criar ambientes que pudessem dar suporte ao planejamento das atividades de estágio como também favorecer a realização da prática pedagógica das disciplinas do Núcleo Comum das Licenciaturas (VARIZO, 2007).

Segundo Lorenzato (2006), o Laboratório de Ensino da Matemática na escola é uma sala-ambiente reservada para que as aulas de matemática aí aconteçam de maneira a estruturar, organizar, planejar e construir o fazer matemático, facilitando tanto para o professor como para o aluno o questionamento, a procura, a experimentação, a análise, a compreensão de conceitos e a conclusão de uma determinada aprendizagem, inclusive com a produção de materiais instrucionais que possam facilitar o aprimoramento da prática pedagógica.

Deve ser um local de referência para as atividades matemáticas, em que os professores possam se empenhar em tornar a matemática mais compreensível para seus alunos. Neste local, o professor poderá também planejar aulas e realizar outras atividades como exposições, olimpíadas, jogos, avaliações, entre outras.

O Laboratório de Matemática numa instituição de Ensino Superior incentiva a melhoria da formação inicial e continuada de professores, promove a integração das ações de ensino, pesquisa e extensão, como também favorece o estreitamento da relação entre a instituição e a comunidade, além de estimular a prática da pesquisa em sala de aula (RÊGO; RÊGO, 2006, p. 41).

A importância deste LEM para os cursos de formação de professores, ao se considerar

> (...) o grande distanciamento entre a teoria e a prática, hoje ainda predominantemente nas salas de aula em todos os níveis de ensino; a baixa conexão entre os conteúdos de Matemática e destes com as aplicações práticas do dia-a-dia e a necessidade de promoção do desenvolvimento da criatividade, da agilidade e da capacidade de organização do pensamento e comunicação de nossos alunos (LORENZATO, 2006, p. 55).

Diante desse contexto, o professor deve atuar como investigador e pesquisador no ambiente da sala de aula e possibilitar estratégias para manter uma visão crítica construtivista da sociedade e do currículo.

### 2.2.2.1 Aspectos Gerais de um Laboratório de Matemática

O laboratório de Matemática pode ser visto como um espaço de construção do conhecimento, tanto individual quanto coletivo. Nesse espaço, professores e alunos podem dar expansão à sua criatividade, enriquecendo o processo ensino-aprendizagem, que diante dessa compreensão, poderão ser capazes de conseguir o que se denomina como pensamento matemático avançado (SILVA; SILVA, 2004). Nesse sentido,

> O laboratório, portanto, é um ambiente propício para estimular no aluno o gosto pela matemática, a perseverança na busca de soluções e a confiança em sua capacidade de aprender e fazer matemática. Além de contribuir para a construção de conceitos, procedimentos e habilidades matemáticas, pode propiciar também a busca de relações, propriedades e regularidades, estimulando o espírito investigativo. Por isso, deve ser neste local da escola onde se respire Matemática o tempo todo e possa ser também um ambiente permanente de busca e descoberta (SILVA; SILVA, 2004, p. 3).

O ambiente deste laboratório proposto deve funcionar, uma vez que, de acordo com Turrioni (2004),

> [...] um centro para discussão e desenvolvimento de novos conhecimentos dentro de um curso de licenciatura em Matemática, contribuindo tanto para o desenvolvimento profissional dos futuros professores como para sua iniciação em atividades de pesquisa. (TURRIONI, 2004, p.7).

Por fim, este laboratório deverá, ainda, no ambiente escolar, pretende-se dotá-lo de um espaço propulsor de recursos adequados ao ensino-aprendizagem da matemática com as seguintes características:

- Promover aulas de acordo com as novas tendências educacionais;
- Possibilitar atividades tanto a nível individual, como em nível de grupos;
- Promover a realização de atividades de investigação e trabalhos de projetos;
- Facilitar o intercâmbio entre os vários níveis de ensino;
- Rentabilizar os equipamentos e materiais didáticos;
- Promover a criação de um espaço para a reflexão sobre o ensino aprendizagem da matemática, com a participação de professores e alunos;
- Contribuir para a formação de um ambiente onde se desenvolvam atividades interativas com materiais didáticos;
- Utilizar a informática como instrumento no ensino da Matemática.

> No que se refere aos alunos, pretende-se que o Laboratório de Matemática seja capaz de auxiliá-los a:
> - Desenvolver a curiosidade e o gosto de aprender matemática;
> - Incrementar uma maior participação;
> - Desenvolver o raciocínio abstrato;
> - Iniciar os alunos na utilização dos computadores;
> - Desenvolver as capacidades de compreensão, análise, aplicação e síntese de software;
> - Promover a compreensão, a interpretação e a utilização de representações matemáticas (tabelas, gráficos, expressões, símbolos,...);
> - Desenvolver o conhecimento do espaço, realizando construções geométricas;
> - Explorar atividades interdisciplinares.
>
> Quanto aos professores, pretende-se que o laboratório de matemática venha possibilitá-los de:
> - Promover a construção e a elaboração de materiais didáticos (jogos, textos, imagens,...);
> - Divulgar e trocar experiências de materiais, atividades, programas e conhecimento diversos;
> - Promover a interdisciplinaridade entre a matemática e as outras disciplinas;
> - Promover a utilização regular de computadores como meio de trabalho de alunos e professores;
> - Promover o intercâmbio de conhecimentos e experiências com outras instituições similares ou com associações de professores (SILVA; SILVA, 2004, p. 3-5).

Nesse contexto, cabe trazer o papel e o significado da didática da matemática para dar sustentação ao processo ensino-aprendizagem. De acordo com Lorenzato (2006), um Laboratório de Ensino de Matemática, de modo geral, pode ter:

- Revistas, jornais e artigos;
- Livros didáticos, paradidáticos e outros;
- Jogos; quebra-cabeça;
- Problemas desafiadores e de lógica;
- Questões de olimpíadas, ENEM e vestibulares;
- Textos sobre história da matemática;
- CDs, transparências, fotos, figuras, sólidos;
- Modelos estáticos ou dinâmicos;

- Materiais didáticos industrializados, instrumentos de medidas;
- Computadores, calculadoras;
- Materiais didáticos construídos pelos alunos e professores;
- Materiais e instrumentos necessários à produção de materiais didáticos e outros.

Dessa forma, apesar das recomendações do autor, este deve estar relacionado às reais necessidades e condições locais de cada instituição, à formação matemática de alunos e professores, ao espaço físico disponível, aos materiais e outros.

# 3 METODOLOGIA ADOTADA

A metodologia conduz e estrutura a pesquisa permitindo avaliar os dados, os questionamentos, as técnicas e os resultados.

Severino (2007) explica que a metodologia para o desenvolvimento de um estudo científico contempla caminhos ou ações que são executadas de forma a permitir o alcance dos objetivos elaborados e a resposta ao problema da pesquisa.

## 3.1 Localização da área de estudo

A Universidade Federal de Mato Grosso (UFMT), com sede e foro em Cuiabá, foi instituída sob a forma de Fundação, por meio da Lei nº. 5647 de 10 de dezembro de 1970, tendo sua origem a partir da fusão do Instituto de Ciências e Letras de Cuiabá, que oferecia os cursos de Pedagogia, Matemática e Economia, com a Faculdade Federal de Direito de Cuiabá.

A partir de 1970, com a implantação da UFMT e seu rápido crescimento, Cuiabá e a região circunvizinha passam a contar com mais de 60 cursos de graduação e pós-graduação, que cobrem praticamente todos os campos do saber humano.

Localizada no centro geodésico da América Latina, num Estado com aproximadamente 881.000 quilômetro, a Universidade traz como temática permanente questões ligadas à fitofisionomia, destacando o tri-ecossistema - cerrado, pantanal e floresta – e a preparação do homem social, sensível aos anseios socioambientais.

Hoje, a UFMT, além do Campus de Cuiabá, desenvolve atividades de ensino, pesquisa e extensão em três *campi* permanentes no interior do estado, a saber: Rondonópolis, Barra do Garças e Sinop. Além disso, oferece cursos de graduação em convênio com prefeituras em Juína, São José dos Quatro Marcos, Mirassol D'Oeste, Água Boa, Torixoréu, Campo Verde e Primavera do Leste, bem como licenciaturas parceladas e ensino à distância, que cobrem a maioria dos municípios mato-grossenses. Compõem, hoje, o conjunto da

UFMT 18 Unidades Acadêmicas de 3ºgrau, um Hospital Universitário e uma Fazenda Experimental.

O *Campus* Universitário de Rondonópolis (CUR) foi criado e homologado em 31 de março de 1976, mediante a Resolução nº. 01/76 do Conselho Universitário pertencente à Universidade Estadual de Mato Grosso, muito embora a Lei Estadual nº 3575 de 2 de dezembro de 1974 já autorizasse a sua criação como Centro Pedagógico de Rondonópolis (CPR).

Oferecendo simultaneamente os Cursos de Ciências e Estudos Sociais, na forma de Licenciatura Curta, o Centro Pedagógico de Rondonópolis iniciou suas atividades em 05 de maio de 1976.

Com a divisão do estado em 1977, deu-se início ao processo de federalização do Centro, integrando-o à Universidade Federal de Mato Grosso, uma vez que o município de Rondonópolis passava a pertencer ao estado de Mato Grosso, agora dividido em duas unidades federativas. De fato, em 5 de julho de 1979, foi instituída a Fundação da Universidade Federal de Mato Grosso do Sul, mediante a Lei Federal nº. 6.674, que em seu artigo 13 transferia para a Universidade Federal do Mato Grosso a responsabilidade pelo Centro Pedagógico de Rondonópolis "[...] o Centro Pedagógico de Rondonópolis, atualmente vinculado à Universidade Estadual de Mato Grosso, passa a integrar com todos os seus bens e direitos, a Universidade Federal de Mato Grosso, com sede em Cuiabá."

Por meio de ato do Conselho Diretor de nº. 05/80, datado de 9 de janeiro de 1980, e com a lotação no quadro de pessoal administrativo (Portaria GR 016/80) e docente (Portaria GR 015/80), dos servidores, o Centro Pedagógico de Rondonópolis integrou-se como Campus à estrutura da Universidade Federal de Mato Grosso. Esta integração evidenciou a necessidade de uma nova adequação à estrutura organizacional da UFMT. Neste sentido, a administração do Centro coube a um coordenador, coadjuvada por seu vice e um Conselho de Departamentos. Assumindo de forma *pro tempore*, o Prof. Etewaldo de Oliveira Borges esteve na Coordenação do Centro, no período de 1979 a 1984.

Os dois cursos que compunham o Centro permaneceram e foram criados dois Departamentos, coordenados por chefes e subchefes, designados pelo Reitor, com base em lista tríplice. De acordo com Maria das Graças Kida (1985), essa estrutura existia apenas no aspecto formal, pois, na realidade, só em agosto de 1983, após cinco anos, com o processo de abertura, discussões e reivindicações, ocorreram eleições, e o Centro passou a contar com vice-coordenador e subchefes de Departamentos. Cada curso organizava-se a partir dos Colegiados de Departamento e de Curso. As necessidades administrativas passaram a contar com uma Secretaria Geral e uma Biblioteca Regional, cuja coordenação, inicialmente, coube ao bibliotecário e professor Javert Melo Vieira.

As demandas da comunidade local e a necessidade de expansão da própria Universidade aceleraram a política de interiorização. Com base nas diretrizes prescritas pelas normas da Universidade e ratificadas pela Resolução nº. CD 04/80, de 8 de maio de 1980, que aprovava a estrutura organizacional do Centro e definia normas sobre os cursos, procedeu-se aos estudos para a elaboração do projeto de criação de novos cursos, já no segundo semestre do mesmo ano.

Tais estudos permitiram a opção por três cursos de graduação a serem oferecidos já no primeiro semestre do ano subsequente, a saber: Ciências Contábeis; Letras, com habilitações em Português e Literatura Portuguesa; e Pedagogia, com habilitações em Supervisão Escolar e Magistério das Matérias Pedagógicas do 2ºGrau.

Aprovados em 27 de janeiro de 1981, por meio da Resolução nº. CD 019/81, esses cursos abrem seus vestibulares em fevereiro do mesmo ano, tendo como limite o número de 30 vagas por curso. As aprovações, diante das instalações disponíveis, revelaram uma questão importante a ser resolvida – o espaço físico. Desde a sua criação, os dois primeiros cursos funcionavam, inicialmente, em algumas salas de aula da Escola Adolfo Augusto de Moraes e no salão paroquial da Igreja Santa Cruz e, posteriormente, na Escola Estadual de 1º e 2º Graus Joaquim Nunes Rocha. O curso de Ciências Contábeis funcionava no prédio da APAE. Os antigos cursos já demandavam espaços maiores, e a criação dos cursos novos, por sua vez, exigiu ainda mais a construção de uma sede própria do Campus, fazendo com que, em abril de 1983, fosse inaugurada a primeira etapa do prédio e feita a transferência dos cursos existentes para as novas instalações, com exceção dos cursos de Ciências Contábeis e Ciências, que continuaram funcionando no prédio da APAE.

O crescimento do município de Rondonópolis e da região sul do Estado exigia a oferta de novos cursos. Tal demanda redundou primeiramente na criação dos referidos cursos de Pedagogia e Letras, em 1981, posteriormente, na implantação dos cursos de História e

Geografia, extinguindo-se, assim, o curso de Estudos Sociais, em 1988, quando também os cursos de Matemática e Biologia substituíram o de Ciências.

Com a Resolução CD nº. 027 de 12 de fevereiro de 1992, que dispôs sobre a reorganização administrativa da UFMT, foi criado o Conselho Administrativo dos Institutos de Rondonópolis (CADIR). Sendo assim, passaram a funcionar, neste Campus, os seguintes Institutos: o Instituto de Ciências Humanas e Sociais (ICHS), que abrange os Departamentos de Educação, Letras, História e Ciências Contábeis; e o Instituto de Ciências Exatas e Naturais (ICEN), compreendendo os Departamentos de Ciências Biológicas, Geografia e Matemática.

Num processo crescente de expansão, o Centro Pedagógico de Rondonópolis desenvolveu o projeto "Unestado", dando sequência à interiorização, iniciada pela UFMT em 1979, mas apenas iniciada neste *Campus* a partir de 1989. Tratava-se de projeto extensionista, com a realização de cursos de atualização em fundamentos didático-pedagógicos para professores da rede pública de ensino das cidades de Pedra Preta, Jaciara, Juscimeira, Poxoréo e Guiratinga.

Esta interiorização teve continuidade com a instalação da Licenciatura Parcelada em Pedagogia no município de Guiratinga a partir de 1995, atendendo a uma clientela específica, composta de cinquenta professores da rede pública de ensino, todos atuantes em municípios circunvizinhos a Rondonópolis. No mesmo ano, foi dado início ao curso de Bacharelado em Ciências Contábeis, no município de Primavera do Leste, atendendo a oitenta alunos daquela região.

As Licenciaturas Parceladas em Pedagogia e Letras, iniciadas em 1996, ano em que ingressaram cento e oitenta alunos, são ministradas no próprio *Campus*, atendendo às demandas dos municípios de Alto Taquari, Campo Verde, Guiratinga, Jaciara, Juscimeira, Paranatinga, Pedra Preta, Poxoréo, Primavera do Leste, São José do Povo e Tesouro.

## 3.2 Abordagem metodológica

Para atingir os objetivos propostos, foram elaboradas quatro etapas metodológicas: pesquisa bibliográfica, Questionários para o aluno, questionários para o professor, simulação do teste Kruskal-Wallis.

A (Figura 4) demonstra as etapas metodológicas do estudo.

**Figura 4 - Fluxograma metodológico**

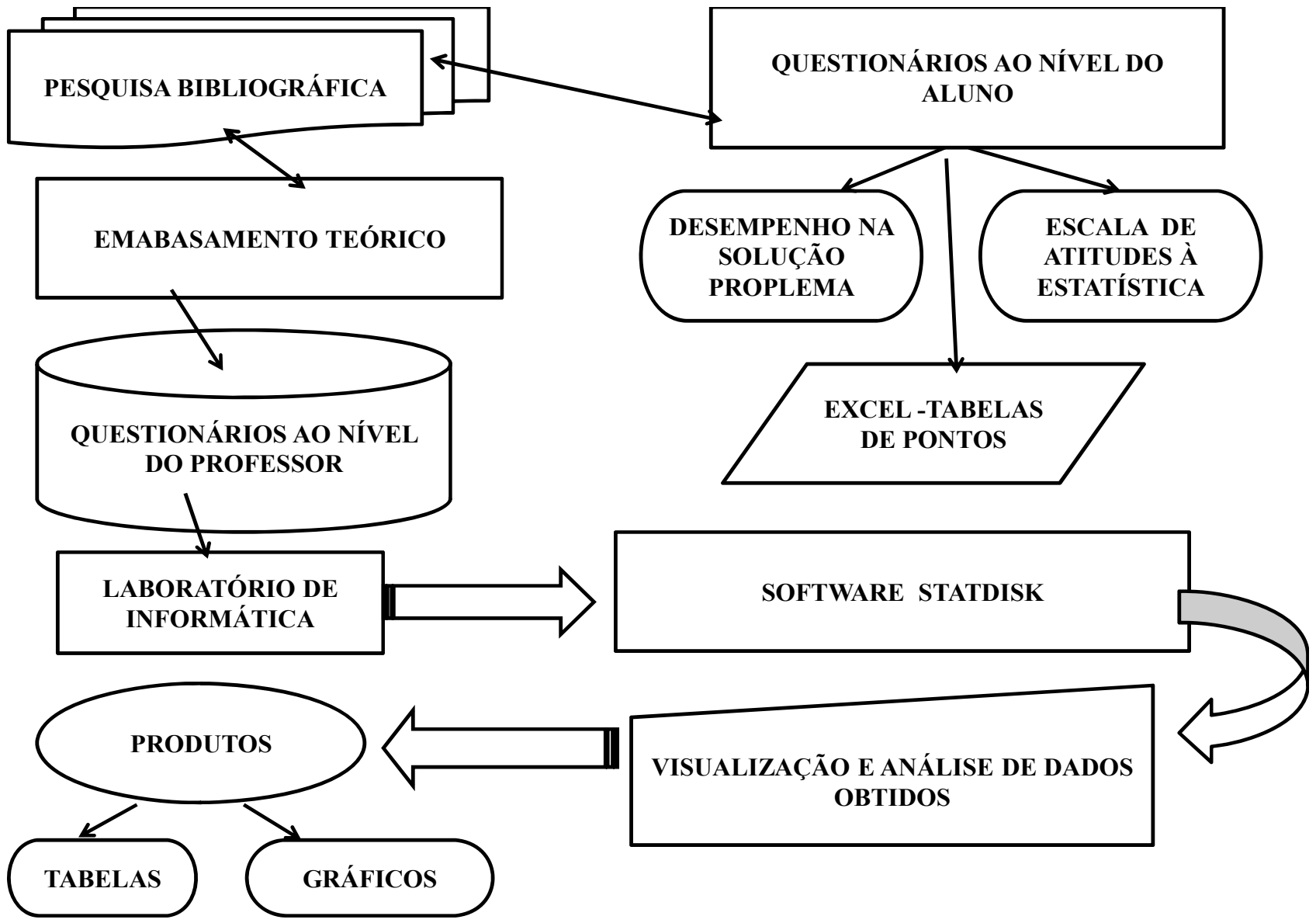

Org.: CAMPOS, Manoel B. N. S. (2008).

De acordo com o fluxograma, a etapa 1 – pesquisa bibliográfica – constituiu no levantamento bibliográfico em livros, dissertações, teses, artigos e no uso da tecnologia na educação. Essa pesquisa possibilitou interação com a temática e com o referencial teórico-metodológico.

> A pesquisa bibliográfica é desenvolvida a partir de material já elaborado, constituído principalmente de livros e artigos científicos. Embora em quase todos os estudos seja exigido algum tipo de trabalho desta natureza, há pesquisas desenvolvidas exclusivamente a partir de fontes bibliográficas. Parte dos estudos exploratórios pode ser definida como pesquisas bibliográficas, assim como certo número de pesquisas desenvolvidas a partir da técnica de análise de conteúdo (GIL, 2007, p. 64).

A fundamentação teórica foi realizada por meio de diversos autores, dentre os quais citamos: Skinner (1972), Bigge (1977), Ponte (1992), Parra (1996), Brito (1998), Silva; Brito (1999), Freire (1999), D'Ambrosio (2000), Vigotsky (2001), Huppes (2002), Novaes (2004) e outros.

A Etapa 2 foi aplicação de questionário aos alunos matriculados na disciplina de Estatística (Apêndice A), para verificar a aceitação e a rejeição da Estatística pelos alunos; e a forma como são dadas as aulas e quais os equipamentos tecnológicos são utilizadas pelos professores para melhorar o processo de ensino-aprendizagem. Nesse sentido, a saber, se o aluno realmente estuda pouco e também achar os motivos da falta de interesse dos mesmos.

A Etapa 3, Questionário aos professores que lecionam a disciplina de Estatística (Apêndice B), objetivou observar a forma de atuação, as metodologias e tecnologias utilizadas pelos docentes. Verificar se as reclamações em relação ao aluno têm realmente fundamento, pois muitos professores reclamam que o aluno não estuda que não presta atenção, que não tem raciocínio, que não tem interesse.

A Etapa 4, Simulação do Teste de Kruskal-Wallis (Apêndice D), foi trabalhada através do software estatístico denominado Statdisk para determinação da significância de uma amostra que foram submetidos a situações de problemas desafiadoras, conseguindo-se, assim, a assimilação do mesmo, necessários para uma aprendizagem efetiva.

## 3.3 Procedimentos metodológicos

Os procedimentos metodológicos deram-se pela construção do marco teórico conceitual, o qual, como define Marconi e Lakatos (2001), possibilita o conhecimento do pesquisador sobre o tema analisado com o aprofundamento teórico. No caso específico desta pesquisa, foram selecionados e analisados estudos que tratavam sobre os temas modelagem, aplicação da estatística, papel do professor e as teorias de aprendizagem.

A pesquisa de campo é a articulação entre a teoria e a prática, e para Marconi e Lakatos (2006, p. 188-189) tem o objetivo de obter "informações e/ou conhecimentos acerca de um problema, para o qual se procura uma resposta, ou de uma hipótese, que se queira comprovar, ou, ainda, descobrir novos fenômenos ou as relações entre eles".

O presente estudo se deu pela forma quali-quantitativa, dos resultados obtidos, o que possibilitou captar a percepção e a atitude do desempenho em relação à estatística.

O tratamento estatístico foi de caráter basicamente descritivo e foram utilizadas: Tabelas de frequência, Diagrama de Dispersão e Reta de Regressão Linear Múltipla – Ajustamento Polinomial, no que diz respeito à força correlacional entre duas variáveis, coeficiente de correlação de Pearson e de Curtose, Teste de significância para os coeficientes de correlação, EXCEL – Análise de variância, sendo que o nível de significância adotado na presente pesquisa foi de p 0,05.

Pesquisa de campo que procurou conhecer e interpretar a realidade do ensino de Estatística numa Instituição de Ensino Superior sem, no entanto, interferir para modificá-la.

### 3.3.1 Universo da população e tamanho da amostra

A pesquisa contou com a participação de dois grupos de atores sociais: os alunos dos cursos de Matemática, Informática, Ciências Contábeis e Enfermagem dos Institutos de Ciências Exata e Naturais/ICEN e Instituto de Ciências Humanas e Sociais/ICHS. Nesse universo, consideram-se 280 alunos matriculados, pois esses cursos são os que ofertam a disciplina de Estatística. Ressaltando que as informações para construir o universo da pesquisa foram fornecidas pelos Departamentos dos referidos cursos.

Os professores do Departamento de Matemática que lecionam a disciplina de Estatística são representados por 20 professores.

Assim de acordo com os dados disponíveis, o número total da população (N) resulta em 300. E o método utilizado, para a elaboração da amostra quantitativa (A), corresponde a 169 pessoas.

O tamanho da população e a amostra foram baseados no método de Krejcie e Morgan (1981), o qual estabelece nível de confiança de 95% e uma margem de erro (E) igual a 5% (Anexo). A população e a mostra foram extraídas por meio das técnicas de Amostragem Casual Simples Estratificada com Partilha Proporcional. A amostragem estratificada de acordo com Berquó (2006, p.137-138), "é aquela em que o pesquisador deseja que as subpopulações sejam representadas na amostra com a mesma proporcionalidade com que compõem a população total".

O (Quadro 1) demonstra como foram extraídas a população e a amostra.

Quadro 1 - Distribuição da Amostragem com Partilha Proporcional

| Estrato | População | Amostra | Relação | Descrição |
|---|---|---|---|---|
| Alunos | 280 | 159 | 0,9408 | Alunos matriculados nos cursos |
| Professores | 20 | 10 | 0,0592 | Professores que lecionam a disciplina de Estatística. |
| Total | 300 | 169 | 0,5633 | - |

Org.: CAMPOS, Manoel B. N. S. (2008)

A entrevista foi executada de forma dirigida através de questionário. Foram entrevistados cento e cinquenta e nove (159) alunos matriculados nos cursos de Matemática, Informática, Ciências Contábeis e Enfermagem e os dez (10) professores do Departamento de Matemática, uma vez que foram selecionados apenas aqueles professores que lecionam a disciplina de Estatística, nos cursos que ofertam a referida disciplina.

### 3.3.2 Instrumentos de coleta de dados

Os instrumentos de coleta de dados e a análise de um grupo específico levaram aos resultados apresentados sem qualquer forma de manipulação por parte do pesquisador, constituindo um estudo que, para Gil (2007), retrata de forma fiel as informações. Gil (1999) define o questionário como uma técnica de investigação composta por um número mais ou menos elevado de questões apresentadas por escrito às pessoas, tendo como objetivo o conhecimento de suas opiniões, suas crenças, seus sentimentos, seus interesses, suas expectativas, suas situações vivenciadas.

Segundo Moroz (2002, p.71),

> A coleta de dados é o momento em que se obtêm as informações necessárias e que será alvo de análise, posteriormente. Deve-se lembrar de que os dados coletados têm uma direção – aquela dada pela questão, enquanto pesquisador pretende-se responder, pelo objetivo que se pretende atingir; mesmo os dados imprevistos só se sabem que são, porque não ocorrem conforme se previa que ocorressem.

A coleta de dados foi desenvolvida em três momentos.

No primeiro momento, foram aplicados questionários para os alunos (Apêndice A) durante os horários de aula, na ausência do professor de Estatística, com o intuito de identificar as percepções dos alunos sobre os aspectos do ensino de Estatística no início e final do semestre.

Mediante uma análise da abordagem do tema em livros didáticos, elaborou-se uma proposta metodológica para ensino de Estatística nos cursos de Matemática, Informática, Ciências Contábeis e Enfermagem. Paralelamente, foram construídos instrumentos didáticos e sequências de atividades para serem aplicados aos alunos. Este material didático foi elaborado e observado, em sua aplicação, em conformidade com as orientações teóricas de estudiosos dos temas de Atividades Investigativas e de Modelagem, pontuadas no decorrer desta tese.

Após a conclusão do trabalho, foi aplicada a prova de Estatística para avaliação de impactos (Apêndice C) e levaram, em média, trinta minutos para a sua realização. Nesse questionário, foi solicitado, também, que os alunos respondessem se a estratégia proposta facilitou a sua compreensão nos conceitos trabalhados em sala de aula. Trata-se de uma prova contendo cinco problemas estatísticos extraídos e adaptados a partir da avaliação semestral feita pelos professores da disciplina Estatística. Todos os problemas admitiam uma estrutura de solução estatística. Essa prova foi aplicada com o objetivo de avaliar também o desempenho dos alunos na solução de problemas.

Quanto à prova de Estatística, conforme já ressaltado, compõe-se de cinco problemas de Estatística em nível superior, envolvendo os seguintes conceitos estatísticos:

1 – Média Aritmética simples e agrupada;

2 – Quartil decil e percentil;

3 – Mediana moda e desvio-padrão;

4 – Coeficiente de variação;

5 – Coeficiente de assimetria.

As questões, por sua vez, foram formuladas junto com o professor de Estatística do curso de Matemática.

No questionário dos alunos, também, contam com a Escala de Likert de quatro pontos no seu grau de concordância ou discordância das declarações relativas à atitude sobre o desempenho das atividades estatísticas (Apêndice A). As perguntas deram aos entrevistados a oportunidade de fornecer respostas claras em vez de respostas neutras e ambíguas, e sua aplicação foi mediante entrevista pessoal, de maneira aleatória intencional. Para a análise Estatística da escala, foi utilizado o coeficiente de correlação de Pearson e de Curtose. E, para a comparação das médias, o Teste-t de Student ao nível de significância de 5%.

> O consumidor constrói níveis de aceitação dos produtos e serviços, conforme suas experiências e influências sociais. Rensis Likert, em 1932, elaborou uma escala para medir esses níveis. As escalas de Likert, ou escalas somadas, requerem que os entrevistados indiquem seu grau de concordância ou discordância com declarações relativas à atitude que está sendo medida. Atribui-se valores numéricos e/ou sinais às respostas para refletir a força e a direção da reação do entrevistado à declaração. As declarações de concordância devem receber valores positivos ou altos enquanto as declarações das quais discordam devem receber valores negativos ou baixos (BRANDALISE, 2005, p. 04).

Neste contexto, Gonçalves (2002, p. 80-81) esclarece a escala proposta por Likert e apresenta os seguintes passos:

> ● Apresentação das afirmações selecionadas a um grupo de pessoas, sendo que deve responder a todos os itens, obedecendo a uma escala de quatro pontos que vai desde o acordo completo à discordância total. Desta forma, as opiniões estão distribuídas entre: Discordo totalmente; Discordo; Concordo totalmente. Isto quer dizer que o sujeito, ao registrar um valor correspondente à sua atitude, não simplesmente concorda ou discorda de uma afirmação, mas indica até que ponto concorda ou discorda em um "continuum" que vai desde o "concordo totalmente" até o "discordo totalmente";
>
> ● Cada uma das afirmações recebe um valor numérico de 1 a 4. Esses valores são substituídos, segundo a direção favorável ou desfavorável de cada afirmação, ou sejas atribui-se às 4 categorias os valores correspondentes a 4, 3, 2 e 1, respectivamente, para os itens favoráveis (atitudes positivas), invertendo-se os resultados para os desfavoráveis (atitudes negativas). O resultado da atitude final de um sujeito é a soma das avaliações isoladas, portanto, a soma do resultado de cada item fornece o resultado final daquele aluno;
>
> ● Realização de uma análise dos itens para verificar quais os que discriminaram mais os sujeitos que obtiveram baixos e elevados resultados na escala total. Um alto escore na escala indica a presença de alto padrão de aceitação da atitude em estudo, enquanto que baixo escore indica o extremo oposto.

Ainda de acordo com o autor, a escala de atitudes com relação à Estatística, utilizada na presente pesquisa, foi validada por Cazorla, Silva, Vendramini e Brito (2000). Assim estão distribuídas as afirmações da referida escala de percepções positiva no (Quadro 2).

**Quadro 2 - A percepção positiva dos alunos**

| Item | Descrição das perguntas |
|---|---|
| 01 | Eu acho a Estatística muito interessante e gosto das aulas de Estatística. |
| 02 | A Estatística é fascinante e divertida. |
| 03 | A Estatística me faz sentir seguro (a) e é, ao mesmo tempo, estimulante. |
| 04 | O sentimento que tenho com relação à Estatística é bom. |
| 05 | A Estatística é algo que eu aprecio grandemente. |
| 06 | Eu gosto realmente da Estatística. |
| 07 | A Estatística é uma das matérias que eu realmente gosto de estudar na faculdade. |
| 08 | Eu fico mais feliz na aula de Estatística que na aula de qualquer outra matéria. |
| 09 | Eu me sinto tranqüilo (a) em Estatística e gosto muito dessa matéria. |
| 10 | Eu tenho uma reação definitivamente positiva com relação à Estatística. Eu gosto e aprecio essa matéria. |

Fonte: Brito (1998)

Dando sequência ao que o autor apresenta, apontamos as percepções negativas Quadro 3.

**Quadro 3 - A percepção negativa dos alunos**

| Item | Descrição das perguntas |
|---|---|
| 01 | Eu fico sempre sob uma terrível tensão na aula de Estatística. |
| 02 | Eu não gosto de Estatística e me assusta ter que fazer essa matéria. |
| 03 | “Dá um branco” na minha cabeça e não consigo pensar claramente quando estuda Estatística. |
| 04 | Eu tenho sensação de insegurança quando me esforço na Estatística. |
| 05 | A Estatística me deixa inquieto (a), descontente, irritado (a) e impaciente. |
| 06 | A Estatística me faz sentir como se estivesse perdido (a) em uma selva de números e sem encontrar a saída. |
| 07 | Quando eu ouço a palavra Estatística, eu tenho um sentimento de aversão. |
| 08 | Eu encaro a Estatística com um sentimento de indecisão, que é resultado do medo de não ser capaz em Estatística. |
| 09 | Pensar sobre a obrigação de resolver um problema estatístico me deixa nervoso (a). |
| 10 | Eu nunca gostei de Estatística e é a matéria que me dá mais medo. |

Fonte: Brito (1998).

Para obter o resultado final da concordância ou discordância das declarações relativas à sua atitude, foram somados todos os pontos que determinaram à média e o desvio padrão dessas declarações. Estas devem oportunizar ao entrevistado expressar respostas claras em vez de respostas neutras, ambíguas.

No segundo momento, foram aplicados questionários para seus professores (Apêndice B). No final do segundo trimestre, numa sala de estudo dos professores do Departamento de Matemática, sempre em horário previamente combinado e de acordo com a disponibilidade dos docentes, o questionário foi respondido por dez professores, cujos dados revelariam o perfil dos que lecionam a disciplina de Estatística.

No terceiro momento, foi à simulação do teste de Kuskal-Wallis (Apêndice D), que foi trabalhado através do software estatístico denominado Statdisk para determinação da significância de uma amostra, a qual foi submetida a situações de problemas desafiadoras, conseguindo-se, assim, a assimilação do mesmo e necessário para uma aprendizagem efetiva.

Reconhece-se a importância de se analisar o que se passa em sala de aula, especialmente em situações de ensino e aprendizagem, visando a sua melhoria.

### 3.3.3 Análise e representação dos dados

Análise dos dados coletados foi processada através da técnica quali-quantitativa apresentando a percepção das habilidades em Matemática e Estatística. De acordo com Marconi e Lakatos (2006, p. 169), "uma vez manipulados os dados e obtidos os resultados, o passo seguinte é a análise e interpretação dos mesmos, constituindo-se ambas no núcleo central da pesquisa".

A representação dos dados se deu em forma de quadros, tabelas, gráficos, figuras, fotos e na apresentação de discussão dos dados coletados (Capítulo 4).

### 3.3.4 Categorias de Análise e Variáveis da Pesquisa

No intuito de responder aos objetivos propostos no presente estudo, foi estabelecida a seguinte categorização de análise: a) a percepção dos alunos no desempenho na solução de problemas e a escala de atitudes em relação à Estatística; b) e a dos professores que lecionam a disciplina de Estatística. Foram feitos contatos com a Coordenação de Ensino dos cursos que ofertam a disciplina para a obtenção de autorização do Departamento, do professor da disciplina e dos alunos selecionados para o desenvolvimento das atividades didáticas.

Segundo Sturza (2006), a percepção é um processo dialético que absorve sujeito (homem) e objeto (lugar), realizando relações entre ambos, dito de outra forma, as interfaces objetivas e subjetivas, expressas ou obscurecidas, entre a globalização e a individualidade. A percepção, a vivência e a memória dos indivíduos e dos grupos sociais são elementos importantes na constituição do saber geográfico, da produção do espaço e da paisagem, que se perfaz a partir do imaginário social.

Cada pessoa interpreta o lugar de acordo com a vivência, envolvimento e situação na temporalidade histórica. A feição ao lugar acontece, pois, conforme Barbosa (2008, p. 4) destaca, os moradores estabelecem a relação entre a infância vivida e as áreas verdes do local "o apego ao lugar, por ser familiar, pela natureza, por representar o passado e pela localização, é o orgulho dos moradores. [...]. A percepção, a atitude e o valor que inferem ao meio ambiente mantêm suas características de visões de mundo muito semelhantes".

De acordo com os aspectos caracterizados, as variáveis da pesquisa são:

- Opinião (positiva ou negativa) em relação ao nível de percepção e do desempenho em Estatística dos alunos;
- A soma dos pontos nas vinte proposições da Escala Likert em relação a atitudes dos alunos em relação à Estatística;
- Características (gênero, idade, grau de escolaridade, renda familiar, atividade didáticas, dentre outros) dos alunos e professores participantes da pesquisa.
  Cabe ressaltar que as variáveis são aspectos, propriedades ou fatores reais ou potencialmente mensuráveis pelos valores que assumem e discerníveis em um objeto de estudo (CERVO; BERVIAN, 2002).

A perspectiva de análise introduzirá os elementos interdisciplinares na discussão do ensino de estatística, enquanto processo e não apenas uma modalidade do ramo da matemática.

# 4 RESULTADO E DISCUSSÃO

## 4.1 Caracterização das estratégias didáticas metodológicas para o ensino de estatística

Para provocar a aprendizagem do tipo reflexivo, o professor deve começar por algum problema, por alguma pergunta, por alguma narração estimuladora da reflexão, por alguma dramatização, por algo que ponha em funcionamento os esquemas mentais de ação com os quais o aluno irá assimilar a teoria a ser estudada.

Diante de tal fato, buscou-se caracterizar por meio de resolução de problemas, o intuito de promover melhor aproveitando, das estratégias didáticas, com descrição das atividades metodológicas, e promovendo a melhoria da qualidade de ensino nos referidos pesquisados.

Apresentamos passo a passo os procedimentos que compõem a proposta Metodológica. A descrição desta proposta metodológica de aprendizagem para o ensino de Estatística está exposta no (Quadro 4).

**Quadro 4 - Descrição da Proposta Metodológica das Atividades Estatística**

| **Nome** | **Descrição metodológica** |
|---|---|
| **Professor** | Deve estar atento para estimular o desejo de aprender do educando. Deve ser competente, dedicado, criativo, modelo, comunicativo, amigo e orientador, elo de compreensão da vida. Deve levar o aluno a pensar em atividades que o desafiem. Deve ter prazer em ensinar, com linguagem simples, clara e objetiva, aliando a teoria à prática, atualizar-se sem parar, abrir-se para as informações que o aluno traz, aprender e interagir com ele.<br>Buscar temas relevantes à realidade do aluno. |
| **Diagnóstico** | O professor deve traçar um diagnóstico da realidade e dos temas de interesse dos educando para trabalhar os conteúdos relativos a esses temas. |
| **Dialogar, cativar, motivar** | Para melhorar a aprendizagem, é necessária uma atmosfera de diálogo que favoreça o ato de pensar. O diálogo é necessário porque a forma de pensar do aluno nem sempre coincide com a forma expressa pelo professor. Cabe ao professor cativar o aluno para uma participação ativa na produção do seu conhecimento, por meio da troca de experiências. Cada um, expondo sua visão sobre os assuntos, sendo tão importante o que o aluno fala e pensa, quanto o que o professor fala e pensa.<br><br>Construir o conhecimento num ambiente que motive o aluno para a exploração, a reflexão e descoberta dos conceitos. |
| **Perguntas, problemas, narrações, Dramatizações** | Cabe ao professor a problematização, buscando superar uma primeira visão mágica por meio de uma visão crítica, numa ida e vinda do concreto ao abstrato e do abstrato ao concreto. Começando por algum problema, pergunta dramatização ou narração, por algo que ponha em funcionamento os esquemas mentais de ação. |
| **Criar conflitos, contradições** | O diálogo gera crítica e problematização. Ao professor cabe então tomar os diferentes significados atribuídos à situação matemática pelos alunos, captando o que realmente o aluno está pensando a partir de sua vivência do passado e do presente. Procurar problematizá-la, criar conflitos, ter como objetivo a exploração de contradições, porque através das dúvidas e contradições se efetiva a aprendizagem. |
| **Levar o mundo do aluno a sala de Aula** | Cabe ao professor devolver ao educando de forma organizada e sistematizada aqueles elementos que este lhe entregou de forma desestruturada. Interagir com o aluno que almeja a aquisição de conhecimentos sistematizados. Matematizar situações reais, que desenvolvam a capacidade de criar teorias adequadas para as situações mais diversas. Descobrir como é a realidade sob o ângulo do |
| | pensamento do aluno. Normalmente o conhecimento não é totalmente novo, partimos do que já foi construído, do que já está disponível, do conhecimento que está diante de nós e o refazemos, o reelaboramos. |
| **Preparação, ativação, criação de esquema** | Sabemos que não há saltos nas assimilações, mas assimilação do novo a partir do antigo. As fontes de estímulo devem ser sensibilizadas, provocando a ativação, estruturação ou criação de esquemas capazes de assimilar os novos objetos, aplicáveis às situações em que os alunos serão colocados. Uma forma de ativação de esquemas é através do uso de elementos mediadores. |
| **Recursos** | São os elementos mediadores da aprendizagem que deverão ser utilizados para ativação, estruturação ou criação de esquemas, que podem ser tecnológicos ou materiais concretos. |
| **Tecnológicos** | Videocassete, televisão, software, Internet, e-mail, utilizados como elementos mediadores da aprendizagem. |
| **Materiais concretos** | Ábaco, tangram, materiais manipuláveis, jogos e brincadeiras que envolvam conceitos e operações matemáticas. |
| **Fazendo, agindo, experimentado, explorando** | O modo mais natural, intuitivo e fácil de aprender é fazendo, agindo, experimentando. Isso é mais do que uma estratégia de aprendizagem: é um modo de ver o ser humano que aprende. Ele aprende pela experimentação e ação sobre o mundo em que vive. |
| **Percepção da nova situação** | O aluno deve perceber, claramente, que está diante de uma situação nova, para a qual necessita buscar esquemas na experiência vivencial, como se a nova situação fosse uma diversificação ou generalização de conhecimentos já dominados. |
| **Acomodação à nova situação** | Combinação de esquemas ou modificação de esquemas para resolver problemas que venham de experiências novas dentro do ambiente. Quando são |
| | estabelecidas ligações entre a nova situação e a vivência do aluno, haverá acomodação dos esquemas que servem à resolução do novo problema. |
| **Assimilação do novo esquema** | Em situações novas em que não há esquemas prévios, o professor deve começar pela preparação de esquemas, através da estruturação de experiências da vida do aluno. Antes de abordar um novo assunto deve criar nos alunos condições de assimilação para os assuntos a serem estudados, pois o aluno só recebe o estímulo se estiver preparado para recebê-lo. |
| **Adaptação** | O aluno deve chegar ao equilíbrio entre acomodação e assimilação. |
| **Aluno** | Ao aluno como sujeito da história deve estar voltada a atenção no ato de ensinar, sendo o alvo de toda aprendizagem. Ele chega à escola com um saber herdado da vivência familiar, da sociedade em que vive, do contato com os meios de comunicação e espera que esse saber seja lapidado e se torne um saber científico. |
| **Objetivo definido** | O aluno deve ter alguma meta definida a ser atingida. É preciso querer, acreditar e perseverar no objetivo a ser atingido. |
| **Desejo e vontade de aprender** | O desejo e vontade decidida de aprender, de descobrir, de ampliar conhecimento e experiência são intrínsecos do homem. |
| **Participativo** | A aprendizagem é facilitada quando há participação efetiva do aluno nesse processo, escolhendo ele mesmo as direções, decidindo a ação a seguir e vivendo as consequências da escolha, por isso é necessário que seja participativo e não um mero objeto da ação educativa. |
| **Não ter medo de errar** | O aluno deve ser acolhido no seu erro. O erro não pode ser fonte de castigo, pois é suporte para a autocompreensão. |
| **Propor a si situações desafiadoras** | O estudante deve estar em confronto direto com problemas práticos e de pesquisa. A atividade que conduz a aprendizagem é a atividade do sujeito construindo seu próprio conhecimento, por isso o aluno deve desafiar-se a si mesmo. |
| **Buscar a compreensão** | O sentido que o indivíduo procura não deve ser dado, imposto ou recebido. Deve ser conquistado. |

| Buscar novas soluções | Levar o aluno a reinventar aquilo de que é capaz.<br><br>Buscar soluções alternativas para os problemas propostos. |
|---|---|
| **Aprendizagem do aluno** | O objetivo final de toda proposta é a aprendizagem do aluno. Para que um novo instrumento lógico se construa, é preciso sempre instrumentos lógicos preliminares. Construção de estruturas através de métodos ativos que envolvam experimentação, reflexão e descoberta. |

Fonte.: CAMPOS, M. B. N. S. (2008)

## 4.2 Atores sociais, alunos e professores

Esta pesquisa contou com a participação de dois grupos de atores sociais: os alunos dos cursos de Matemática, Informática, Ciências Contábeis e Enfermagem dos Institutos de Ciências Exatas e Naturais/ICEN e Institutos de Ciências Humanas e Sociais/ICHS, os professores do Departamento de Matemática que lecionam a disciplina de Estatística.

A análise do (Gráfico 2) nos mostra que a maioria dos alunos está na faixa etária entre dezenove a vinte e quatro anos de idade, seguido de longe pelos alunos cuja faixa etária se situa entre vinte e cinco a trinta anos. Ou seja, a Universidade está a cada dia recebendo alunos mais jovens.

**Gráfico 2 - Faixa etária dos alunos**

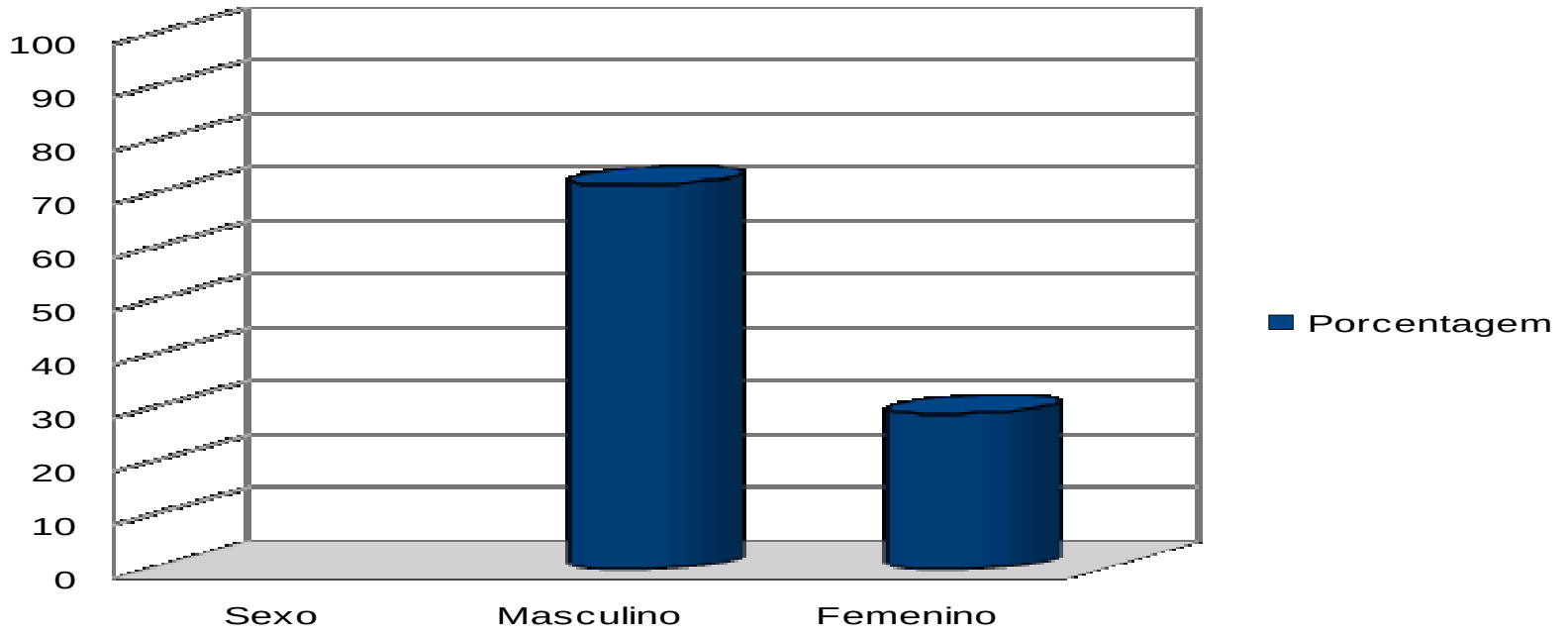

Fonte: Construção do autor

No (Gráfico 3), verifica-se que o percentual de homens é bem superior a 71,90% ao das mulheres. Isso pode ser explicado pela característica dos cursos das Ciências Exatas e Naturais, em que os homens têm mais predisposição em cursá-los.

**Gráfico 3 - Sexo dos alunos**

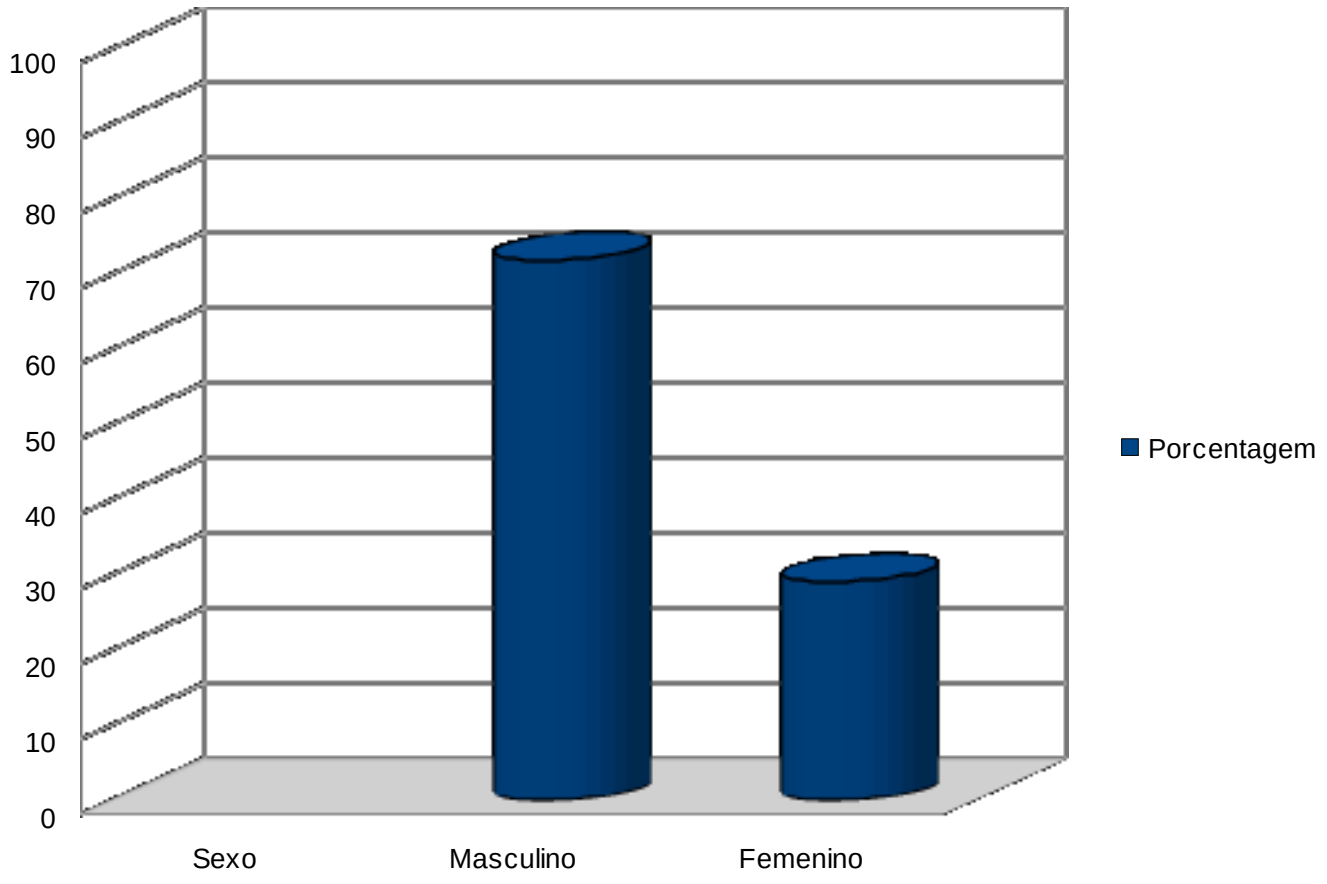

Fonte: Construção do autor

Verifica-se que não há diferenças significativas em relação à renda familiar. A baixa renda familiar, certamente, interfere na aprendizagem do aluno, pois estes têm acesso restrito ao conhecimento escrito através de revistas, jornais, periódicos e ao conhecimento disponível nas tecnologias em emergência, principalmente, o computador e a Internet, que são de grande importância quando usados para pesquisa. Esses alunos dependem quase que, exclusivamente, dos materiais disponíveis na biblioteca do *Campus* (Gráfico 4).

**Gráfico 4 - Renda familiar dos alunos**

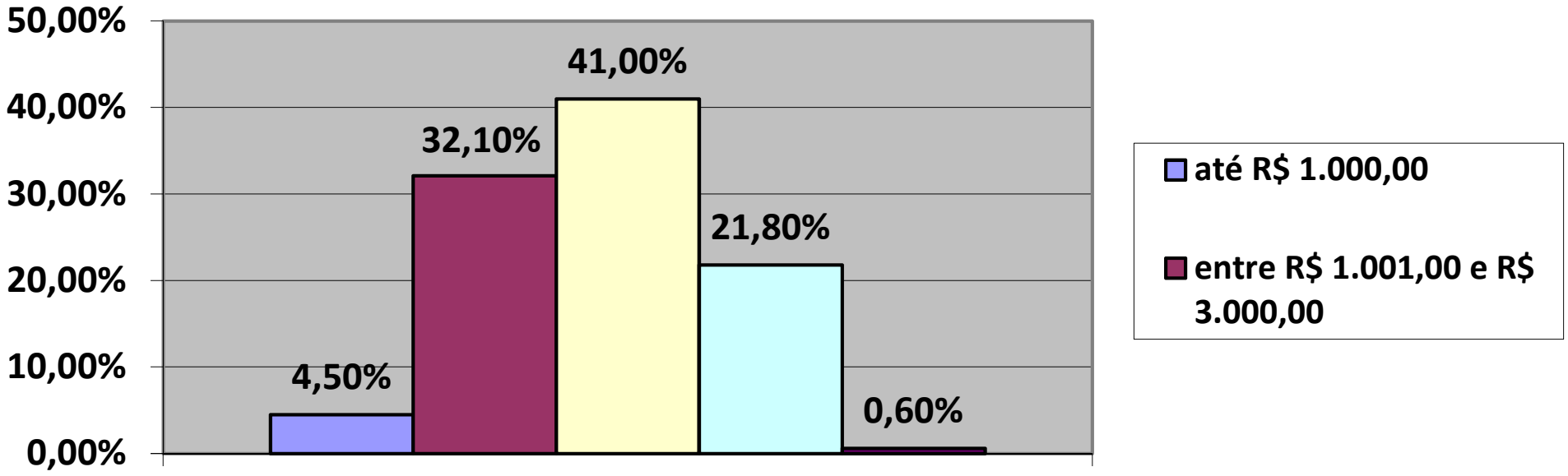

Fonte: Construção do autor

Quanto ao índice do grau de escolaridade familiar, observou-se a predominância do ensino superior completo, o que pode revelar que o ensino é desenvolvido por pessoas com mais elevado nível de conhecimento acadêmico. Essa alta escolaridade contradiz a baixa renda familiar que muitas famílias têm, conforme (Gráfico 5).

**Gráfico 5 - Grau de escolaridade familiar dos alunos**

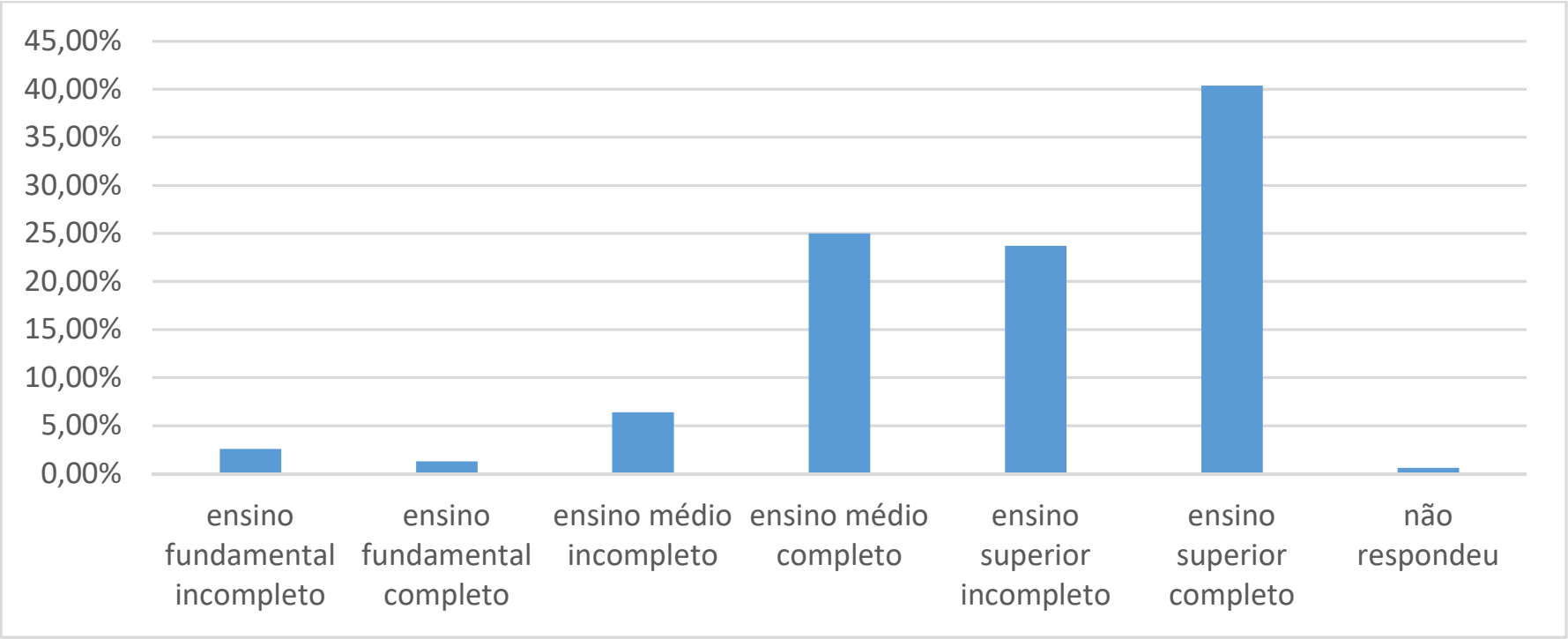

Fonte: Construção do autor

Da população investigada, a maioria 54,55% encontra-se cursando o terceiro ano. Isso pode ser explicado pelo fato de que o objetivo da pesquisa seria entrevistar alunos que estão cursando a disciplina (Gráfico 6).

**Gráfico 6 - Ciclo**

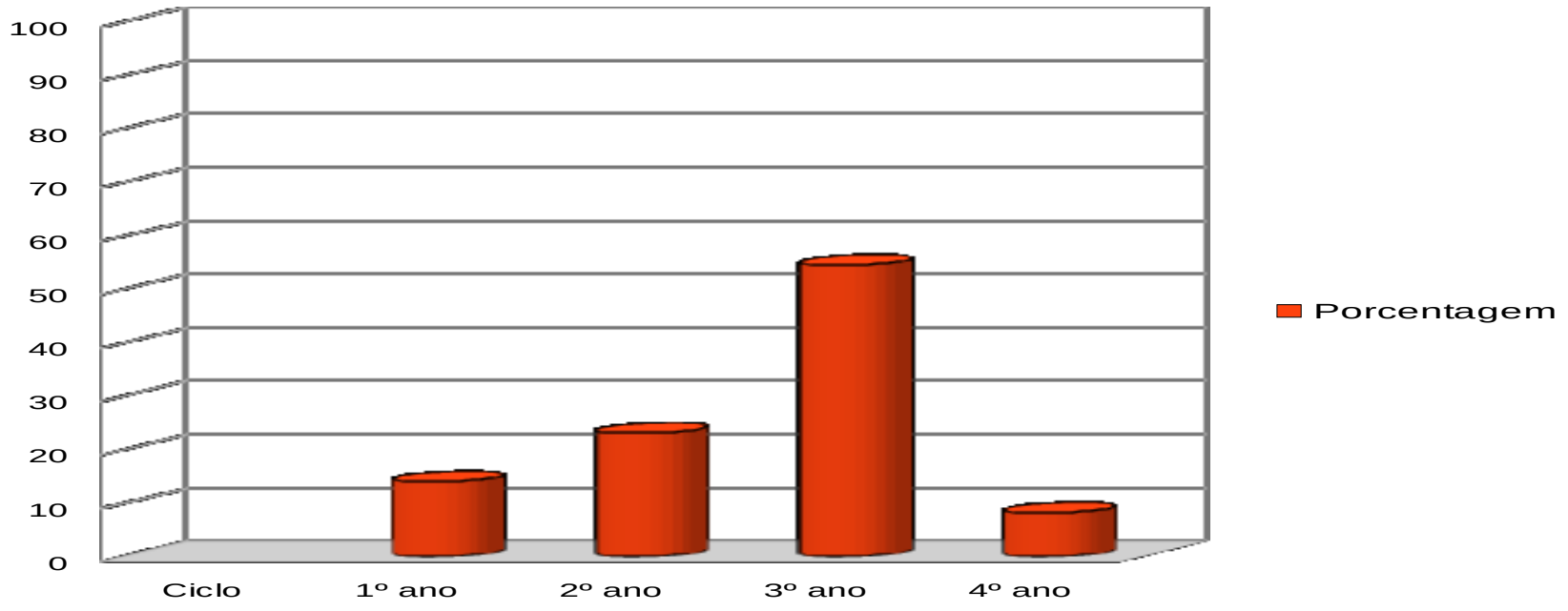

Fonte: Construção do autor

Quando questionados se o professor havia apresentado o plano de ensino, a maioria respondeu que não ou que não sabia, para 50,41%. Esta é uma situação preocupante, pois é o plano de ensino que norteará todo o desenvolvimento da disciplina (Gráfico 7).

**Gráfico 7 - O professor apresentou e discutiu o plano de ensino com os alunos?**

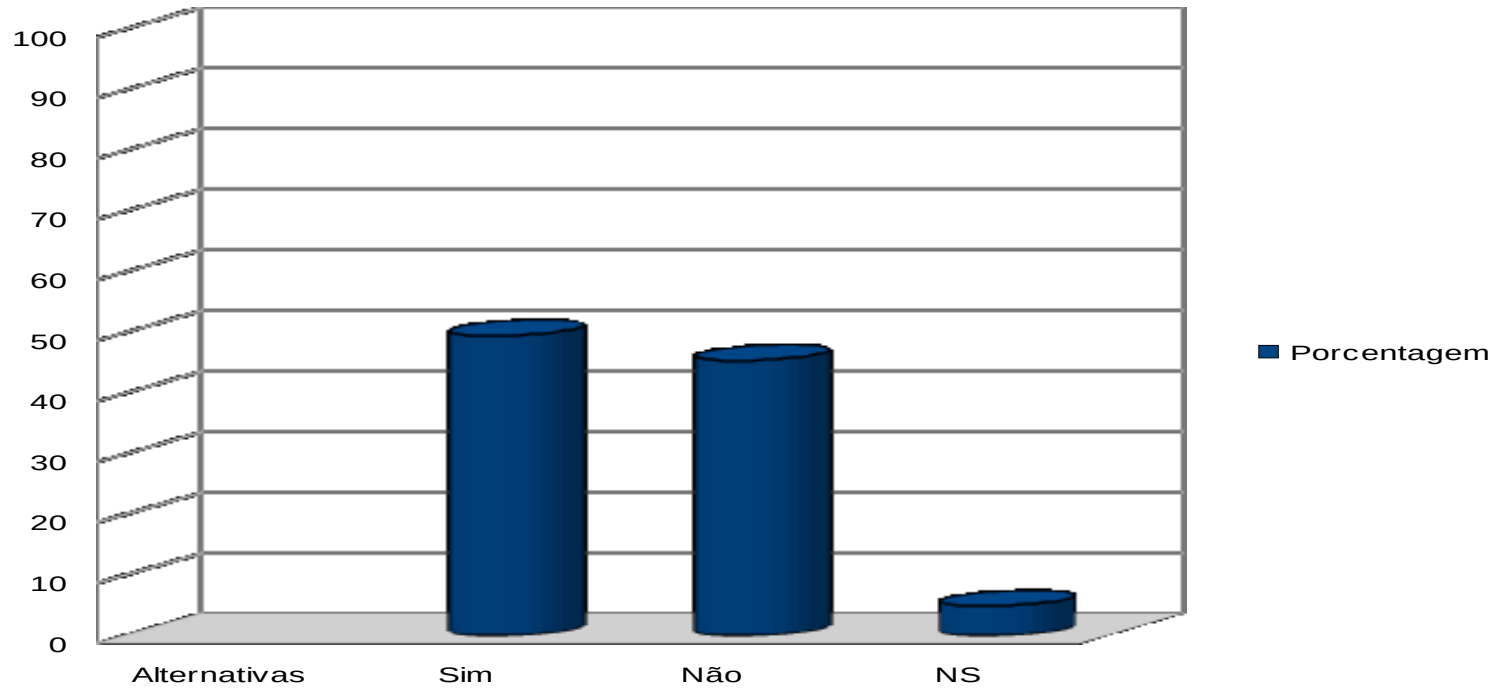

Fonte: Construção do autor

O (Gráfico 8) mostra um dado preocupante, pois, para 48,76% dos entrevistados, o professor não fez a articulação entre a teoria e prática, e isso pode comprometer o interesse dos alunos pela disciplina, pois eles não estarão vendo onde a estatística é empregada, podendo até fazer com que eles não dêem a devida atenção para a disciplina.

**Gráfico 8 - Houve articulação entre a teoria e prática para o desenvolvimento do processo de aprendizagem?**

Fonte: Construção do autor

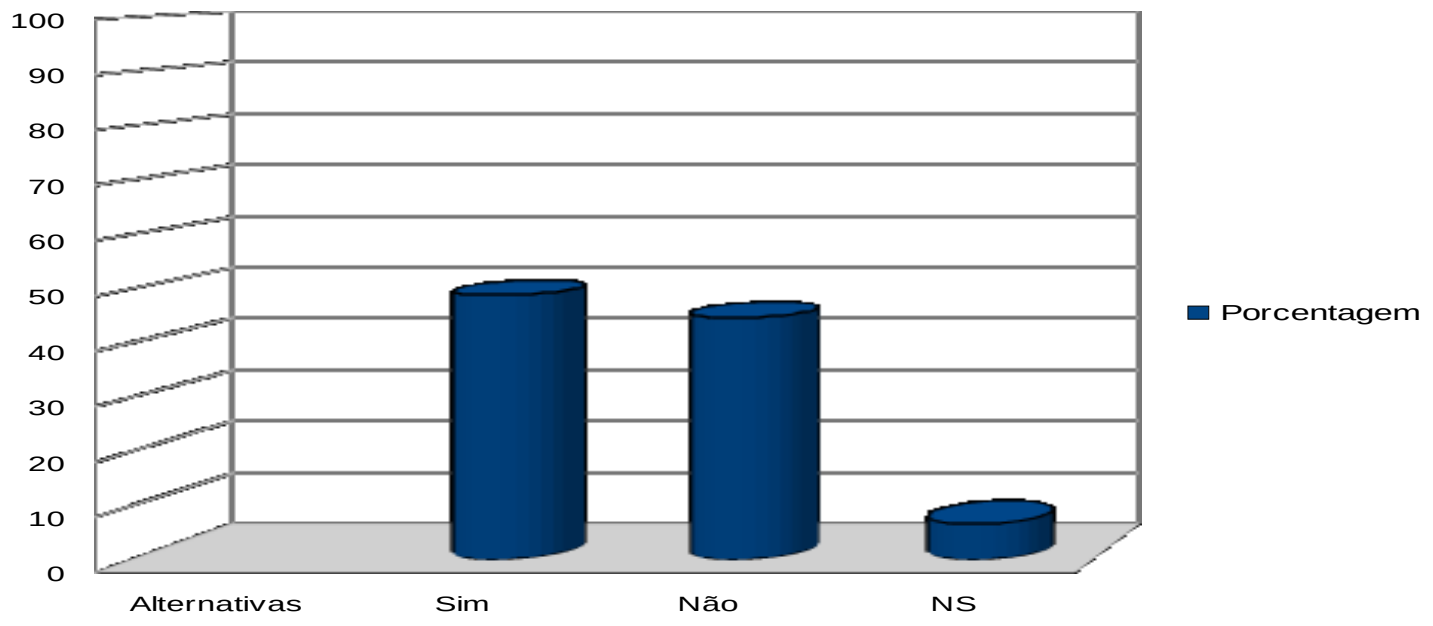

Quando indagados se os objetivos da disciplina foram alcançados, mais da metade deles 50,41% afirma que não ou não soube responder. É um dado preocupante também e deverá ser discutido com os professores que ministram a disciplina, visando à melhoria do processo ensino-aprendizagem (Gráfico 9).

Gráfico 9 - Os objetivos da disciplina foram alcançados?

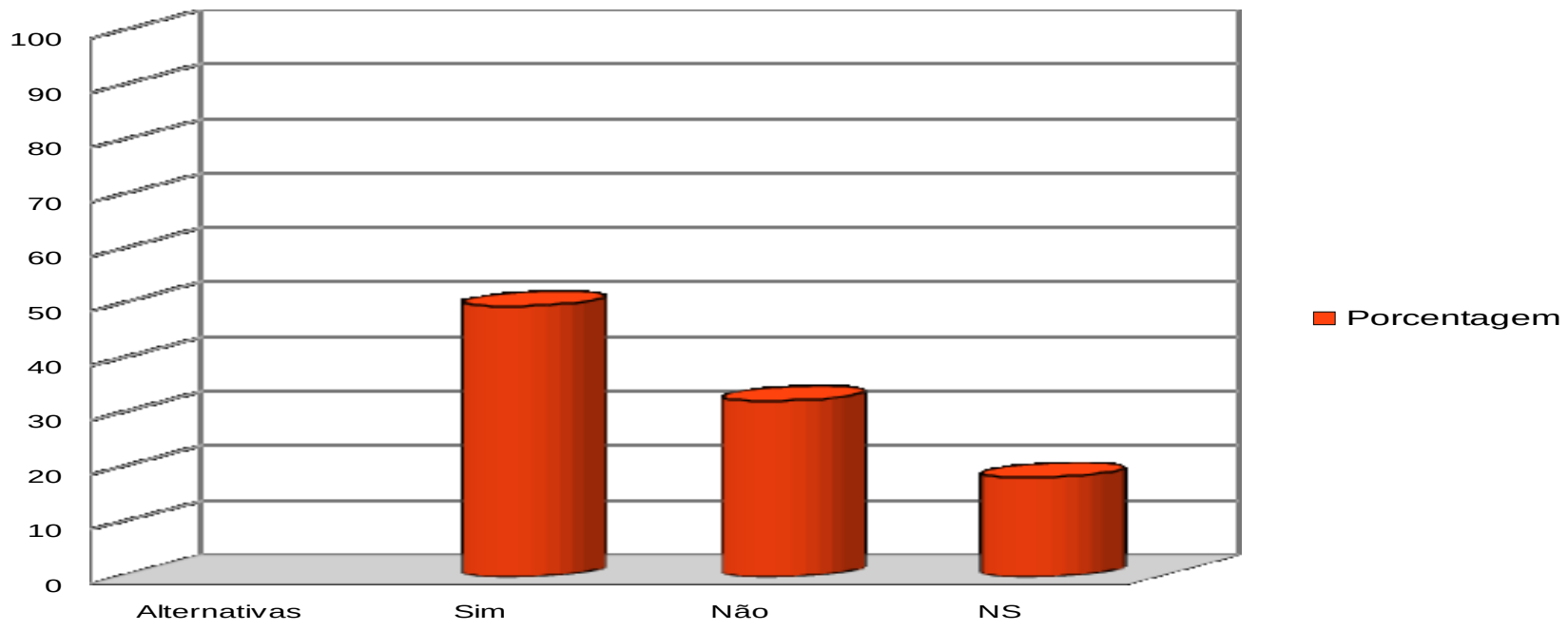

Fonte: Construção do autor

Para a maioria, 51,24%, a Estatística incentivou a participação dos alunos em sala. O processo de ensino-aprendizagem não deve simplesmente transmitir técnicas e conceitos para os alunos, mas incentivá-los a serem críticos, participativos, cientes da realidade da qual fazem parte (Gráfico 10).

**Gráfico 10 - A disciplina de Estatística incentivou a participação dos alunos em sala de aula**

Fonte: Construção do autor

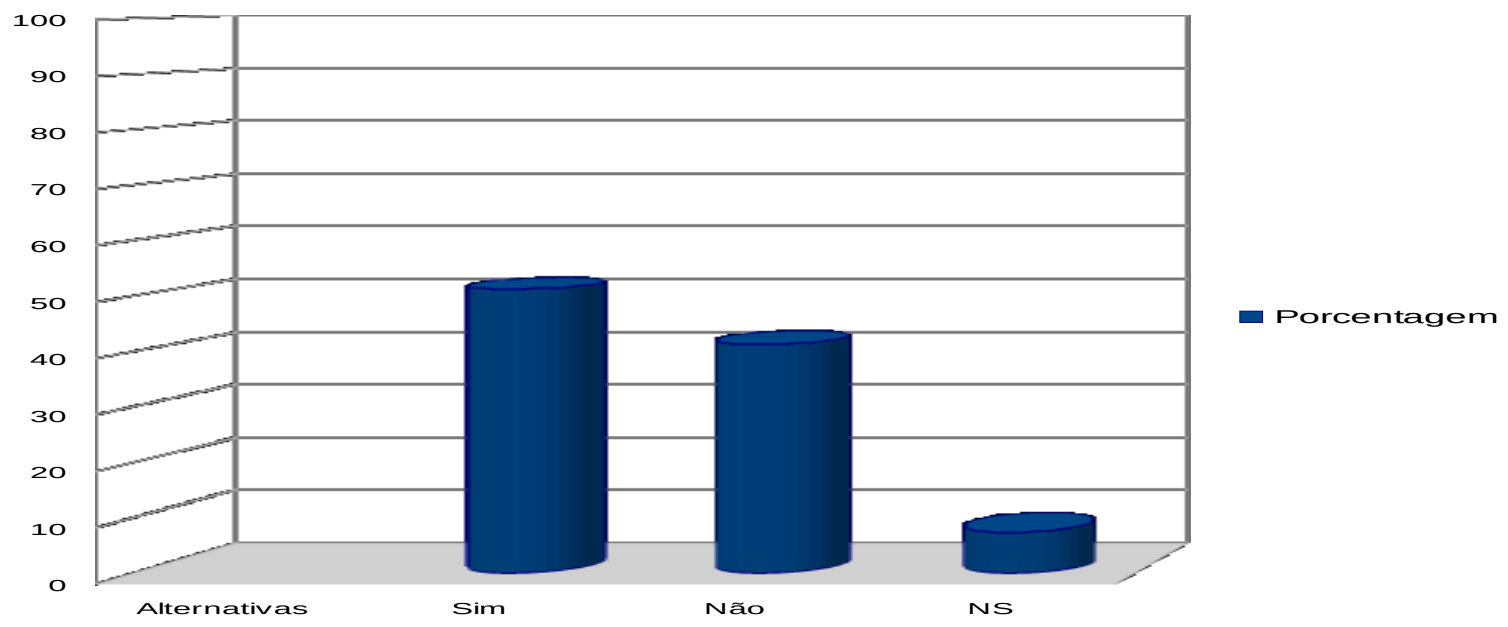

Em relação à estrutura física, o (Gráfico 11) representa a distribuição da estrutura física. Conforme os dados, 83,47% de nossos alunos disseram que a estrutura física interfere no processo de ensino-aprendizagem, o que demonstra que eles precisam ter boas condições físicas para que tenham um bom rendimento no processo.

**Gráfico 11 - Você acha que a estrutura física interfere no processo ensino-aprendizagem?**

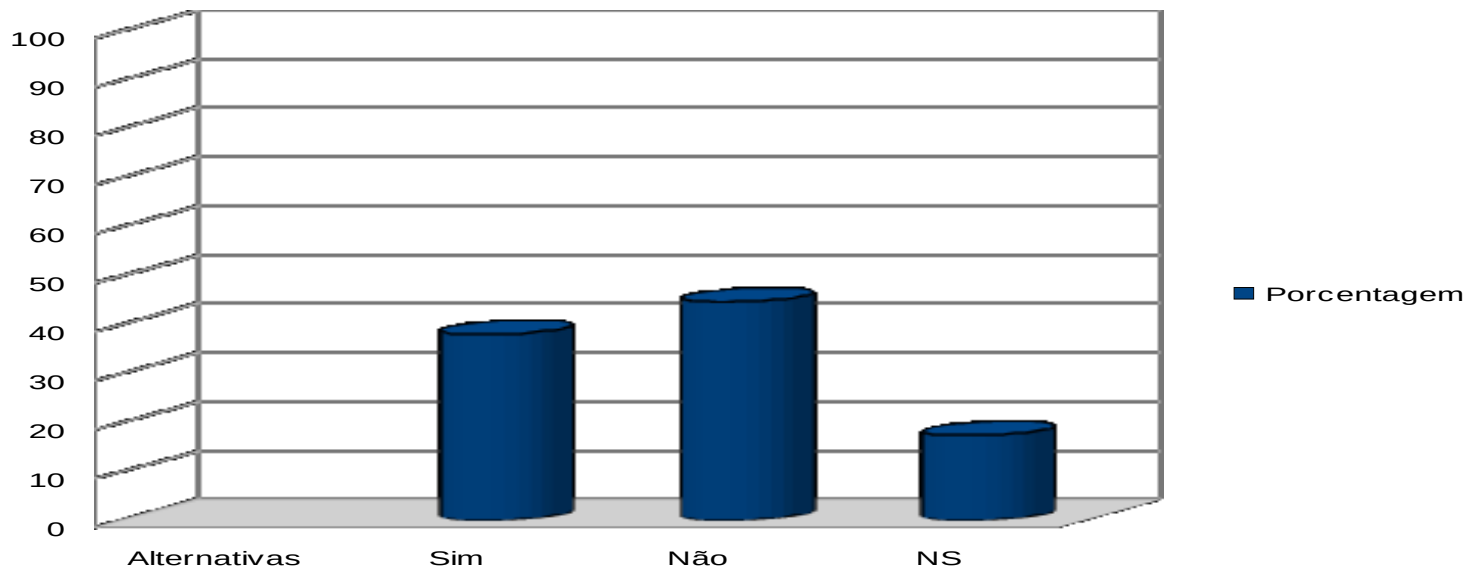

Fonte: Construção do autor

Para a maioria, 44,63%, não houve diversificação das avaliações por partes dos professores. É fundamental hoje em dia que seja propiciado aos alunos diversificação de atividades em sala de aula, e estas podem ser avaliadas à sua maneira, fazendo com que os alunos conheçam diferentes técnicas de aprendizado e, por consequência, diversas maneiras de serem avaliados (Gráfico 12).

**Gráfico 12 - Houve diversificação dos tipos de avaliação de modo a promover aprendizagem?**

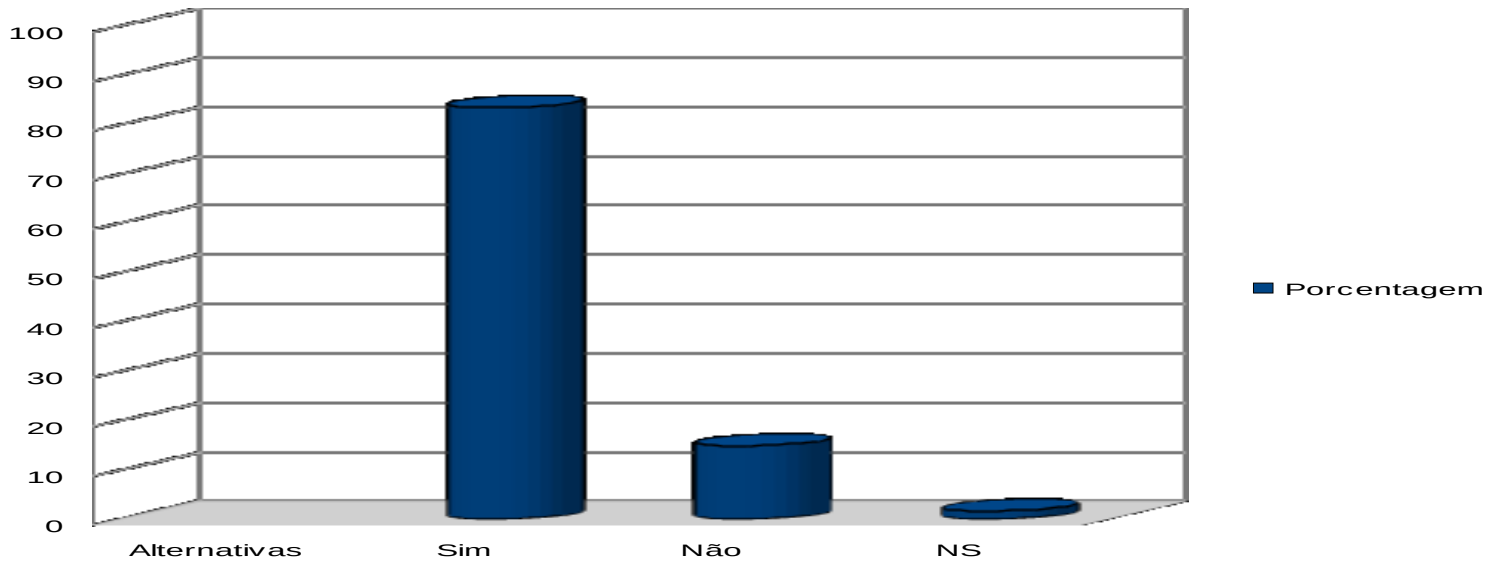

Fonte: Construção do autor

O (Gráfico 13) mostra que a maioria dos alunos não busca complementar seus estudos com outras fontes, se não as que são indicadas pelo professor. Isso pode ser uma falha por parte dos alunos, pois quanto mais informações tiverem acesso, melhor será seu desempenho na disciplina e por consequência mais sucesso ao solucionar problemas que envolvam a estatística.

Gráfico 13 - Além da bibliografia recomendada pelo professor, você buscava outras fontes para complementar os estudos

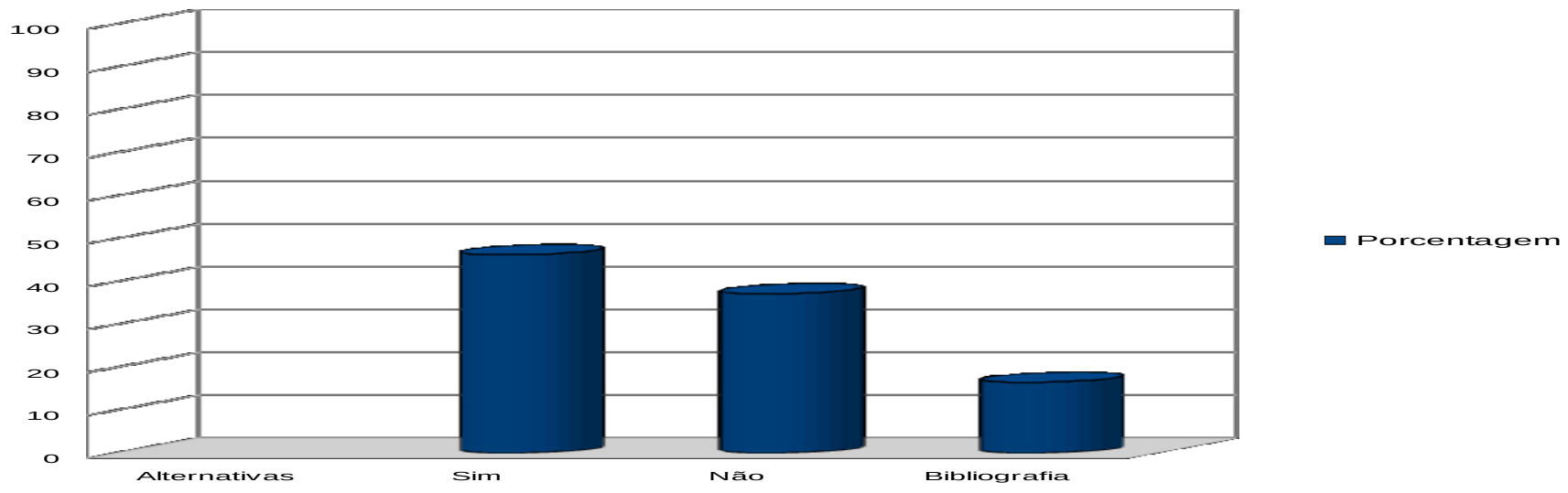

Fonte: Construção do autor

Detecta-se que 68,60% dos alunos afirmaram ser assíduos nas aulas de estatística. Isso leva a uma reflexão, como pode um aluno aprender se não está presente para receber os conhecimentos, principalmente em uma disciplina de exatas, em que os conteúdos são, em boa medida, interligados. E se o aluno não esteve presente durante a exposição do conteúdo, possivelmente, no próximo, terá dificuldades no entendimento (Gráfico 14).

**Gráfico 14 - Você era assíduo às aulas de Estatística?**

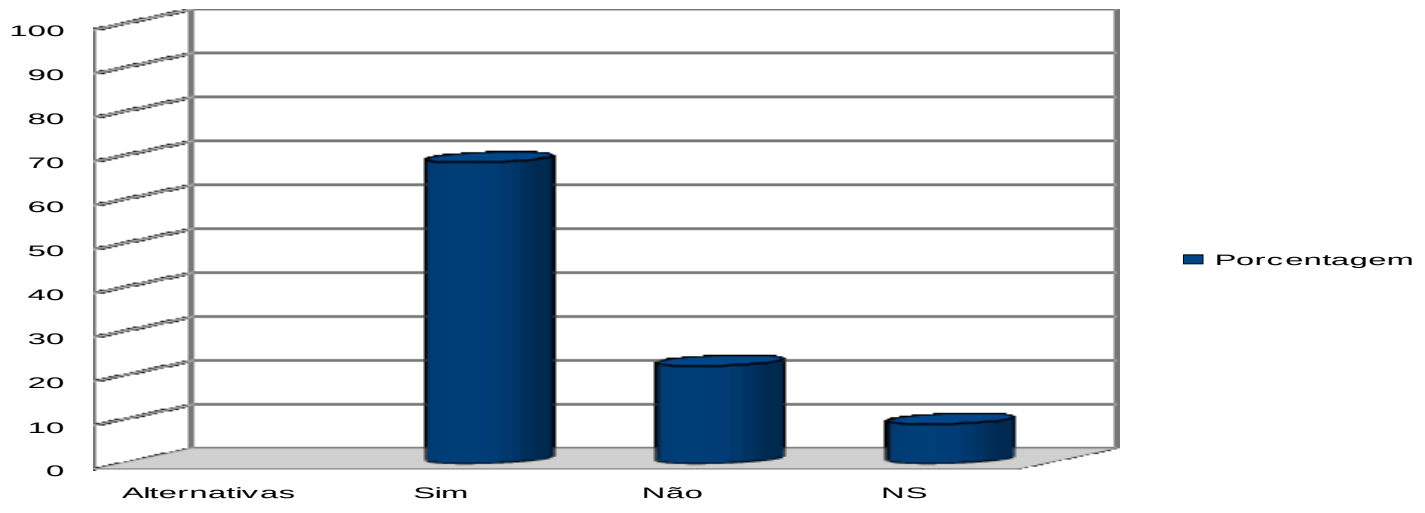

Fonte: Construção do autor

Segundo a maioria, 44,63%, o professor não discutia os resultados das avaliações. Essa informação é algo preocupante, pois a avaliação pode ser usada para detectar possíveis falhas tanto do professor quanto dos alunos, e detectando-as, deverão corrigi-las da melhor maneira possível (Gráfico15).

**Gráfico 15 - O professor discutia os resultados das avaliações?**

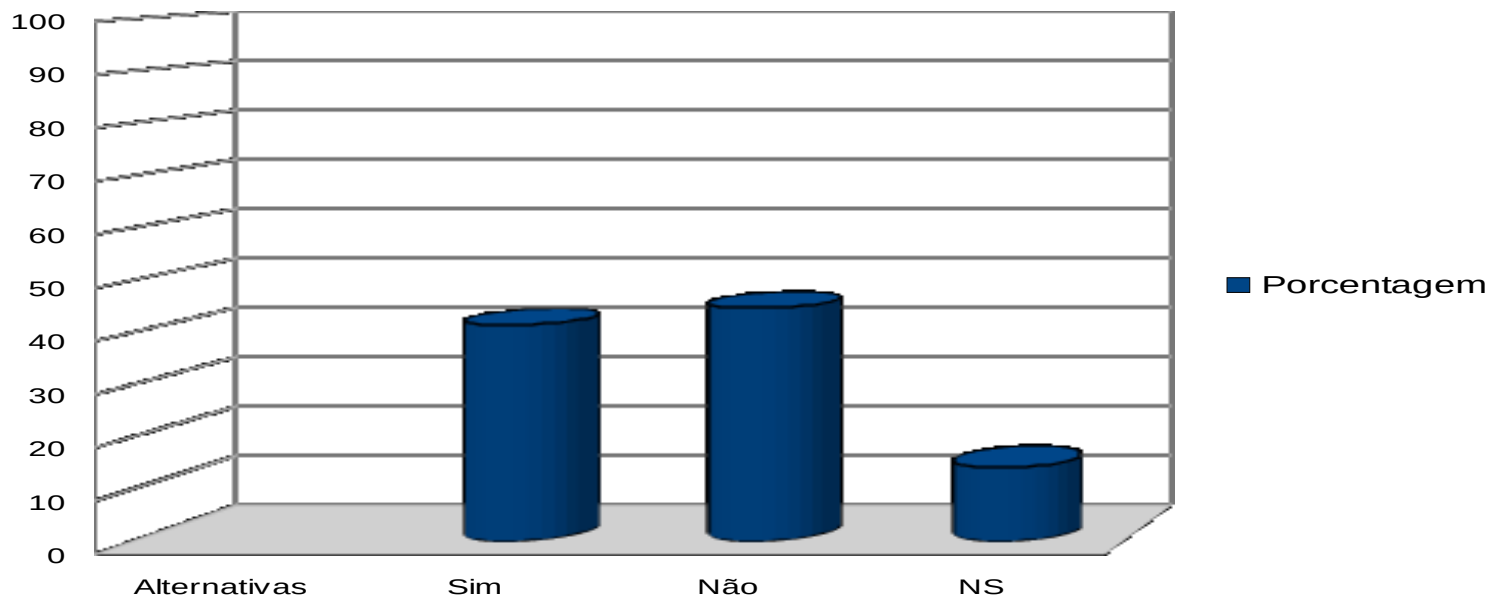

Fonte: Construção do autor

O (Gráfico 16) mostra que apenas 2,48% dos alunos não procuram sanar suas dúvidas quando não entenderam algum conteúdo. O restante procura o professor ou seus colegas para solucionar as dúvidas.

**Gráfico 16 - Quando não entendia um conteúdo, você...**

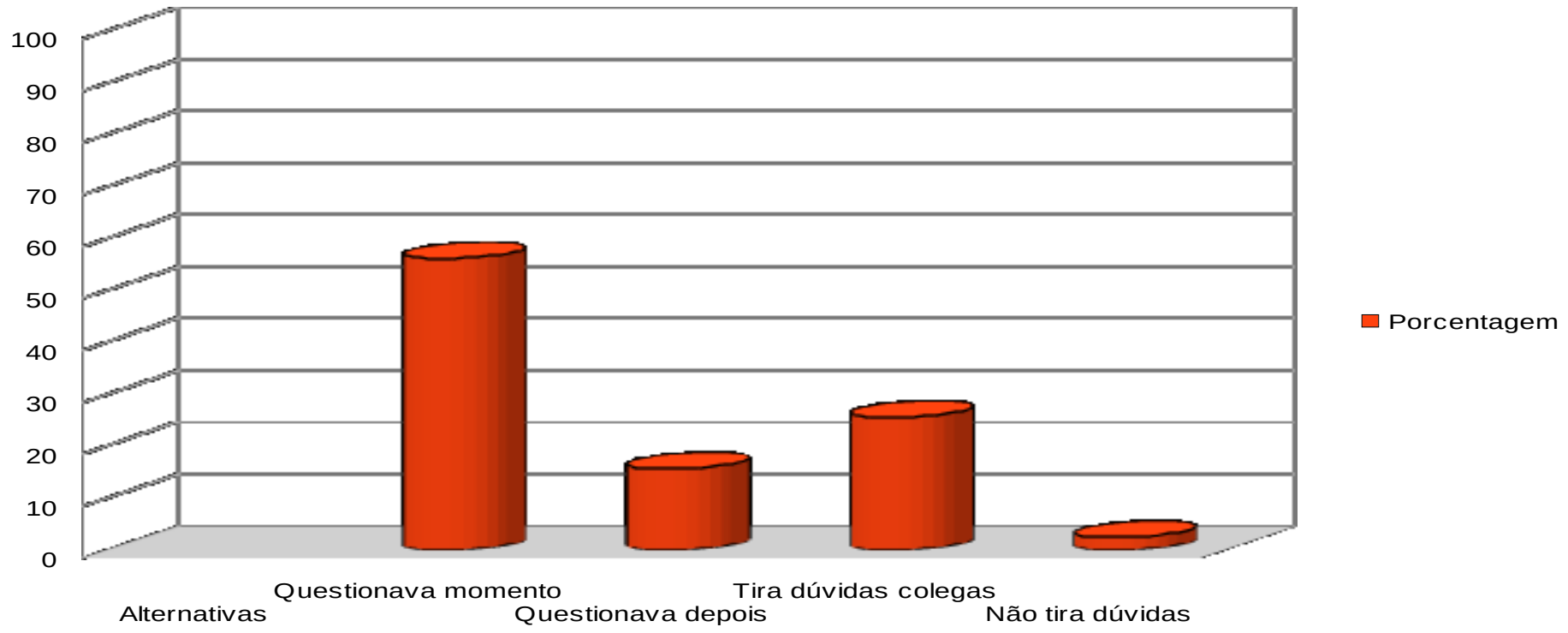

Fonte: Construção do autor

Apenas, para 2,50% dos alunos, o Curso contribui com formação humanística, enquanto que, para 58,00%, não há essa formação (Gráfico 17).

**Gráfico 17 - O curso oferece uma formação humanista além dos conteúdos técnico/científico?**

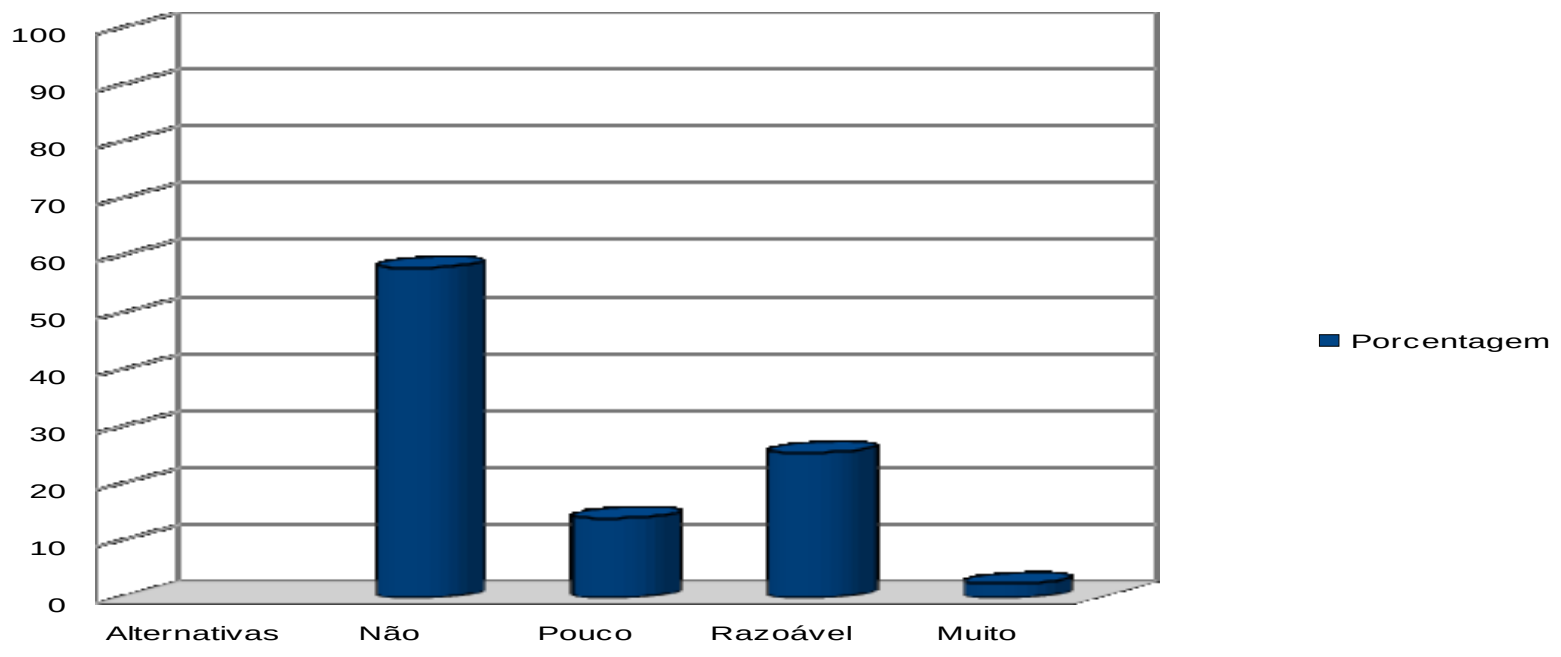

Fonte: Construção do autor

Perguntado como o aluno aprende estatística, a resposta escolhida surpreendeu a equipe de pesquisa, pois um percentual de mais 52,89% aprendeu a estatística, facilmente, despendendo algum tempo e esforço (Gráfico 18).

**Gráfico 18 - Eu aprendo Estatística**

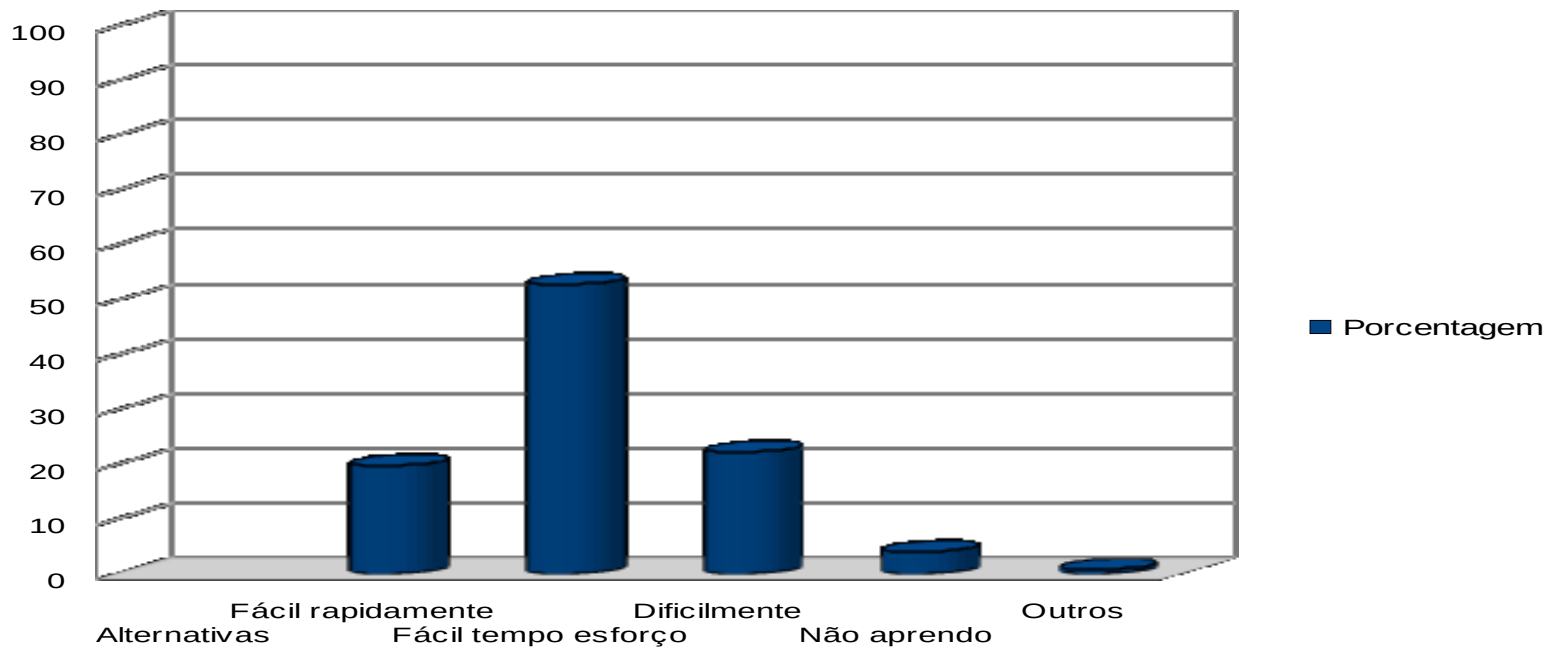

Fonte: Construção do autor

Na sala de aula, a figura do professor é preponderante, independente de seu estilo de atuação. Logo, foi perguntado aos alunos, se ele compreende as explicações do professor. 61,16% responderam que, após resolver alguns exercícios sobre assunto; e apenas 4,96% dos alunos afirmam não compreender as explicações do professor em hipótese alguma (Gráfico19).

**Gráfico 19 – Eu compreendo as explicações do professor**

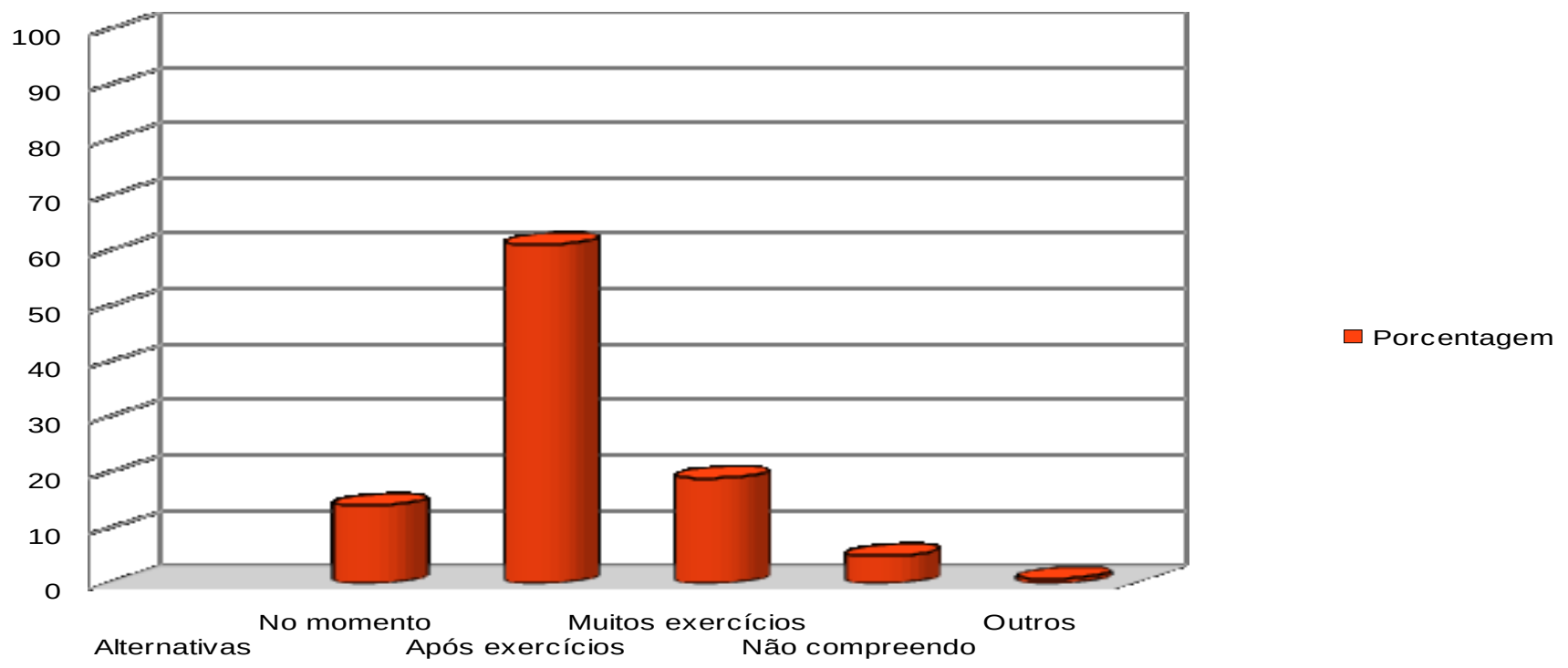

Fonte: Construção do autor

Com a intenção de confrontar as respostas sobre os problemas estatísticos, 2,48% dos alunos afirmaram não serem capazes de resolvê-los, e 24,79% têm dificuldades (Gráfico 20).

**Gráfico 20 - Diante de problemas estatísticos novos para mim**

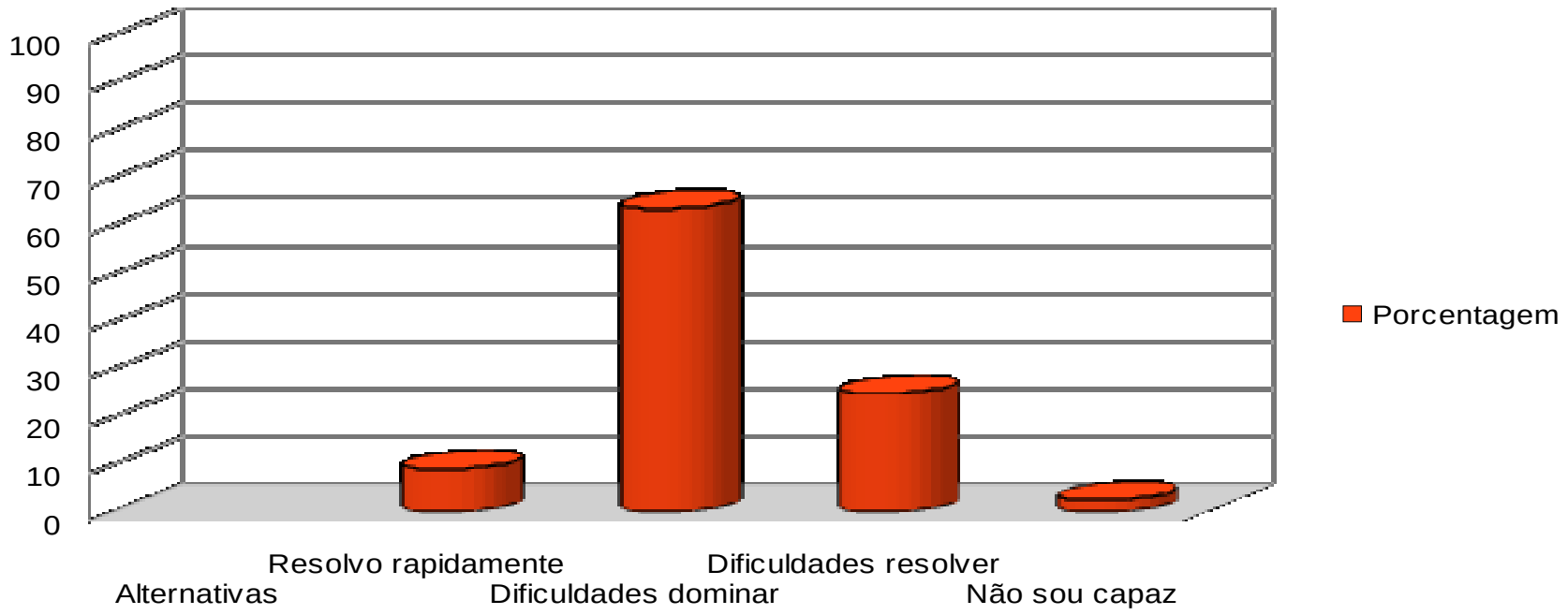

Fonte: Construção do autor

Ao solucionar problemas estatísticos, a maioria, 85,12%, faz uso de calculadora ou papel (Gráfico 21).

**Gráfico 21 - Ao solucionar problemas estatísticos**

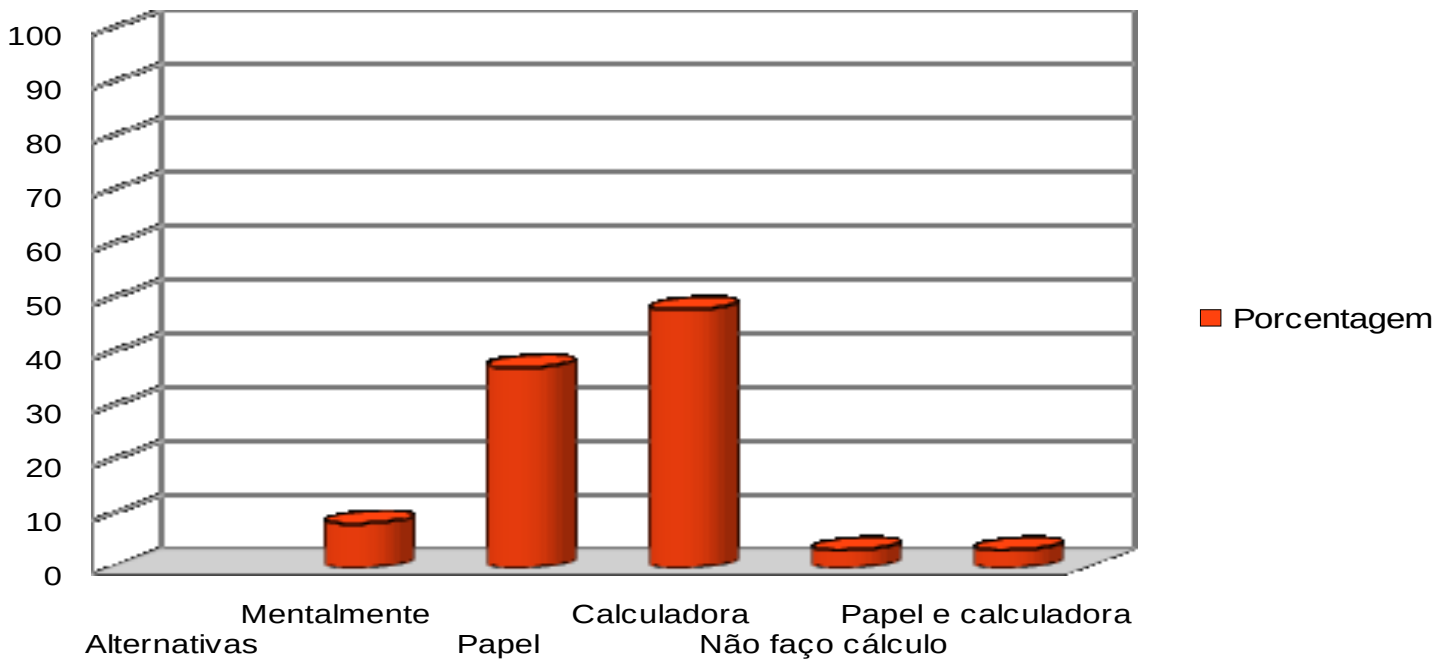

Fonte: Construção do autor

O (Gráfico 22) nos mostra que a maioria, (38,02%), sente-se motivada a resolver os exercícios. A resolução de exercícios permite que os alunos sanem possíveis dúvidas, quando há explicação do professor.

**Gráfico 22 - Durante as aulas de exercícios de Estatística**

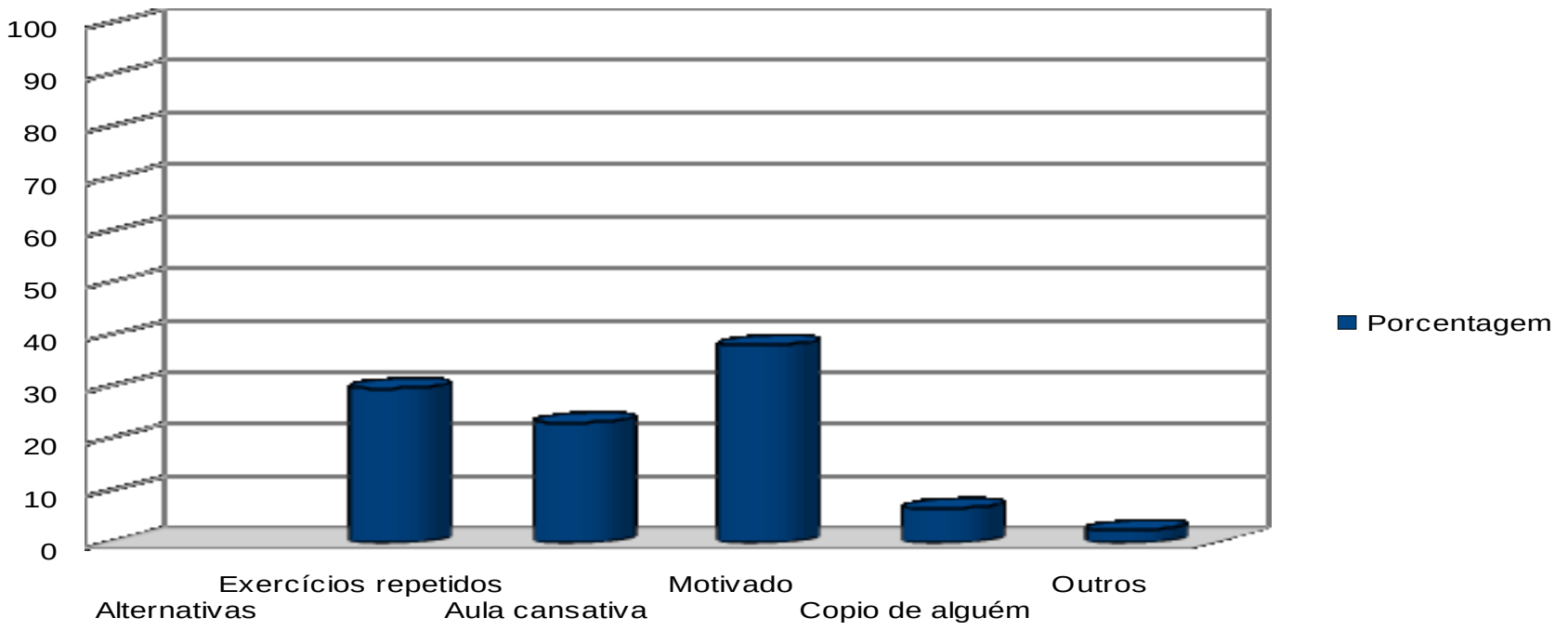

Fonte: Construção do autor

O (Gráfico 23) nos mostra que a maior dificuldade em aprender estatística para a maioria, (31,93%), é a base fraca em matemática, e 5,04% deles afirmam não ter dificuldades em aprender.

**Gráfico 23 - Qual o motivo maior que dificultou o aprendizado em Estatística?**

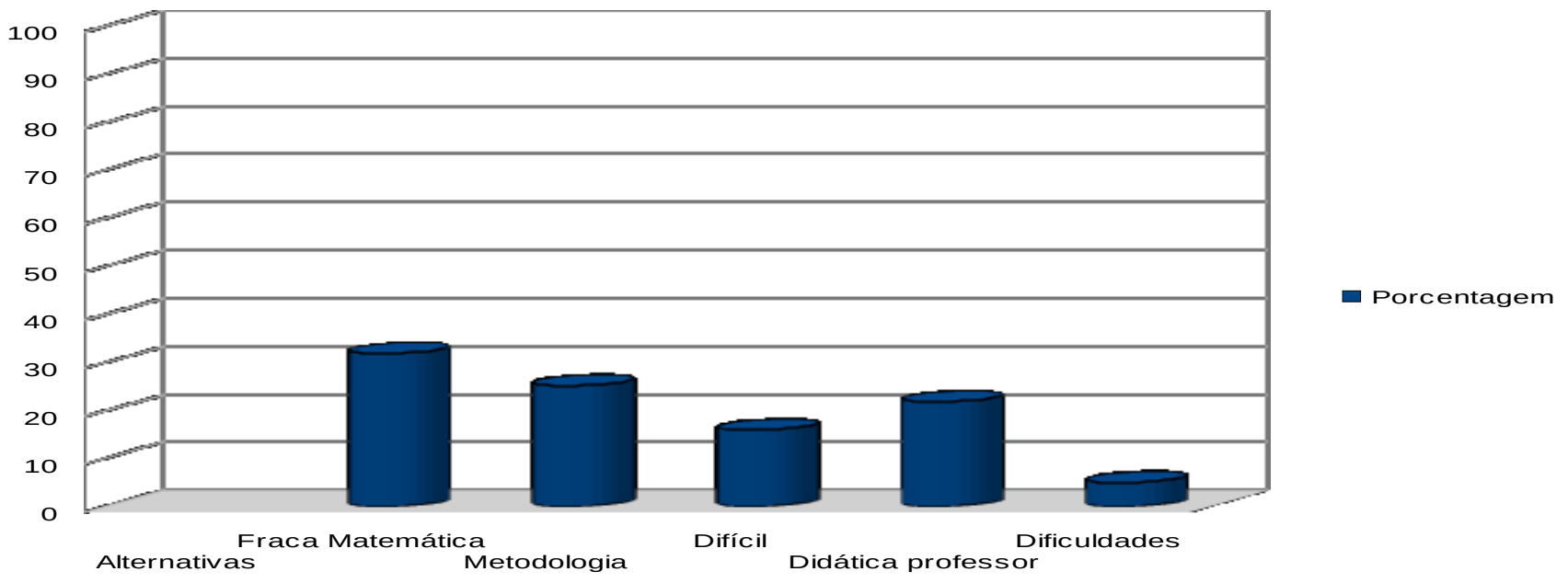

Fonte: Construção do autor

A maioria dos alunos (72,50%) disse que o professor não apresentou e nem fez uso de nenhum software estatístico (Gráfico 24).

**Gráfico 24 - O professor apresentou ou fez uso de algum software estatístico?**

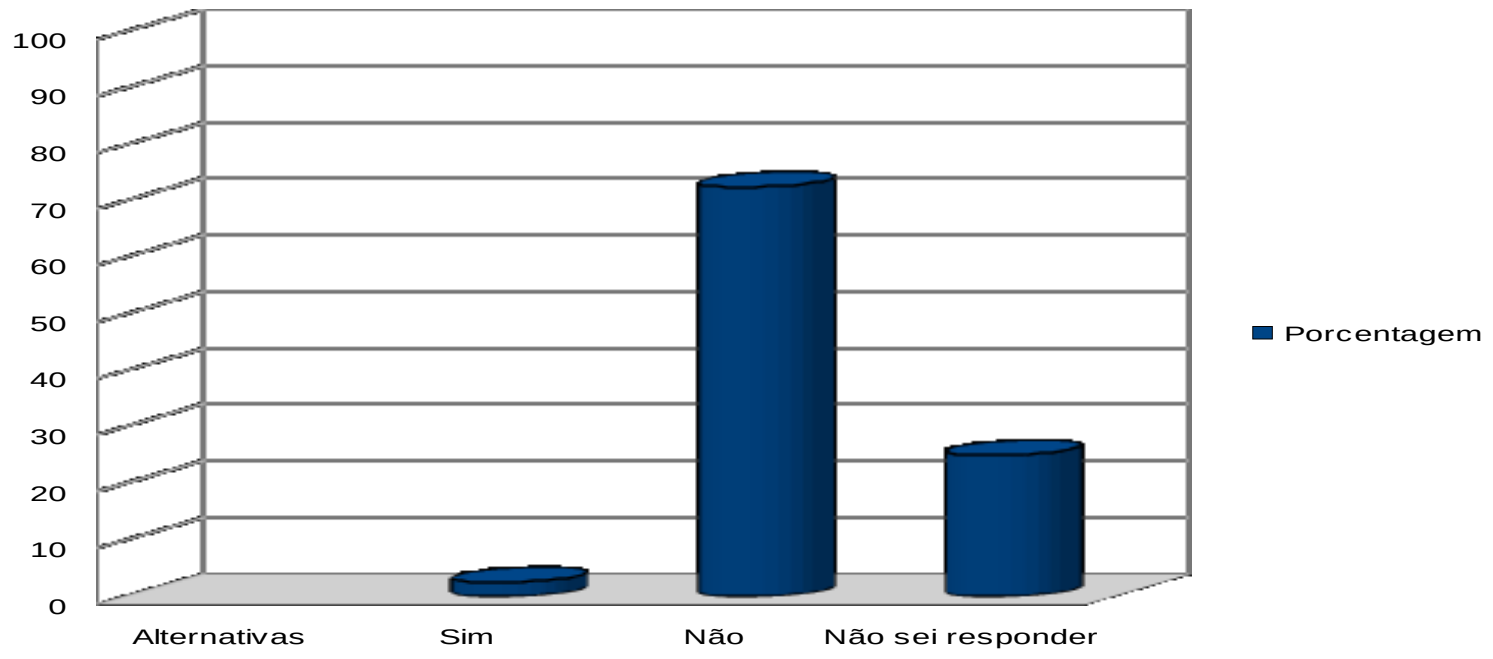

Fonte: Construção do autor

A análise dos dados obtidos por meio da EAE – Escala de Atitudes em relação à Estatística, cuja elaboração e adaptação já foram descritas na (Tabela 1), nos apresenta alguns dados positivos.

Dentre as afirmações positivas, destacam-se estas: *"*A Estatística me faz sentir seguro", "A Estatística é uma das matérias que eu mais gosto de estudar" *e* "Eu fico mais feliz na aula de Estatística". Estas foram aquelas que apresentaram o maior índice (14,00%) com relação à opção Concordo totalmente. Para "O sentimento que tenho em relação à Estatística é bom", obteve-se o maior índice, (51,20%), na opção Concordo. Já "A Estatística é uma das matérias que eu realmente gosto de estudar" apresentou o maior índice, (21,00%), para a opção Discordo totalmente. Em *"*A Estatística me faz sentir seguro (a) e é, ao mesmo tempo, estimulante", obteve o melhor resultado, (67,3%) com relação à opção Discordo.

**Tabela 2 - Distribuição das frequências segundo as afirmações positivas na EAE**

| Afirmações positivas | Discordo totalmente | Discordo | Concordo | Concordo totalmente |
|---|---|---|---|---|
| Eu acho a Estatística muito interessante e gosto das aulas de Estatística. | 11,60% | 46,50% | 34,90% | 7,00% |
| A Estatística é fascinante e divertida. | 14,00% | 48,80% | 30,20% | 7,00% |
| A Estatística me faz sentir seguro (a) e é, ao mesmo tempo, estimulante. | 4,70% | 67,30% | 14,00% | 14,00% |
| O sentimento que tenho com relação à Estatística é bom. | 11,60% | 32,60% | 51,20% | 4,60% |
| A Estatística é algo que eu aprecio grandemente. | 11,60% | 46,50% | 34,90% | 7,00% |
| Eu gosto realmente da Estatística. | 14,00% | 58,10% | 18,60% | 9,30% |
| A Estatística é uma das matérias que eu realmente gosto de estudar na faculdade. | 21,00% | 55,70% | 9,30% | 14,00% |
| Eu fico mais feliz na aula de Estatística que na aula de qualquer outra matéria. | 20,90% | 55,80% | 9,30% | 14,00% |
| Eu me sinto tranqüilo (a) em Estatística e gosto muito dessa matéria. | 14,00% | 48,80% | 25,60% | 11,60% |
| Eu tenho uma reação definitivamente positiva com relação à Estatística. Eu gosto e aprecio essa matéria. | 7,00% | 44,20% | 39,50% | 9,30% |
| N = 159 | | | | |

Fonte: Construção do autor

Das afirmações negativas, *"*Eu fico sempre sob uma terrível tensão na aula de Estatística"*,* apresentou o maior índice (34,90%) para a opção Discordo totalmente; "Eu nunca gostei de Estatística e é a matéria que me dá mais medo" apresentou o melhor resultado (67,50%) com relação à opção Discordo; "Quando eu ouço a palavra Estatística, eu tenho um sentimento de aversão" apresentou o maior índice (41,90%) para a opção Concordo; "Eu tenho a sensação de insegurança quando me esforça na estatística" teve o maior índice (11,60%) para a opção Concordo totalmente.

**Tabela 3 - Distribuição das frequências segundo as afirmações negativas na EAE**

| Afirmações negativas | Discordo totalmente | Discordo | Concordo | Concordo totalmente |
|---|---|---|---|---|
| Eu fico sempre sob uma terrível tensão na aula de Estatística. | 34,90% | 32,60% | 30,20% | 2,30% |
| Eu não gosto de Estatística e me assusta ter que fazer essa matéria. | 27,90% | 55,80% | 14,00% | 2,30% |
| “Dá um branco” na minha cabeça e não consigo pensar claramente quando estuda Estatística. | 16,30% | 55,80% | 27,90% | 0,00% |
| Eu tenho sensação de insegurança quando me esforço na Estatística. | 14,00% | 44,20% | 30,20% | 11,60% |
| A Estatística me deixa inquieto (a), descontente, irritado (a) e impaciente. | 9,30% | 51,20% | 32,50% | 7,00% |
| A Estatística me faz sentir como se estivesse perdido (a) em uma selva de números e sem encontrar a saída. | 11,60% | 51,20% | 30,20% | 7,00% |
| Quando eu ouço a palavra Estatística, eu tenho um sentimento de aversão. | 14,00% | 41,90% | 41,90% | 2,20% |
| Eu encaro a Estatística com um sentimento de indecisão, que é resultado do medo de não ser capaz em Estatística. | 16,30% | 55,70% | 23,30% | 4,70% |
| Pensar sobre a obrigação de resolver um problema estatístico me deixa nervoso (a). | 14,00% | 44,20% | 32,60% | 9,20% |
| Eu nunca gostei de Estatística e é a matéria que me dá mais medo. | 14,00% | 67,50% | 7,00% | 11,50% |
| Eu fico sempre sob uma terrível tensão na aula de Estatística. | 34,90% | 32,60% | 30,20% | 2,30% |
| N = 159 | | | | |

Fonte: Construção do autor

De acordo com a (Tabela 04), percebem-se as afirmações positivas, por exemplo, “Eu tenho uma reação definitivamente positiva com relação à Estatística”, “Eu gosto e aprecio essa matéria”. As respostas positivas apresentaram a maior média, 2,51%, e as que obtiveram significativas médias foram “A Estatística é algo que aprecio grandemente” e “Eu fico mais feliz na aula de Estatística que na aula de qualquer outra manteria”, 2,16. A média das médias das pontuações das afirmações positivas foi de 2,33.

**Tabela 4 - Distribuição dos valores referentes à Estatística Descritiva de cada uma das afirmações positivas da EAE**

| Afirmações positivas | Mínimo | Máximo | Soma | Média | Desvio Padrão |
|---|---|---|---|---|---|
| Eu acho a Estatística muito interessante e gosto das aulas de Estatística. | 1 | 4 | 102 | 2,37 | 0,79 |
| A Estatística é fascinante e divertida. | 1 | 4 | 99 | 2,37 | 0,80 |
| A Estatística me faz sentir seguro (a) e é, ao mesmo tempo, estimulante. | 1 | 4 | 102 | 2,37 | 0,79 |
| O sentimento que tenho com relação à Estatística é bom. | 1 | 4 | 107 | 2,49 | 0,77 |
| A Estatística é algo que eu aprecio grandemente. | 1 | 4 | 93 | 2,16 | 0,75 |
| Eu gosto realmente da Estatística. | 1 | 4 | 102 | 2,37 | 0,79 |
| A Estatística é uma das matérias que eu realmente gosto de estudar na faculdade. | 1 | 4 | 96 | 2,23 | 0,81 |
| Eu fico mais feliz na aula de Estatística que na aula de qualquer outra matéria. | 1 | 4 | 93 | 2,16 | 0,92 |
| Eu me sinto tranqüilo (a) em Estatística e gosto muito dessa matéria. | 1 | 4 | 101 | 2,35 | 0,87 |
| Eu tenho uma reação definitivamente positiva com relação à Estatística. Eu gosto e aprecio essa matéria. | 1 | 4 | 108 | 2,51 | 0,77 |

Fonte: Construção do autor

Na (Tabela 05), percebem-se as afirmações negativas, "Eu tenho a sensação de insegurança quando me esforça na estatística", obteve a maior média: 2,40, e "Eu não gosto de Estatística e me assusta ter que fazer essa matéria", menor média, 1,91.

Já na (Tabela 06) é visualizada a média das médias das pontuações das afirmações negativas, cujo resultado foi de 2,22.

**Tabela 5 - Distribuição dos valores referentes à Estatística Descritiva de cada uma das afirmações negativas na EAE**

| Afirmações negativas | Mínimo | Máximo | Soma | Média | Desvio Padrão |
|---|---|---|---|---|---|
| Eu fico sempre sob uma terrível tensão na aula de Estatística. | 1 | 4 | 86 | 2,00 | 0,87 |
| Eu não gosto de Estatística e me assusta ter que fazer essa matéria. | 1 | 4 | 82 | 1,91 | 0,72 |
| "Dá um branco" na minha cabeça e não consigo pensar claramente quando estuda Estatística. | 1 | 4 | 91 | 2,12 | 0,66 |
| Eu tenho sensação de insegurança quando me esforço na Estatística. | 1 | 4 | 103 | 2,40 | 0,88 |
| A Estatística me deixa inquieto (a), descontente, irritado (a) e impaciente. | 1 | 4 | 102 | 2,37 | 0,76 |
| A Estatística me faz sentir como se estivesse perdido (a) em uma selva de números e sem encontrar a saída. | 1 | 4 | 100 | 2,33 | 0,78 |
| Quando eu ouço a palavra Estatística, eu tenho um sentimento de aversão. | 1 | 4 | 100 | 2,33 | 0,75 |
| Eu encaro a Estatística com um sentimento de indecisão, que é resultado do medo de não ser capaz em Estatística. | 1 | 4 | 93 | 2,16 | 0,75 |
| Pensar sobre a obrigação de resolver um problema estatístico me deixa nervoso (a). | 1 | 4 | 102 | 2,37 | 0,85 |
| Eu nunca gostei de Estatística e é a matéria que me dá mais medo. | 1 | 4 | 93 | 2,16 | 0,81 |

Fonte: Construção do autor

**Tabela 6** - **Distribuição das médias segundo o conjunto das afirmações positivas e negativas na EAE**

| Afirmações | N | Mínimo | Máximo | Soma | Média | Desvio Padrão |
|---|---|---|---|---|---|---|
| Positivas | 430 | 1 | 4 | 1003 | 2,33 | 0,121 |
| Negativas | 430 | 1 | 4 | 953 | 2,22 | 0,171 |
| **Total** | **860** | **1** | **4** | **1956** | **2,27** | **0,802** |

Fonte: Construção do autor

Assim na (Tabela 07), é possível visualizar a média das pontuações do conjunto das afirmações positivas e negativas. O resultado mostra que é superior à média das pontuações do conjunto das afirmações negativas, podendo ser confirmada pela tabela abaixo que mostra o teste de comparação de médias, em que as variâncias não foram consideradas homogêneas (P(F>Fc) = 0,000), conforme.

**Tabela 7 - Teste de comparação de médias, segundo o conjunto das afirmações positivas e negativas, em que as variâncias não foram consideradas homogêneas**

| | |
|---|---|
| Estimativa por ponto de m1-m2: | 0,11 |
| Graus de liberdade: | 772 |
| Nível de significância: | 5% |
| Teste da hipótese: | Ho: m1-m2 = 0,00 |
| t calculado: | tc = 10,89 |
| Probabilidade: | P(t>\|tc\|)= 0,000 |
| Decisão: | Rejeita-se Ho: nível de 5,00% de probabilidade |

Fonte: Construção do autor

Como observado na pesquisa e na codificação das fases **(Tabela 08),** a Estática recebe valoração positiva**.** Além disso, os dados mostram as correlações entre as afirmações positivas, que foram codificadas.

**Tabela 8 - Codificação para as afirmações positivas**

| **Codificação** | **Frases** |
|---|---|
| Frase 3 | Eu acho a Estatística muito interessante e gosto das aulas de Estatística. |
| Frase 4 | A Estatística é fascinante e divertida. |
| Frase 5 | A Estatística me faz sentir seguro (a) e é, ao mesmo tempo, estimulante. |
| Frase 9 | O sentimento que tenho com relação à Estatística é bom. |
| Frase 11 | A Estatística é algo que eu aprecio grandemente. |
| Frase 14 | Eu gosto realmente da Estatística. |
| Frase 15 | A Estatística é uma das matérias que eu realmente gosto de estudar na faculdade. |
| Frase 18 | Eu fico mais feliz na aula de Estatística que na aula de qualquer outra matéria. |
| Frase 19 | Eu me sinto tranquilo (a) em Estatística e gosto muito dessa matéria. |
| Frase 20 | Eu tenho uma reação definitivamente positiva com relação à Estatística. Eu gosto e aprecio essa matéria. |

Fonte: Construção do autor

A partir dos dados expostos, verifica-se que a frase 9 se correlaciona apenas com a frase 4, isso quer dizer que a estatística se torna fascinante e divertida à medida que se tem uma boa relação com ela. Percebe-se, claramente, que as demais correlações existentes estão intimamente ligadas ao gosto pela estatística, e isso traz segurança aos alunos, fazendo com que a disciplina seja estimulante, divertida e também muito interessante e, de certa maneira, faz com que os alunos gostem mais da estatística do que qualquer outra disciplina. Em linhas gerais, se o aluno gosta da estatística, tudo se torna mais fácil, ele consegue se mantiver tranquilo e isso lhe proporciona segurança ao trabalhar os conteúdos (Tabela 09).

**Tabela 9 - Correlações entre as afirmações positivas (... Continuação)**

| | Frase 3 | Frase 4 | Frase 5 | Frase 9 | Frase 11 |
|---|---|---|---|---|---|
| Frase 3 | | | | | |
| Frase 4 | 0,270 | | | | |
| Frase 9 | 0,125 | 0,373* | 0,125 | | |
| Frase 11 | 0,216 | 0,192 | -0,024 | 0,024 | |
| Frase 14 | 0,194 | 0,345* | 0,309* | 0,204 | 0,136 |
| Frase 15 | 0,420** | 0,438** | 0,048 | 0,196 | 0,092 |
| Frase 18 | 0,340* | 0,446** | -0,151 | 0,053 | -0,039 |
| Frase 19 | 0,154 | 0,118 | 0,015 | 0,238 | 0,057 |
| Frase 20 | 0,268 | 0,013 | 0,308* | 0,091 | 0,223 |

| | Frase 14 | Frase 15 | Frase 18 | Frase 19 |
|---|---|---|---|---|
| Frase 3 | | | | |
| Frase 4 | | | | |
| Frase 9 | | | | |
| Frase 11 | | | | |
| Frase 14 | | | | |
| Frase 15 | 0,159 | | | |
| Frase 18 | 0,471** | 0,266 | | |
| Frase 19 | 0,223 | 0,118 | 0,224 | |
| Frase 20 | 0,111 | 0,148 | 0,215 | 0,440* |

Fonte: Construção do autor

A partir dos dados, percebe-se que a frase 1 se correlaciona apenas com a frase 2, o que leva a refletir que o não gostar da disciplina acarreta numa situação de tensão nas aulas de estatística. A frase 2 se correlaciona com as frases 10 e 16, o que indica que o não gostar da disciplina estatística e a obrigação de ter que cursá-la fazem com que os alunos se sintam perdidos e nervosos ao terem que resolver questões pertinentes ao assunto. A frase 6 se correlaciona com as frases 12 e 13, e leva a pensar que o medo da disciplina desperta um sentimento de indecisão e aversão, fazendo com que os alunos não consigam raciocinar claramente sobre o objeto de estudo.

Como se observa nas (Tabelas 10 e 11), ao longo das frases 8, percebe-se que elas se correlacionam apenas com a frase 16, isso quer dizer que o nervosismo ocasionado por uma problemática encontrada, muitas vezes, deixa o aluno em situação de desconforto, ocasionando-lhe um sentimento de inquietude, irritabilidade, descontententamento e impaciência. A frase 12 se correlaciona com as frases 13 e 17, confirmando que o medo deixa os alunos indecisos, e isso desperta neles um sentimento de repúdio em relação à disciplina.

**Tabela 10 - Codificação para as afirmações negativas**

| Codificação | Frases |
|---|---|
| Frase 1 | Eu fico sempre sob uma terrível tensão na aula de Estatística. |
| Frase 2 | Eu não gosto de Estatística e me assusta ter que fazer essa matéria. |
| Frase 6 | “Dá um branco” na minha cabeça e não consigo pensar claramente quando estuda Estatística. |
| Frase 7 | Eu tenho sensação de insegurança quando me esforço na Estatística. |
| Frase 8 | A Estatística me deixa inquieto (a), descontente, irritado (a) e impaciente. |
| Frase 10 | A Estatística me faz sentir como se estivesse perdido (a) em uma selva de números e sem encontrar a saída. |
| Frase 12 | Quando eu ouço a palavra Estatística, eu tenho um sentimento de aversão. |
| Frase 13 | Eu encaro a Estatística com um sentimento de indecisão, que é resultado do medo de não ser capaz em Estatística. |
| Frase 16 | Pensar sobre a obrigação de resolver um problema estatístico me deixa nervoso (a). |
| Frase 17 | Eu nunca gostei de Estatística e é a matéria que me dá mais medo. |

Fonte: Construção do autor

**Tabela 11 - Correlações entre as afirmações negativas (... Continuação)**

| | Frase 1 | Frase 2 | Frase 6 | Frase 7 | Frase 8 |
|---|---|---|---|---|---|
| Frase 2 | 0,418** | | | | |
| Frase 6 | 0,247 | 0,274 | | | |
| Frase 7 | -0,062 | 0,022 | 0,247 | | |
| Frase 8 | 0,036 | 0,241 | 0,149 | 0,096 | |
| Frase 10 | 0,245 | 0,311* | 0,202 | -0,123 | 0,194 |
| Frase 12 | 0,110 | -0,075 | 0,307* | 0,162 | 0,118 |
| Frase 13 | 0,290 | 0,117 | 0,438** | 0,225 | 0,142 |
| Frase 16 | 0,129 | 0,529** | 0,006 | 0,022 | 0,337* |
| Frase 17 | 0,268 | 0,230 | 0,229 | 0,174 | 0,093 |

| | Frase 10 | Frase 12 | Frase 13 | Frase 16 |
|---|---|---|---|---|
| Frase 2 | | | | |
| Frase 6 | | | | |
| Frase 7 | | | | |
| Frase 8 | | | | |
| Frase 10 | | | | |
| Frase 12 | 0,059 | | | |
| Frase 13 | 0,151 | 0,411** | | |
| Frase 16 | 0,318 | 0,030 | 0,052 | |
| Frase 17 | 0,290 | 0,380* | 0,538* | 0,359* |

Fonte: Construção do autor

Em linhas gerais, percebe-se, claramente, que o não gostar da estatística e o sentimento de aversão à disciplina são ocasionados pelo medo e pelo nervosismo quando se deparam com assuntos ligados à Estatística.

### 4.2.1 Análise dos resultados da prova de Estatística

Visto que a presente pesquisa tinha como uma das preocupações verificar o conhecimento estatístico adquirido durante o semestre pelos 159 alunos dos cursos que ofertam a disciplina de Estatística, aplicou-se uma prova no final do semestre de 2009, composta de 5 (cinco) questões que abordaram conhecimentos básicos de Estatística trabalhados pelos alunos desde o início até o final do mencionado período.

Para Crespo (2000, p. 79), são conteúdos essenciais de Estatística que os alunos não dominam tais como média aritmética, problemas envolvendo Quartil, Decil, Percentil, Moda, Mediana, Desvio-padrão e Coeficiente de variações.

De acordo com os resultados da prova realizada com os alunos, o problema número 01 apresentou o maior índice de erro (93,0%) e também o menor índice de acerto (7,0%). O aluno deveria apresentar conhecimentos sobre média, desvio-padrão e coeficiente de variação. Um dos motivos identificados para a dificuldade dos alunos, segundo pesquisas anteriores, segundo Batanero (2001a), deve-se a obstáculos didáticos, criados pela forma de ensino sugerida pelos livros, bem como à falta de formação adequada dos professores nessa área. Outros obstáculos são de natureza epistemológica, como é o caso da confusão entre média, mediana e moda. A seguir são apresentados os resultados obtidos em cada questão da prova (Tabela 12).

**Tabela 12 - Distribuição de frequência segundo os acertos e erros dos problemas de Estatística**

| Problemas | Corretos | f% | Errados | f% |
|---|---|---|---|---|
| Problema 1 | 11 | 7,0% | 148 | 93,0% |
| Problema 2 | 48 | 30,2% | 111 | 69,8% |
| Problema 3 | 141 | 88,4% | 18 | 11,6% |
| Problema 4 | 133 | 83,7% | 26 | 16,3% |
| Problema 5 | 152 | 95,4% | 7 | 4,6% |

Fonte: Construção do autor

A normalidade das variáveis utilizadas é comprovada por meio dos coeficientes de assimetria e curtose, cujos valores variam aproximadamente de -3 a +3 (BERQUÓ, 2006), conforme (Tabela 13).

**Tabela 13 - Comparação da prova de Estatística de acordo com as medidas de posições, dispersões e valores dos coeficientes de assimetria e curtose**

| **Errada** | | **Correta** | |
|---|---|---|---|
| Média | 16,8 | Média | 26,2 |
| Erro-padrão | 7,638062582 | Erro-padrão | 7,638062582 |
| Mediana | 7 | Mediana | 36 |
| Moda | #N/D | Moda | #N/D |
| Desvio-padrão | 17,07922715 | Desvio-padrão | 17,07922715 |
| Variância da amostra | 291,7 | Variância da amostra | 291,7 |
| Curtose | 2,17297816 | Curtose | -2,17297816 |
| Assimetria | 0,749456461 | Assimetria | -0,74945646 |
| Intervalo | 38 | Intervalo | 38 |
| Mínimo | 2 | Mínimo | 3 |
| Máximo | 40 | Máximo | 41 |
| Soma | 84 | Soma | 131 |
| Contagem | 5 | Contagem | 5 |
| Nível de confiança (95,0%) | 21,20670539 | Nível de confiança (95,0%) | 21,20670539 |

Fonte: Construção do autor

O (Gráfico 25) aborda a construção da Modelagem Matemática do modelo de sistemas lineares, fenômeno com característica que possui uma descrição gráfica polinomial.

Os dados do gráfico indicam que os problemas de Estatística estão bem correlacionados ao desempenho nessa disciplina. As correlações variam no que diz respeito à força e ao sentido. É por meio do diagrama de dispersão que se pode verificar como elas se distribuem no plano cartesiano, que é um gráfico capaz de mostrar a maneira pela qual as duas variáveis, X (problemas apresentados aos alunos) e Y (notas dos alunos em Estatística) convergem. Os pontos representam o cruzamento dos valores da variável X com a variável Y. Quanto mais concentrados os pontos em torno da linha reta, maior é a força correlacional entre as duas variáveis. Com relação ao sentido, pode-se afirmar que a variável errada é uma fraca relação negativa, e a variável correta é uma forte relação positiva.

**Gráfico 25 – Diagrama de dispersão - Ajustamento polinomial às notas dos alunos em Estatística em função dos acertos e erros**

Notas dos Alunos

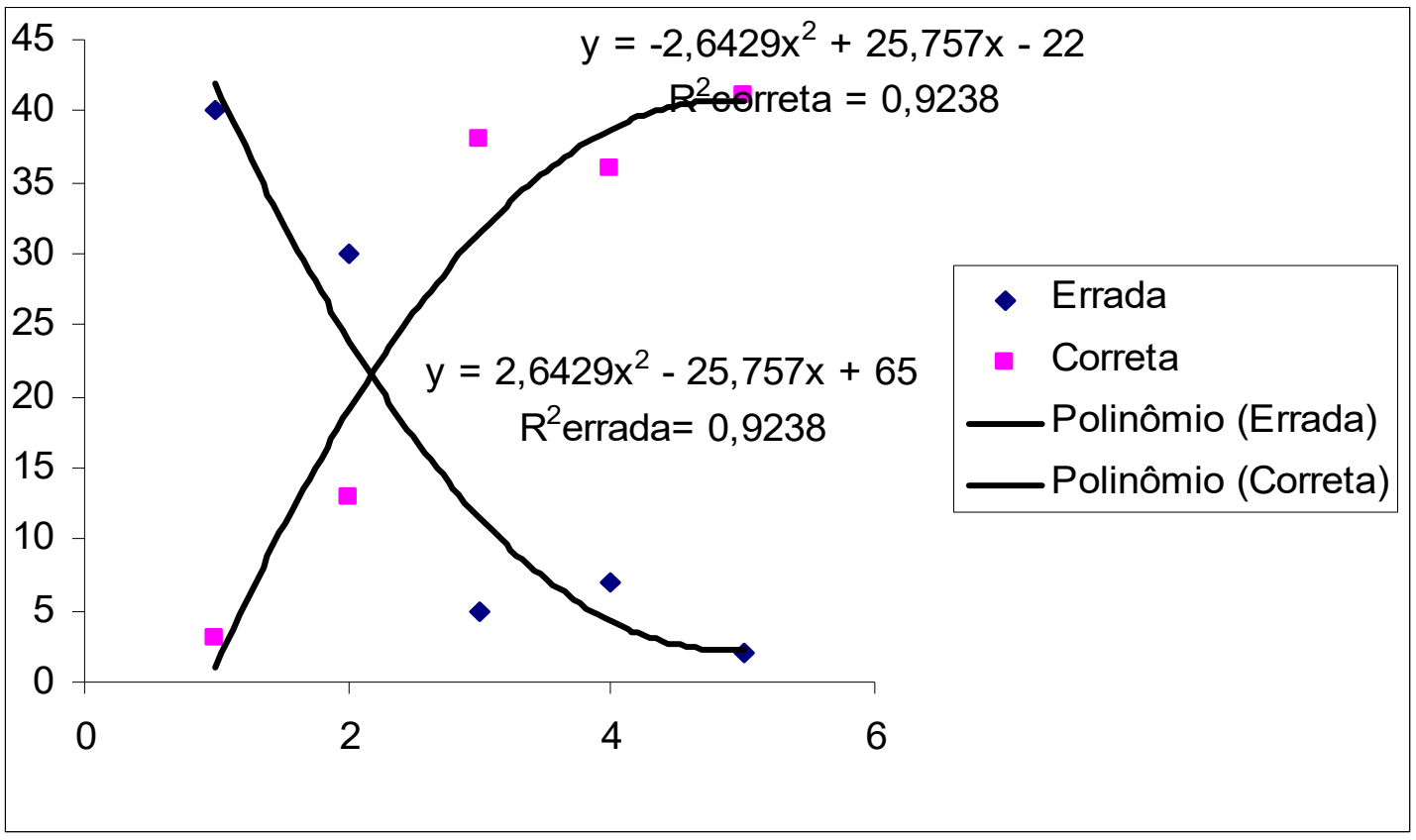

Fonte: Construção do autor

Partindo dessa premissa, faz-se importante ter conhecimento do ajustamento polinomial das notas referente acertos e erros cometidos pelos alunos, como se deve agir e interagir com as atividades de estatística. Com esta atividade, os professores puderam constatar que, através da modelagem Matemática, é possível construir com os alunos em salas de aula algumas fórmulas existentes em livros didáticos, não sendo necessários oferecê-las prontas.

### 4.2.2 Análise dos problemas determinados pela prova de Estatística

Como foi citada anteriormente, a prova de Estatística foi aplicada durante o horário de aula, na ausência do professor de Estatística, e os alunos levaram em média cinquenta minutos para a sua realização. A seguir, são apresentados, na (Tabela 14), os resultados de acordo com a Análise de Confiabilidade de ALFA DE CRONBACH.

**Tabela 14 – Índices de confiabilidade alfa cronbach da prova de Estatística de acordo com os problemas oferecidos para os alunos**

| Problemas de Estatística | Errada | Correta | Alfa Cronbach |
|---|---|---|---|
| Problema 1 | 156 | 3 | -0,30329 |
| Problema 2 | 90 | 69 | 0,289249 |
| Problema 3 | 5 | 154 | -0,22174 |
| Problema 4 | 7 | 152 | -0,12333 |
| Problema 5 | 2 | 157 | -0,33862 |
| **Alfa** | 0,563327 | 0,686931 | |
| **Total** | 0,64697 | | -0,19801 |

Fonte: Construção do autor

Conforme o entendimento de Hair Junior et al. (2005), confiabilidade é o grau em que um conjunto de indicadores é consistente em suas mensurações. Os resultados indicaram que a prova de Estatística apresentou uma consistência $\alpha$ de Cronbach geral de 0,64697 com valores variando de 0,563327 a 0,686931. O valor ($\alpha$) encontrado representa um resultado aceitável para o tipo de análise em estudo.

### 4.2.3 Análises dos valores das médias das notas bimestrais

Para analisar o desempenho dos alunos durante o período letivo em que a pesquisa foi realizada, recorreu-se às notas bimestrais que são obtidas por meio de provas e trabalhos. A seguir, a (Tabela 15) mostra a distribuição segundo as médias das notas dos alunos por bimestre no ano de 2009.

Observa-se que as médias das notas foram aumentando ao longo dos bimestres, conforme mostra a tabela, porém a nota média do primeiro bimestre não difere estatisticamente do segundo (P(t>tc) = 0,494). Já a comparação entre o segundo e terceiro bimestre e entre o terceiro e quarto bimestre, as notas médias diferem estatisticamente, cujas probabilidades são (P(t>tc) = 0,024) e (P(t>tc) = 0,006), respectivamente, o que pode sugerir que uma didática melhor trabalhada, com a criação de instrumento facilitador do processo ensino-aprendizagem, tenha levado os alunos a um melhor desempenho na disciplina.

**Tabela 15 - Distribuição segundo as médias das notas dos alunos por bimestres**

| | N | Mínimo | Máximo | Média | Desvio-Padrão |
|---|---|---|---|---|---|
| (B1) Notas do primeiro bimestre | 159 | 1,0 | 9,0 | 5,7 | 1,97 |
| (B2) Notas do segundo bimestre | 159 | 2,0 | 9,5 | 6,0 | 2,08 |
| (B3) Notas do terceiro bimestre | 159 | 4,0 | 9,5 | 6,9 | 1,53 |
| (B4) Notas do quarto bimestre | 159 | 6,0 | 10,0 | 7,7 | 1,10 |

Fonte: Construção do autor

A (Tabela 16) mostra o teste de comparação de médias entre o primeiro e segundo bimestres, em que as variantes foram consideradas homogêneas (P(F>Fc) = 0,363).

**Tabela 16 - Teste de comparação de médias entre o primeiro e segundo bimestres, em que as variantes foram consideradas homogêneas**

| | |
|---|---|
| Estimativa por ponto de m1-m2: | -0,3 |
| Graus de liberdade: | 84 |
| Nível de significância: | 5% |
| Teste da hipótese: | Ho: m1-m2 = 0,00 |
| t calculado: | tc = -0,68 |
| Probabilidade: | P(t>\|tc\|)= 0,494 |
| Decisão: | Não Rejeita Ho: nível de 5% de probabilidade |

Fonte: Construção do autor

Na (Tabela 17) mostra o teste de comparação de médias entre o segundo e terceiro bimestres, em que as variantes foram consideradas homogêneas (P(F>Fc) = 0,024).

**Tabela 17 - Teste de comparação de médias entre o segundo e terceiro bimestres, em que as variantes não foram consideradas homogêneas**

| | |
|---|---|
| Estimativa por ponto de m1-m2: | - 0,9 |
| Graus de liberdade: | 84 |
| Nível de significância: | 5% |
| Teste da hipótese: | Ho: m1-m2 = 0,00 |
| t calculado: | tc = - 2,28 |
| Probabilidade: | P(t>\|tc\|)= 0,025 |
| Decisão: | Rejeita-se Ho: nível de 5% de probabilidade |

Fonte: Construção do autor

A (Tabela 18) mostra o teste de comparação de médias entre o terceiro e quarto bimestres, em que as variâncias não foram consideradas homogêneas (P(F>Fc) = 0,017).

**Tabela 18 - Teste de comparação de médias entre o terceiro e quarto bimestres, em que as variantes não foram consideradas homogêneas**

| | |
|---|---|
| Estimativa por ponto de m1-m2: | - 0,8 |
| Graus de liberdade: | 84 |
| Nível de significância: | 5% |
| Teste da hipótese: | Ho: m1-m2 = 0,00 |
| t calculado: | tc = - 2,78 |
| Probabilidade: | P(t>\|tc\|)= 0,006 |
| Decisão: | Rejeita-se Ho: nível de 5% de probabilidade |

Fonte: Construção do autor

Pela comparação dos bimestres, os dados mostram a necessidade e a importância de uma metodologia de aprendizagem diferenciada que apresente estratégia didática que possa desenvolver atividade estatística para a construção do conhecimento matemático e estatístico.

### 4.2.4 Resultado da pesquisa com os professores que leciona Estatística

Conforme o Gráfico 26, há a predominância do gênero masculino, normalmente, as mulheres possuem uma maior capacidade de distinção de ambiente e, em geral, são mais detalhistas que os homens, o que leva suas opiniões a serem mais criteriosas na avaliação dos mínimos detalhes existentes, principalmente nas atividades didáticas.

**Gráfico 26- Sexo dos professores**

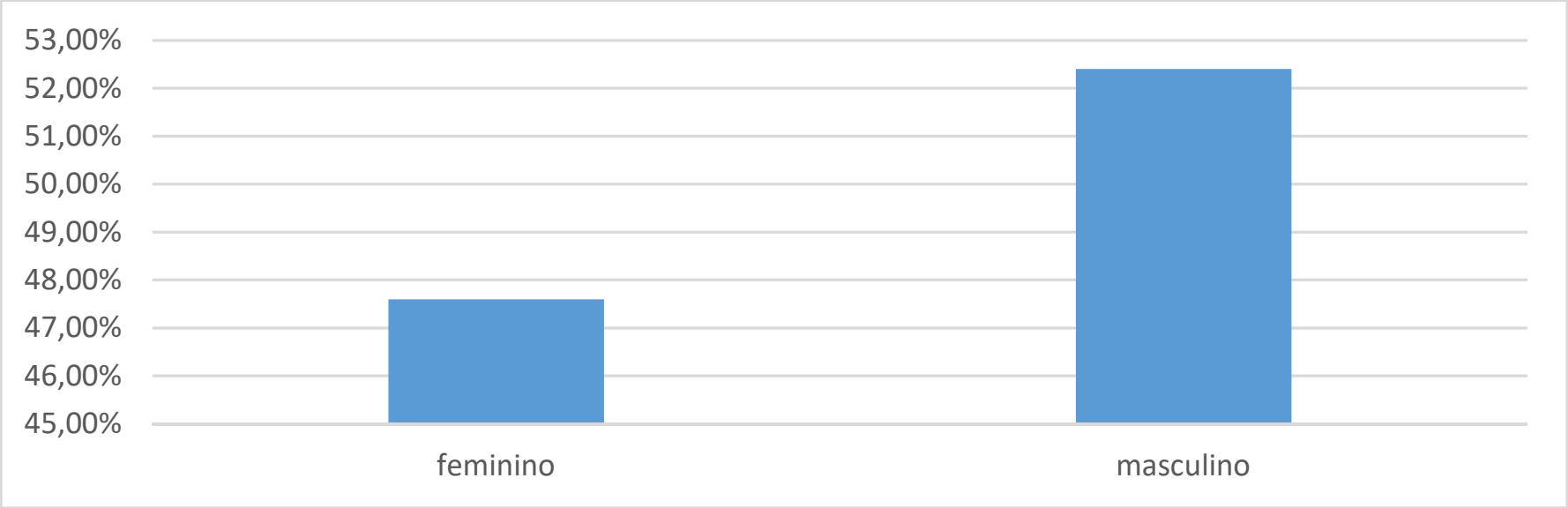

Fonte: Construção do autor

A faixa etária dos professores figura uma pequena predominância de pessoas entre 36 e acima de 45 anos (Gráfico 27), o que revela que certa maturidade dos professores, pois todos dizem gostar de lecionar matemática.

**Gráfico 27 - Faixa etária dos professores**

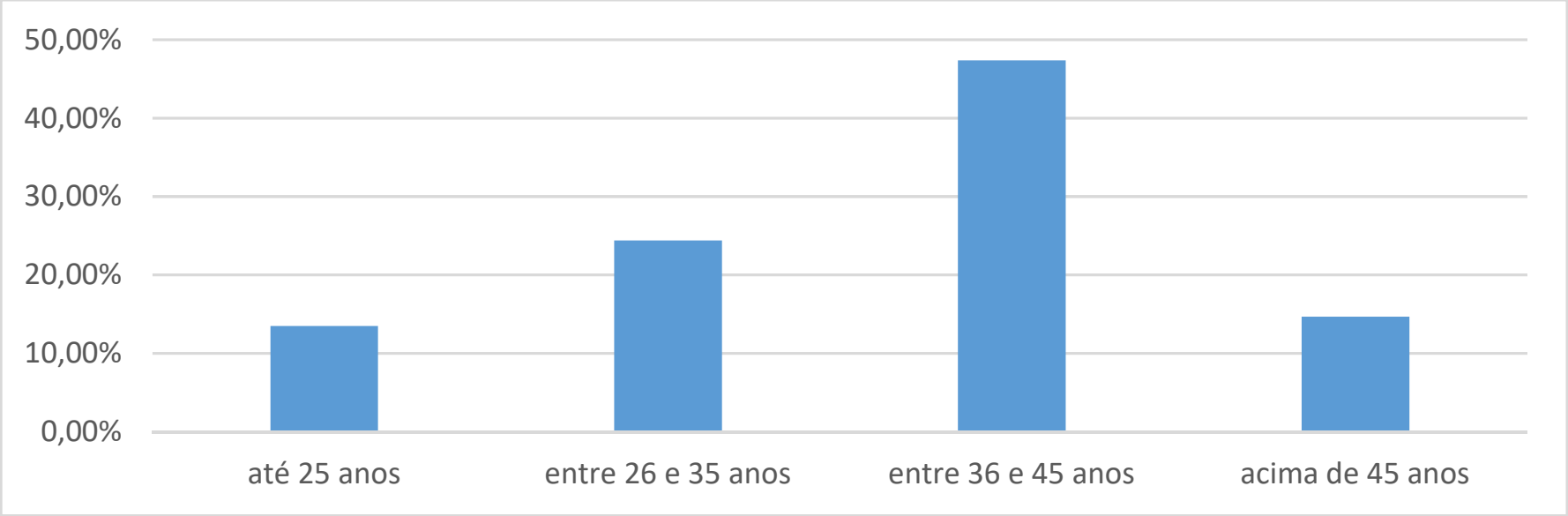

Fonte: Construção do autor

**Tabela 19 - Tempo que ministra a disciplina de Estatística**

| Tempo que ministra a disciplina Estatística | % |
|---|---|
| Menos de 2 anos | 37,50 |
| 2 a 7 | 37,50 |
| 14 a 19 | 25,00 |
| Total | 100,00 |

Fonte: Construção do autor

**Tabela 20 - Tipo de vínculo**

| Tipo de Vínculo | % |
|---|---|
| Efetivo | 75,00 |
| Substituto | 25,00 |
| Total | 100,00 |

Fonte: Construção do autor

**Tabela 21 - Tipo de habilitação**

| Tipo de habilitação | % |
|---|---|
| Bacharelado | 37,50 |
| Licenciado | 62,50 |
| Total | 100,00 |

Fonte: Construção do autor

**Tabela 22 - Titulação**

| Titularidade | % |
|---|---|
| Graduado | - |
| Especialista | - |
| Mestre | 75,00 |
| Doutor | 25,00 |
| Total | 100,00 |

Fonte: Construção do autor

Quando questionados se a disciplina Estatística é compatível com a sua formação (Tabela 23), 87,50% dos professores afirmam que sim, totalmente ou parcialmente. Às vezes, os professores se sentem obrigados a assumirem disciplinas que não têm afinidades e/ou formação compatíveis, devido ao fato de não haver professor disponível para aquela disciplina, ou até mesmo para complementar a sua carga horária semanal de trabalho.

**Tabela 23 - Você acha que a disciplina Estatística é compatível com sua formação?**

| Acha? | % |
|---|---|
| Sim, totalmente | 50,00 |
| Sim, parcialmente | 37,50 |
| Não | 12,50 |
| Total | 100,00 |

Fonte: Construção do autor

A (Tabela 24) mostra um dado preocupante, pois 25% dos professores afirmaram não estar preparados ou capacitados apenas para trabalharem alguns conteúdos da disciplina estatística, podendo comprometer substancialmente o processo ensino-aprendizagem.

**Tabela 24 - Você se sente seguro e preparado/capacitado para trabalhar os conteúdos da disciplina Estatística?**

| Se sente? | % |
|---|---|
| Sim, em todos | 37,50 |
| Sim, na maioria | 37,50 |
| Sim, em apenas alguns | 12,50 |
| Não | 12,50 |
| Total | 100,00 |

Fonte: Construção do autor

No que diz respeito à carga horária (Tabela 25), 62,50% dos professores acham que não está adequada ou está parcialmente adequada quanto aos objetivos da disciplina, nesse caso, deve-se verificar qual é o número de horas-aulas necessárias para o bom desenvolvimento da disciplina.

**Tabela 25 - Você acha que a carga horária é adequada para alcançar os objetivos da disciplina?**

| Acha? | % |
|---|---|
| Sim, totalmente | 37,50 |
| Sim, parcialmente | 25,00 |
| Não | 37,50 |
| Total | 100,00 |

Fonte: Construção do autor

A (Tabela 26) nos mostra que todo o professor consegue articular a teoria com a prática, mas não em todos os conteúdos.

**Tabela 26 - Você consegue fazer a articulação entre a teoria e a prática?**

| Consegue? | % |
|---|---|
| Sim, em todos os conteúdos | 25,00 |
| Sim, na maioria dos conteúdos | 37,50 |
| Sim, em alguns conteúdos apenas | 37,50 |
| Total | 100,00 |

Fonte: Construção do autor

Para 37,50% dos professores, a disciplina estatística não está articulada com nenhuma outra disciplina do curso (Tabela 27). Possivelmente, esses professores não conhecem o Projeto Pedagógico do Curso em que estão inseridos, até mesmo pelo fato de estarem trabalhando há pouco tempo no Curso, pois, com certeza, a Estatística é articulada com outras disciplinas.

**Tabela 27 - Existe articulação com outras disciplinas do(s) curso(s) em que trabalha?**

| Existe? | % |
|---|---|
| Sim | 62,50 |
| Não | 37,50 |
| Total | 100,00 |

Fonte: pesquisa (2009)

Para 37,50% dos professores, é possível dialogar com outras áreas do saber a fim de ter uma visão humanista para as disciplinas da área das Exatas (Tabela 28).

**Tabela 28 - Você procura introduzir no conteúdo técnico/científico de suas disciplinas uma visão humanista quanto à sua aplicação?**

| Existe? | % |
|---|---|
| Sim | 62,50 |
| Não | 37,50 |
| Total | 100,00 |

Fonte: Construção do autor

Para 62,50% dos professores sempre conseguem diversificar as atividades (Tabela 29). Essa medida se faz necessária para que as aulas não se tornem monótonas e repetitivas, despertando nos alunos o interesse pela disciplina.

**Tabela 29 - Você consegue diversificar as atividades, visando à melhoria do processo ensino-aprendizagem?**

| Consegue? | % |
|---|---|
| Sempre | 50,00 |
| Freqüentemente | 12,50 |
| Raramente | 25,00 |
| Não sei responder | 12,50 |
| Total | 100,00 |

Fonte: Construção do autor

Metade dos professores avaliou os recursos audiovisuais como regulares (Tabela 30) e 87,50% deles afirmam não utilizar softwares estatísticos (Tabela 31). O uso de softwares é indispensável hoje em dia no ensino de estatística, para que os alunos saibam utilizar as ferramentas disponíveis nesses pacotes computacionais.

**Tabela 30 - Como você avalia os recursos audiovisuais disponíveis para uso na disciplina?**

| Avaliação | % |
|---|---|
| Ótimos | 12,50 |
| Bons | 12,50 |
| Regulares | 50,00 |
| Péssimos | 12,50 |
| Não utilizo | 12,50 |
| Total | 100,00 |

Fonte: Construção do autor

Interessante salientar que Huppes (2002) menciona que as atividades com equipamentos tecnológico melhoram a aprendizagem dos alunos.

**Tabela 31 - Você já utilizou algum software estatístico em suas aulas?**

| Já utilizou? | % |
|---|---|
| Sim | 12,50 |
| Não | 87,50 |
| Total | 100,00 |

Fonte: Construção do autor

Quanto às avaliações, 12,5% dos professores afirmaram que nem sempre dialogam com os alunos sobre os critérios de avaliação (Tabela 32), e 25% deles nem sempre discutem e analisam os resultados das avaliações (Tabela 33). Há erro por parte dos professores, pois tem que haver esclarecimento dos critérios adotados para todo e qualquer tipo de avaliação, desde uma resolução de exercícios, passando por uma apresentação de seminário e chegando à prova escrita, e, posteriormente, a esses, devem-se analisar os resultados para que os possíveis erros encontrados durante o processo sejam corrigidos.

**Tabela 32 - Você dialoga com os alunos sobre os critérios de avaliação?**

| Dialoga? | % |
|---|---|
| Sim | 87,5 |
| Nem sempre | 12,5 |
| Total | 100 |

Fonte: Construção do autor

Tabela 33 - Você discute e analisa os resultados das avaliações com os alunos?

| Discute? | % |
|---|---|
| Sim | 75,00 |
| Nem sempre | 25,00 |
| Total | 100,00 |

Fonte: Construção do autor

A (Tabela 34) mostra que a maioria dos professores (87,50%) exige dos alunos na medida certa.

**Tabela 34 - Qual o nível de exigência na disciplina?**

| Nível de exigência | % |
|---|---|
| Exijo muito | 12,50 |
| Exijo na medida | 87,50 |
| Total | 100,00 |

Fonte: Construção do autor

Quando questionados se eles, no primeiro dia de aula, discutem e apresentam o plano de ensino (Tabela 35), metade dos professores afirma que Sim.

**Tabela 35 - Você no primeiro dia de aula apresenta e discute o plano de ensino com os alunos?**

| Discute? | % |
|---|---|
| Sim | 50,00 |
| Não sei responde | 10,00 |
| Nem Sempre | 40,00 |
| Não | - |
| Total | 100,00 |

Fonte: Construção do autor

A avaliação dos alunos quanto aos conteúdos trazidos do ensino médio (Tabela 36) não foi positiva, pois metade dos professores a classificou como regular e metade como ruim.

**Tabela 36 - Como você avalia os alunos quanto aos conteúdos trazidos do ensino médio?**

| Avaliação | % |
|---|---|
| Regular | 50,00 |
| Ruim | 50,00 |
| Total | 100,00 |

Fonte: Construção do autor

### 4.2.5 Comparativo dos resultados de alunos e professores

Quanto à realização das atividades acadêmicas em laboratórios de Informática, verificou-se que houve discrepância em relação ao uso de praticamente todos os equipamentos tecnológicos. Para o material concreto, os professores dizem que usam muito, enquanto para os alunos prevalece o resultado, às vezes.

Verifica-se que, realmente, os equipamentos tecnológicos são pouco utilizados nos cursos pesquisados, com exceção do material concreto. Nas respostas dadas pelos alunos, prevalece o resultado nunca, com exceção do material concreto. Nas respostas dadas pelos professores, com exceção do vídeo e material concreto, também prevalece o resultado nunca (Tabela 37).

Acredita-se que os resultados desastrosos da Estatística e Matemática se devem em parte ao fato de se usar poucos elementos mediadores que envolvam conceitos estatísticos e operações matemáticas, para enriquecer os ambientes de aprendizagem.

**Tabela 37 - Uso de Equipamento Tecnológico**

| | Retroprojetor | | Vídeo | | Slides | | Material concreto | | Computador | |
|---|---|---|---|---|---|---|---|---|---|---|
| | Aluno % | Professor % | Aluno | Professor | Aluno | Professor | Aluno | Professor % | Aluno % | Professor % |
| Muito | 4 | - | 2 | 8 | 4 | - | 18 | 82 | 2 | 10 |
| Às vezes | 25 | 34 | 25 | 42 | 7 | - | 50 | 18 | 15 | 27 |
| Raramente | 7 | 24 | 14 | 38 | 9 | 8 | 12 | - | 2 | 1 |
| Nunca | 64 | 42 | 59 | 12 | 80 | 92 | 20 | - | 81 | 62 |

Fonte: Construção do autor

Como esclarece Huppes (2002), há que se considerar que ambos estão fortemente interligados e fazem parte de um mesmo contexto de aproveitamento de equipamentos tecnológicos para o desenvolvimento do processo ensino-aprendizagem, pois devem ser inovadores, criativos, saber e conseguir romper com o óbvio. Ser capaz de formular a pergunta que ninguém ousaria propor o que ninguém proporia. Para desenvolver essa criatividade, é preciso se desapegar da acomodação, ter coragem de enfrentar resistências e não ter medo de errar. Criatividade depende, antes de tudo, de autoconfiança e confiança no outro.

O aluno deve confiar em suas potencialidades e confiar no professor e o professor confiar nas capacidades do aluno. Muitas vezes, um instante de inspiração, leva a um ano inteiro de trabalho duro, e isso assusta a muitos.

### 4.2.6 Resultado da Simulação do Teste de Krukal-Wallis

Para a atividade proposta, utilizou-se um microcomputador, para que fosse possível fazer uso do "PowerPoint", e os alunos pudessem visualizar as atividades propostas nesta sessão. Acreditamos, também, que seja uma maneira de se introduzir com facilidade o software estatístico denominado Statdisk. Apresentou-se, no "PowerPoint", o exemplo do Teste de Kruskal-Wallis. Trata-se de teste extremamente útil para decidir se K amostras (K > 2) independentes provêm de populações com médias iguais. Poderá ser aplicado para variáveis intervalares ou ordinais.

Testar, no nível de erro 5%, a hipótese da igualdade das médias para os três grupos de alunos que foram submetidos a esquemas diferenciados de aulas. Foram registradas as notas obtidas para uma mesma prova, conforme a distribuição da (Tabela 38). Esta simulação foi inspirada em uma situação didática conforme Fonseca e Martins (1996). Adaptamos seus enunciados e acrescentamos alguns tópicos para tentarmos obter uma reflexão por parte dos alunos.

**Tabela 38 - Orientações da simulação - Teste de Kruskal-Wallis para o nível de erro 5%**

| Aulas Expositivas | Aulas com recursos audiovisuais | Aulas através de ensino programado |
|---|---|---|
| 6,5 | 6,0 | 6,1 |
| 6,2 | 7,1 | 6,9 |
| 6,8 | 6,6 | 6,7 |
| 7,0 | 6,3 | 7,2 |
| 6,0 | 6,4 | 7,4 |
| | 5,9 | |

Fonte: Construção do autor

Em seguida, foi trabalhado o teste através do software *Statdisk*, no qual é possível obter os resultados das distribuições, sendo necessário apenas que o usuário deste software dê, como entrada, a *Significance*, $\alpha = 0{,}05$ e o valor das notas no *Untitled* e *Columns* 1-3 of 9. O software faz o processamento destes dados iniciais e apresenta como saída as conclusões:

Se $H$ □ $\chi^2_{sup.}$, não se pode rejeitar $H_0$

Se $H$ □ $\chi^2_{sup.}$, rejeita-se $H_0$ concluindo-se com o risco α que há diferença entre médias dos *K* grupos.

Para sua comprovação, basta escrever no espaço em branco os valores de α igual a 0,05 e o valor das notas 65, 62, 68, 70, 60, 60, 71, 66, 63, 64, 59, 61, 69, 67, 72, 74. E clicando o mouse no botão *Evaluate*, o programa projetou os seguintes dados (Figura 05).

**Figura 5- Resultados das análises do teste de Kruskal-Wallis sobre as igualdades das médias, obtidos pelo programa do software Statdisk**

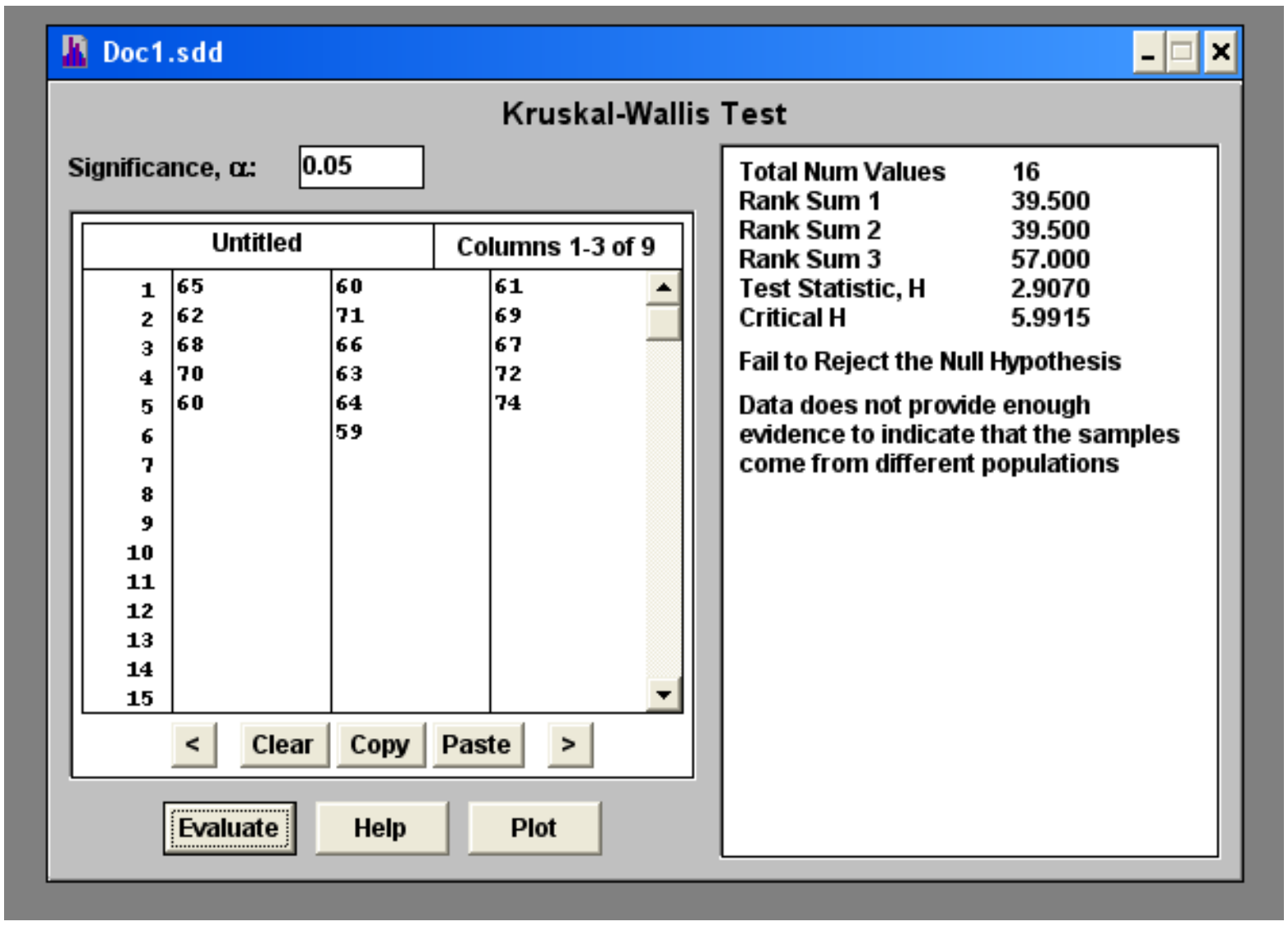

Org.: CAMPOS, M. B. N. S (2008)

As distribuições das hipóteses da igualdade das médias para os três grupos de alunos que foram, submetidos a esquemas diferenciados de aulas, uma aceitação da hipótese testada.

Tal fato foi contatado na análise descritiva que, o *Total Num Values* igual 16, o *Rank Sum 1* igual 39.500, *Rank Sum 2* igual 39.500, *Rank Sum 3* igual 57.000. Assim o cálculo do *Test Statístic, H* igual a 2,9070 e o valor crítico de t para o teste é t *Critical H* igual 5,9955 este valor será testado para a conclusão do teste.

O resultado apresentado pelo programa foi como H < 5,9955, não se pode rejeita $H_0$ . Confirmando o resultado, as notas médias dos alunos podem ser consideradas iguais, ao nível de erro 5%.

# 5 CONSIDERAÇÕES FINAIS

Para alcançar um elevado nível de qualidade na educação, é necessário aprimorar o conhecimento sobre esse o processo. O curso de Informática deverá trabalhar mais a capacidade de aprender do aluno do que a informação pronta.

A prática usual de integrar apenas uma disciplina de Estatística aos currículos dos cursos de Licenciatura em Informática, talvez, seja insuficiente para abordar os principais conceitos de Estatísticas. A formação dos egressos pode ser prejudicada, especialmente, se a disciplina for trabalhada com interdisciplinaridade junto às demais disciplinas do curso, como ocorre em muitos casos, e com as cargas horárias atualmente previstas.

De acordo com as teorias de aprendizagem, Skinner (1972) nos ensina que a aprendizagem deriva do conhecimento e treinamento, para Vygotsky (2001), o desenvolvimento da pessoa se realiza através de estímulos auxiliares retirados de cultura e pela interação social através do uso de signos. Os métodos que favorecem o desenvolvimento mental são os que levam o aluno a pensar.

O educador deve olhar os tempos atuais com uma nova perspectiva e se habilitar ao uso das novas ferramentas tecnológicas. A tecnologia sozinha não garante a participação real e pode servir de apoio ao projeto pedagógico que foca a aprendizagem ligada à vida (MORAN, 1998).

De acordo com Piaget (1987), o desenvolvimento da aprendizagem resulta da interação do homem com o ambiente, que consiste em adaptar-se por meio de acomodações e assimilações. Assimilação do novo a partir do antigo através da sensibilização dos esquemas existentes. O conhecimento é o resultado de constante construção do indivíduo, processada pelos esquemas mentais e agregada a esses esquemas.

Os resultados aqui relatados indicam a necessidade de outros estudos, com a inclusão de novas fórmulas de estatística para se chegar a uma metodologia diferente da convencional, que apresente índices mais elevados de confiabilidade e validade para ser utilizado no contexto educacional, revelando o raciocínio estatístico dos investigadores sociais com uma visão humanista, no que diz respeito à coleta e leitura de dados apresentados em gráficos e tabelas estatísticas.

O ensino para ser bem sucedido deve utilizar métodos e técnicas com objetivos mais direcionados aos alunos. Para uma maior eficiência, deve-se buscar o desenvolvimento integral do aluno.

Os cidadãos do futuro necessitam saber lidar com desafios, com problemas inesperados e, para que isso ocorra, é necessário compreender a situação, conceber um plano de ação e executar esse plano (HAPPES, 2002).

Em síntese, observa-se que a metodologia proposta para aprendizagem pode efetivamente ampliar os horizontes, não só pela flexibilidade, mas, acima de tudo, ao proporcionar novas competências e novas formas de aprendizado.

Por fim, como iniciei esta tese apresentando inquietações em minha prática docente, encerrá-la-ei apontando algumas implicações que esta pesquisa pode vir a ter, não apenas em minha prática docente, mas também em futuras pesquisas.

O mesmo não ocorreu quando a negociação de significados se relacionava com a situação da realidade. Por duas vezes, ao tentarem propor diálogos que envolviam a situação da realidade, os alunos tiveram seus convites rejeitados, não estabelecendo, portanto, cenários para investigação. Esse fato merece ainda maior destaque por estarmos nos referindo à proposição de uma atividade de Modelagem Matemática, a qual pressupõe a busca de uma situação do dia a dia.

Por outro lado, foi solicitado aos alunos que buscassem situações do cotidiano para serem estudadas, e que utilizassem, para isso, o conteúdo estatístico e o computador, mas quando surgiu à possibilidade da constituição de cenários para investigação envolvendo essas situações do cotidiano, obstáculos foram colocados. Apresentamos, a seguir, uma abordagem dessa análise, dividindo-a em duas partes: discussões e sugestões para trabalhos futuros.

**Discussões**

Um aspecto que se destacou nesta pesquisa foi à negociação de significados entre os alunos (e entre os alunos e mim). Às vezes, surgiam obstáculos às negociações, noutras, elas ocorriam de forma livre, constituindo cenários para investigação.

De acordo com Stieler (2007), é interessante destacar como o convite às negociações foi aceito, ou seja, os cenários para investigação foram constituídos, quando o tema em debate era estatístico. Tentativas de se iniciarem negociações de significados, relacionados com a situação real dos projetos de Modelagem Matemática, depararam-se com obstáculos. Como podemos compreender essa situação?

A presença do computador, talvez, tenha sido importante para o estabelecimento desses cenários para investigação. A imprevisibilidade dos acontecimentos, quando se trabalha em ambientes informatizados, abre possibilidades para que investigações aconteçam. Mas a simples presença dos computadores não garante a existência de investigações: é importante que os alunos aceitem o convite às investigações, seja ele feito pelo professor ou pelos próprios alunos diante das possibilidades abertas pelo computador.

**Sugestões para trabalhos futuros**

As sugestões aqui apresentadas podem representar significativa contribuição na formação do profissional, pois minimizam os impactos negativos gerados pelo desenvolvimento das atividades didáticas na resolução de problema na área de Estatística. Nesse sentido, considera-se que a metodologia de forma diferenciada em sala de aula garante a melhoria do ensino em longo prazo e, por essa razão, devem ser integradas em um contexto que envolve os professores, os alunos, a Coordenação de Curso e a Instituição.

Sobre minha prática docente, as questões que agora me instigam dizem respeito em como colocar em prática a Modelagem Matemática em conjunto com as Tecnologias Informáticas nas aulas de Estatística, levando em conta as ideias aqui defendidas. Sei que o contexto que me espera poderá oferecer obstáculos à minha prática, mas esta pesquisa me dá suporte para enfrentar tais obstáculos por meio do diálogo, respeitando sempre as perspectivas de meus colegas. Assim, minha prática docente provocou inquietações que me impulsionaram a desenvolver uma pesquisa e esta, por sua vez, devolve-me à prática com uma quantidade ainda maior de inquietações.

Durante o desenvolvimento da pesquisa, surgiram algumas questões que se relacionavam com o tema, porém não puderam ser aqui tratadas. Colocamos essas questões como sugestões para outros trabalhos.

a) Deve prevê na grade do curso pelo menos duas disciplinas obrigatórias, de pelo menos 60 horas (sessenta) cada, uma para a Introdução da Estatística, tratando da parte Descritiva e Inferencial, e outra específica, para as Distribuições de Probabilidades e Estatísticas não Paramétricas. Além disso, uma disciplina específica de Metodologia da Pesquisa e Desenvolvimento poderia ser incorporada no currículo, prática adotada em universidades dos Estados Unidos.

b) Sobre os testes não paramétricos e seu uso, quais as dificuldades dos alunos na montagem dos procedimentos estatísticos? Que tipo de software pode minimizar as dificuldades encontradas?

c) Privilegiar situações de aprendizagem que promovam atitudes criativas, críticas e transformadoras, para que possam ser profissionais criativos, críticos e transformadores da realidade onde se encontram sua área de atuação.

d) A aplicação da metodologia em outras disciplinas também constitui numa possibilidade para futuros trabalhos.

Desse modo, os tópicos citados, anteriormente, constituem temas promissores para pesquisas futuras e o seu desenvolvimento pode proporcionar uma grande melhoria no ensino da Estatística proposto nesta pesquisa.

Chegando ao fim desta tese, acredito que seja importante retomar as inquietações que me impulsionaram a desenvolver a pesquisa. Todas as ideias aqui apresentadas e discutidas não estavam presentes enquanto vivia minha prática docente, como professor de Estatística do Departamento de Matemática, responsável praticamente por quase todas as disciplinas ofertadas pelos cursos. Agora, talvez, seja o momento de vivenciar, na prática, todas essas ideias que experienciei aqui em um nível teórico.

Enfatizo aqui que a janela foi construída por mim, de acordo com minha capacidade de construtor, e que ela está ajustada à altura de meu corpo. Além disso, ao olhar por essa janela, posso não perceber aspectos da paisagem que se destacariam para outras pessoas (ARAÚJO, 2002).

Deixo, então, ao leitor um convite para refletir sobre a paisagem que ele enxerga. Deixo-o também à vontade para promover alguma reforma na janela, se assim ele julgar importante.

## REFERÊNCIAS

ALMEIDA, L. M. W. Modelagem Matemática e Formação de Professores. In: **Seminário de Pesquisa em Educação da Região Sul,** 5, 2004, Curitiba. Anais. Curitiba: Universidade Católica do Paraná, 2004.

ALVES, Érica Valéria. **Um estudo exploratório dos componentes da habilidade matemática requeridos na solução de problemas aritméticos por estudantes do Ensino Médio**. Dissertação de Mestrado. Faculdade de Educação/ UNICAMP, Campinas, 1999.

ARAÚJO, Jussara de Loiola. **Cálculo, Tecnologias e Modelagem Matemática:** As Discussões dos Alunos. Tese de Doutorado elaborada junto ao Programa de Pós-Graduação em Educação Matemática. Rio Claro, São Paulo, 2002.

ASSOCIAÇÃO DE PROFESSORES DE MATEMÁTICA. **História da Estatística.** Disponível em: <http://www.apm.pt/pa/index.cap?accao=showtext&id=2750>Acesso em: 04 fev. 2005.

BARBOSA, J. C. **O que pensam os professores sobre a modelagem matemática?** Zetetiké, Campinas, v. 7, n. 11, p. 67-85, 1999.

BARBOSA, Letícia Maria. Topofilia, **Memória e Identidade na Vila do IAPI em Porto Alegre**. Universidade Federal do Rio Grande do Sul/UFRGS, Rio Grande do Sul, 2008.

BASSAB, Wilton O. **Estatística básica.** 4. ed. São Paulo: Atual, 1987.

BASSANEZI, R. C.; FERREIRA, W. C. **Equações Diferenciais com Aplicações**. São Paulo: Harbra Ltda, 1988.

BASSANEZI, R. C. Modelling as a Teaching-Learning Strategy. **For the Learning of Mathematics**, Vancouver, v. 14, n. 2, p. 31-35, june, 1994.

BASSANEZI, Rodney Carlos. **Ensino-aprendizagem com Modelagem Matemática**. São Paulo: Ed. Contexto, 2004.

BASSANEZI, Rodney Carlos. **Ensino-aprendizagem com Modelagem Matemática:** uma nova estratégia. 3. Ed. São Paulo: Contexto, 2006. 389 p. ISBN 8572442073

BATANERO, C. **Didática de la Estatística**.Granada: Reprografia de la Facultad de Ciencias/Universidad de Granada, 2001a.

BATANERO, C.; GODINO, J. D. **Análises de Dados y su didática**: Reprografia de la Faculdade de Ciencias/Universidad de Granada, 2001b.

BECKER, Fernando. **Da ação à operação:** o caminho da aprendizagem em J. Piaget e P. Freire. 2. ed. Rio de Janeiro: DP&A Editora. 1997.

BERQUÓ, Elza Salvatori. **Bioestatística/** José Maria Pacheco de Souza, Sabina Lea Davidson Gotlieb. 2 ed. São Paulo: EPU, 2006.

BIEMBENGUT, Maria Sallet; HEIN, Nelson. **Modelagem Matemática no Ensino.** São Paulo: Contexto, 2005.

BIGGE, Morris L. **Teorias da aprendizagem para professores**. São Paulo: Editora da Universidade de São Paulo. 1977.

BISOGNIN, E. Modelagem Matemática na Escola. In: **I Congresso Nacional de Educação Matemática, VII Encontro Regional de Educação Matemática:** A Gestão da Sala de Aula, Perspectivas e Desafios. Anais. Ijuí, RS: Unijui, 2008, 1, CD-ROM.

BOYER, C. B. **História da Matemática**. São Paulo: Egar Blücher, 1998.

BRASIL. **Parâmetros Curriculares Nacionais:** Matemática. Brasília: MEC/SEF, 1998.

BRASIL. **Parâmetros Curriculares Nacionais**: Ensino Médio. Brasília: MEC/SEF, 1999.

BRASIL. **Parâmetros Curriculares Nacionais**. Matemática. Brasília: MEC/SEF, 1997.

BRITO, M.R.F. **Psicologia e educação matemática**. Revista Educação Matemática da SBEM, SP, ano 1 (1):31-62, 1993.

BRITO, M.R.F. **Atitude em relação à Matemática em estudantes de 1º e 2º graus**. Tese de Livre Docência: Faculdade de Educação/UNICAMP. Campinas, 1996.

BRITO, M.R.F. **Adaptação e validação de uma escala de atitudes em relação à Matemática**. Zetétike. 1998. p. 6,9, 109-166.

BUENO, B. W.; VENDRAMINI, C. M. M.; GHIRALDELLI, C.; et al. **Prova de raciocínio estatístico para diferentes níveis de ensino**. Pôster apresentado no II Encontro de Pesquisadores em Educação. Universidade São Francisco, Itatiba- São Paulo, 2003.

BUENO, B. W.; VENDRAMINI, C. M. M.; SILVA, M. C.; et al. Avaliação do raciocínio estatístico pela teoria de resposta ao item. Em I Congresso Nacional de Avaliação Psicológica e IX Conferência Internacional de Avaliação Psicológica**: Formas e Contextos**: Livro de resumos. Campinas: Instituto Brasileiro de Avaliação Psicológica, 2003, p.97.

BURAK, D. **Modelagem Matemática:** uma metodologia alternativa para o ensino Matemática – IGCE. Universidade Estadual Paulista Julio Mesquita Filho - Unesp.

BUSINESS SCHOOL. **European Monagement Journal**: vol.21, n. 1, pp. 1-10, fevereiro de 2003.

CARRETERO, Mário. **Construtivismo e educação**. Porto Alegre: Artes Médicas, 1997.

CARVALHO, Dione Lucchesi de. **Metodologia do ensino de Matemática**. São Paulo: Cortez, 1991.

CAZORLA, I. M. **A relação entre habilidade viso-pictórica e o domínio de conceitos estatísticos na leitura de gráficos**. Tese de Doutorado. Faculdade de Educação, Universidade Estadual de Campinas, 2002.

CAZORLA, I. M.; SILVA, C. B.; VENDRAMINI, C. M. M.; et al. Adaptação e validação de uma escala de atitudes em relação à estatística. In: Conferência Internacional Experiências e Perspectivas do Ensino da Estatística - Desafios para o século XXI. Florianópolis**, Anais**. Florianópolis: União Europeia, 1999, pp.45-57.

CORDANI, L.K. *O* **ensino de Estatística na universidade e a controvérsia sobre os fundamentos da inferência**. Tese de Doutorado: FE/USP. São Paulo, 2001, p. 11.

COSTA, E. R. **As estratégias de aprendizagem e a ansiedade de alunos do ensino fundamental:** implicações para a prática educacional. Dissertação (Mestrado em Educação) FE/UNICAMP. Campinas, 2000.

COSTA, Wilse Arena da. **A construção social do conceito de bom professor**. Dissertação de Mestrado. UFMT, 1998.

CERVO, A.L.; BERVIAN, P. A. **Metodologia científica.** 5. ed. São Paulo: Printice Hall, 2002.

CHAUÍ, M. **Convite à Filosofia**. São Paulo: Ática, 1994.

CRESPO, Antonio Arnot**. Estatística fácil**. 17. ed. São Paulo: Saraiva, 2000.

D' AMBRÓSIO, U. **Os novos paradigmas e seus reflexos na destruição de certos mitos hoje prevalentes na Educação**. Ciências, Informação e Sociedade. Brasília: Universidade de Brasília, 1994.

D'AMBRÓSIO, Ubiratan. **Da Realidade à Ação: Reflexões sobre Educação Matemática**. Campinas, SP: Editora da UNICAMP, 1986.

D'AMBRÓSIO, Ubiratan. **Educação Matemática:** da teoria à prática. 10. ed. Campinas, SP: Papirus, 1996, 120p

D'AMBROSIO, Ubiratan. **Educação Matemática da Teoria à Prática**. 6. ed. Campinas, SP: Papirus, 2000.

DUARTE, Newton**. O ensino da Matemática na educação adulto**. 7. ed. São Paulo: Cortez. 1995.

DUBROVINA, I.V. **As Study of Matematical abilities in children in the primary grades-** Saviet studies in school Mathematics education, 83-96. IV, 1992.

FERREIRA, A. B. H. **Novo Aurélio:** O Dicionário da Língua Portuguesa. Rio de Janeiro: Lexikon Informática Ltda. Rio de Janeiro: Nova Fronteira, 2000. 1 CD-ROM. Windows 95/98.

FREIRE, Paulo. GXFDomR_ H_ PXGDQoD. 12. ed. São Paulo: Editora Paz e Terra, 1986.

FREIRE, Paulo. **Pedagogia da autonomia** – saberes necessários à prática educativa. Rio de Janeiro: Paz e Terra, 1997.

FREIRE, Paulo. **Pedagogia da Autonomia: saberes necessários à prática educativa.** 10. ed. São Paulo: Editora Paz e Terra, 1999.

FLORINI, José Valdir. **Professor e pesquisador**: (exemplificação apoiada na Matemática). 2. ed. Blumenau: FURB, 2000.

FONSECA, Jairo Simon da; MARTINS, Gilberto de Andrade. **Curso de Estatística**. 3. ed. São Paulo: Atlas, 1982.

FONSECA, Jairo Simon da; MARTINS, Gilberto de Andrade. **Curso de Estatística**. 6. ed. São Paulo: Atlas, 1996.

GADOTTI, Moacir. **Educação e Compromisso**. 5. ed. Campinas, SP: Papirus Editora, 1995.

GARDNER, Howard. Estruturas da Mente e a teoria das Inteligências Múltiplas. Porto Alegre. RS. Artes Médicas Sul. 1994b.

GIL, Antonio Carlos. **Projetos de pesquisa**. São Paulo: Atlas, 1987.

GIL, A.C. **Métodos e técnicas de pesquisa social**. São Paulo: Atlas, 1999.

GONÇALEZ, Norival. **Atitudes dos alunos do curso de Pedagogia com a relação à disciplina de Estatística no Laboratório de Informática**. Tese de Doutorado, Faculdade de Educação/Universidade Estadual de Campinas. p. 80-81, 129-136, 2002.

GREYTON, Lynda; GHOSHAL, Sumantra. **Administrando o capital humano pessoal**: nova ética para o emprego "voluntário". London Business School Tradução de Thaís Sato (IED-UFSC). European Management Journal Vol. 21, Nº 1,pp.1-10, fevereiro de 2003.

GUIMARÃES, Paulo Ricardo B. Coordenador do curso de Bacharelado em Estatística da Universidade Federal do Paraná-UFPR. Disponível em: <http://www.est.ufpr.br/curso/detrault.htm>. Acesso em: 08 fev. 2005.

HAIR JUNIOR, F.; ANDERSON, R. E.; TATHAM, R. L.; BLACK, W. C. **Análise multivariada de dados.** Porto Alegre: Bookman, 2005. 600 p.

HUPPES ,Roque. **Uma proposta de melhoria do ensino-aprendizagem da matemática.** Programa de Pós-graduação em Engenharia de Produção da Universidade Federal de Santa Catarina. Florianópolis, 2002.

KIDA, Maria das Graças. **Institucionalização e normalização das ações do Centro Universitário de Rondonópolis***,* FUFMT. Monografia de Especialização, 1985.

KREJCIE; MORGAN. 1970. In: GERARDI, L.; SILVA, B. **Quantificação em geografia**. São Paulo: Difusão Editorial, 1981.

LABES, Emerson Moisés. **Questionário***:* do planejamento à aplicação na pesquisa. Chapecó: Grifos, 1998.

LOPES, Celi Aparecida Espasandin. **A Probabilidade e a Estatística no Ensino Fundamental***:* uma análise curricular. Dissertação de Mestrado. Faculdade de Educação/UNICAMP. Campinas, 1998.

LORENZATO, S. Laboratório de ensino de matemática e materiais didáticos manipuláveis. In: LORENZATO, Sérgio. **Laboratório de Ensino de Matemática na formação de professores.** Campinas: Autores Associados, 2006, p.3-38.

LOUREIRO, OLIVEIRA e BRUNHEIRA. **Ensino e Aprendizagem da Estatística**. Lisboa: GRAFIS, Cooperativa de Artes Gráficas CRL, 2000.

LUDKE, Menga; ANDRÉ, Marli E. D. **Pesquisa em educação***:* abordagens qualitativas. São Paulo: EPU, 1986.

MACHADO, N. J. **Ensaios transversais**: cidadania e educação. São Paulo: Escrituras, 1997.

MACHADO, N. J. **Matemática e Linguagem de** Matemática - Análise de uma impregnação mútua. São Paulo: Cortez, 1987.

MACIEL, Domício Magalhães. **A avaliação no processo ensino-aprendizagem de Matemática, no ensino médio***:* uma abordagem sócio-cognitivista. Dissertação (Mestrado em Educação), FE/UNICAMP. Campinas, 2003.

MANDIM, Daniel. **Estatística Descomplicada**. Brasília, DF: 3. ed. Editora Vestcon, 1995.

MARCONI, Marina de Andrade; LAKATOS, Eva Maria. **Técnicas de pesquisas**. 3. ed. São Paulo: Atlas. 1996.

MARCONDES; GENTIL; SÉRGIO. **Matemática para o Ensino Médio**. Volume único. São Paulo: Ática, 1998.

MARTINS, Gilberto de Andrade. **Manual para elaboração de monografias e dissertações**. São Paulo: Atlas, 2002.

MARTINS, Gilberto de A; DENAIRE, Denis. **Princípios de Estatística**. São Paulo: Atlas, 1990.

MATTAR, Fanze Najib. **Pesquisa de Marketing**: metodologia, planejamento. São Paulo: Atlas, v.2, 1992, p. 64.

MCLED e ADAMS, **Assect and Matematical problem. solving**. A new perspective: New Yorkspringer,1989.

MELO, Elisabete Carvalho de. **A escrita da prática pedagógica como estratégia metodológica de formação.** IX Congresso Estadual Paulista sobre formação de educadores, 2007, UNESP, Universidade Estadual Paulista.

MORAN, José Manuel. Mudar a forma de aprender e ensinar com a Internet. **In: Salto para o futuro: TV e Informática na educação.** Brasília. Secretaria de Educação a Distância. Ministério da Educação e do Desporto. 1998.

MOORE, D. S. New Pedagogy and new content: the case of Statistcs. **International Statistical Review**, 65(2), 123-137, 1997.

MOROZ, Melania; GIANFALDONI, Mônica Helena T. A. **O processo de pesquisa**: iniciação. Brasília: Plano Editora, 2002.

NOVAES, Diva Valério**. A mobilização de conceitos estatísticos**: estudo exploratório com alunos de um curso de Tecnologia em Turismo. Tese de Mestrado. Pontifícia Universidade Católica de São Paulo. 2004, pp. 24-28.

OLIVEIRA, Francisco Estevam Martins de**. Estatística e Probabilidade**. 2. ed. São Paulo: Atlas, 1999.

OLIVEIRA, Marta Kohl de. Pensar em educação. In.: CASTORINA, José Antonio et al. **Piaget-Vygotsky Novas contribuições para o debate**. São Paulo: Ática, 1997. p. 51-81.

PARRA, Cecília (Org.). **Didática de Matemática:** Reflexões psicopedagógicas. Porto Alegre. Artes Médicas, 1996.

PILETTI, Claudino. **Didática Geral**. São Paulo: Ática, 1991.

POLYA, George. **A arte de resolver problemas**. Trad. Heitor Lisboa de Araújo. Rio de Janeiro: Interciência, 1978.

PONTE, João Pedro da. **A modelação no processo de Aprendizagem**. Educação e Matemática nº 23. 1992.

RAGAZZI, N. **Uma escala de atitude em relação à Matemática**. Dissertação de Mestrado. IP/USP-São Paulo, 1976.

RÊGO, R.M.; RÊGO, R.G. Desenvolvimento e uso de materiais didáticos no ensino de matemática. In: LORENZATO, Sérgio. **Laboratório de Ensino de Matemática na formação de professores.** Campinas: Autores Associados, 2006, p.39-56.

ROBERT, A. Ontils d' Analyse des Contenus mathématiques à enseigner au Iycée et à I' Université. **Recherches em didactique des matématiques**, vol.18,nº2,pp.139-190, 1998.

SAMPIERI, Roberto Hernández, et al. **Metodologia de la investigación**. 2ª edição, Mcgraw – Hill. Interamericano Editores, S.ª de C. V. 2001.

SÁNCHEZ, J.; ENCINAS, L. H.; COMNDE, M. J. **Análisis comparada del currículo de matemática (nível médio) em Iberoamérica**. Madrid: Maré Nostrum, 1992.

SAVANI, Dermeval. **A nova lei da educação:** trajetórias, limites e perspectivas. 7.ed. Campinas: Autores Associados, 1997.

SELLTIZ, WRIGHTSMAN e COOK. **Métodos de Pesquisa nas Relações Sociais**. Trad. De Maria Martha Hubner d' Oliveira e Miriam Marinotte del Rey. Edit. E.P.U., 1987.

SEVERINO, Antônio Joaquim. **Metodologia do trabalho científico.** 23. ed. Revista e Atualizada. São Paulo: Cortez, 2007.

SKOVSMOSE, Ole. **Educação matemática crítica:** a questão da democracia. Campinas: Papirus, 2001 (Coleção Perspectivas em Educação Matemática).

SILVA, C.; CARZOLA I.; BRITO, M. **Concepções e atitudes em relação à Estatística.** Anais da Conferência Internacional Experiências e Perspectivas do Ensino de Estática, Desafios para o século XXI. Florianópolis, 1999, p. 18-29.

SILVA, C.B**. Atitudes em relação à Estatística**: um estudo com alunos de graduação. Dissertação de Mestrado. UNICAMP/SP, 2000.

SILVA, Ermes Medeiros da, et al. **Estatística I.** 2. ed. São Paulo: Atlas, 1996 (Volume I e II).

SILVA, Raquel Correira da; SILVA, José Roberto da. **O papel do Laboratório no Ensino de Matemática.** VIII encontro Nacional de Educação Matemática, UFPE – Recife-PE, 2004.

SOUZA, Carlos Henrique Medeiros de. **Comunicação, educação e novas tecnologias**. Campos de Goytacazes, RJ: Editora FFFIC, 2003, p.47-50.

SKINNER, Burrhus Frederic. **Tecnologia do ensino.** 2.a.reimpressão. São Paulo: Editora da Universidade de São Paulo. 1972.

STURZA, José Adolfo Iriam. . O sentido de lugar em Rondonópolis-MT e o topocídio do cerrado: uma contribuição aos estudos de cognição ambiental. In: GERARDI, Lucia H. e CARVALHO, Pompeu F. **Geografia:** ações e reflexões. Rio Claro: UNESP/IGCE: AGETEO, 2006. pp. 341-358.

STIELER, Marinez Cargnin. **Compreensão de conceitos de Matemática e Estatística na perspectiva da Modelagem Matemática:** Caminhos para uma aprendizagem significativa e contextualizada no ensino superior. Programa de Pós-Graduação do Centro Universitário Franciscano. Santa Maria, RS, 2007.

TAJRA, Sanmya Feitosa. **Novas ferramentas pedagógicas para o professor da atualidade**. São Paulo: Érica, 2001.

TURRIONI, A.M.S. **O Laboratório de Educação Matemática na formação inicial de professores**. Dissertação (Pós-graduação em Educação Matemática e seus fundamentos Filosóficos-Científicos) Universidade Estadual Paulista, Rio Claro-SP. Orientador: Geraldo Perez. 2004, 168p.

THIOLLENT, M**. Metodologia da Pesquisa-Ação**. São Paulo: Cortez, 1983.

VARIZO, Z.C.M. **O Laboratório de Educação Matemática do IME/UFG:** Do sonho a realidade. In: ENEM, 10, Belo Horizonte. **Anais...** Belo Horizonte, 2007, p.1-12.

VALENTE, José Armando. Análise dos diferentes tipos de software usados na educação. In: **Salto para o Futuro:** TV e Informática na Educação. Brasília. Secretaria de Educação a Distância. Ministério da Educação e do desporto. 1998. p. 91-112.

VENDRAMINI, C.M.M. **Implicações das atitudes e das habilidades matemáticas na aprendizagem dos conceitos de Estatística**. Tese de Doutorado. UNICAMP/SP, 2000.

VENDRAMINI, C. M. M. Análise de itens de uma prova de raciocínio estatístico. In: IX Seminário IASI de Estatística Aplicada: Estatística na Educação e Educação em Estatística [CD-ROM**]. Comunicações orais e Resumo das comunicações da Sessão Pôster**. Rio de Janeiro: Instituto Brasileiro de Geografia e Estatística, 2003, p.15.

VENDRAMINI, C. M. M. Contribuições da Educação Estatística para a Educação Matemática. Artigo submetido à revista **Revista Pedagógico**, Universidade do Oeste de Santa Catarina. Número especial, 2004.

VENDRAMINI, C. M. M.; BRITO, M. R. F. Relações entre atitude, conceito e utilidade da Estatística. **Psicologia Escolar e Educacional,** *5*(1), 59-73, 2001.

VENDRAMINI, C. M. M.; BRITO, M. R. F. Implicações das habilidades matemáticas e das atitudes na aprendizagem dos conceitos de Estatística [CD-ROM]. **Anais do II Seminário Internacional de Pesquisa em Educação Matemática.**. São Paulo: Sociedade Brasileira de Educação Matemática, 2003.

VENDRAMINI, C. M. M.; CHENTA, V. C.; SILVA, L. S. **Leitura de dados estatísticos**: Um estudo com alunos do ensino fundamental (texto não publicado). 2004.

VENDRAMINI, C. M. M.; SILVA, M. C.; BUENO, B. W.; SILVA, L. S. **Avaliação da dimensionalidade de uma prova de desempenho estatístico**. Pôster apresentado no II

Encontro de Pesquisadores em Educação. Universidade São Francisco, Itatiba- São Paulo, 2003.

VENDRAMINI, C. M. M.; SILVA, M. C.; CANALE, M. Análise de itens de uma prova de raciocínio estatístico. Artigo submetido à revista **Psicologia em Estudo**, Maringá, 2003.

VERGNAUD, G. La theorie des champs conceptual. In: BRUN, J. (org.) **Didactique des Mathématiques. Lousanne**, Paris: Delachaux et Miestlé, 1996 a.

VIEIRA, Sônia. **Elementos de Estatística**. 3. ed. São Paulo: Atlas, 1999.

VIGOTSKI, Lev Semenovich; LURIA, Alexander Romanovich; LEONTIEV, Alexis N. **Linguagem, Desenvolvimento e aprendizagem.** 7ª Edição. São Paulo. Ícone Editora Ltda. 2001.

VIECILI, Cláudia R. Confortin. **Modelagem Matemática:** uma proposta para o ensino de Matemática. 2006. 119 f, Dissertação (Mestrado em Ciência da educação e Matemática) – PUC, Porto Alegre, 2006.

WADA, R. (1996). **Estatística e ensino:** um estudo sobre as representações de professores de 3º grau. Tese de doutorado, UNICAMP, Campinas, SP.

WINFRIED, Bohm**. Escuela y Sociedade Leay**. Ver. Papiro, Bs. As. Ano VI, n. 19, Wurzsurg, pp. 27-28, 1981.

# APÊNDICES

## Apêndice A – Questionário dos alunos matriculados na disciplina de estatística

Caro aluno

Este questionário faz parte de uma pesquisa a respeito da "Uma proposta metodológica para a aprendizagem: Reflexão sobre as práticas pedagógicas da Estatística ao elaborar os instrumentos de pesquisa sociais". Essa pesquisa está sendo realizado pelo professor Manoel Benedito Nirdo da Silva Campos, que resultará em uma tese de Doutorado.

Obrigado pela atenção e disponibilidade.

**I. CARACTERÍSTICAS SOCIOECONÔMICAS**

Idade:

( ) Até 18 anos ( ) 19 a 24 anos ( ) 25 a 30 anos ( ) Mais de 30 anos

Sexo:

( ) Masculino ( ) Feminino

Escolaridade:

( ) ensino fundamental incompleto

( ) ensino fundamental completo

( ) ensino médio incompleto

( ) ensino médio completo

( ) ensino superior incompleto

( ) ensino superior completo

Ciclo:

( ) 1º ano ( ) 2º ano ( ) 3º ano ( ) 4º ano

Período:

( ) Vespertino ( ) Noturno ( ) matutino ( ) Integral

**NO QUESTIONÁRIO ABAIXO, ASSINALE A ALTERNATIVA QUE MAIS SE APROXIMA DE SUA REALIDADE.**

1 O professor apresentou e discutiu o plano de ensino com os alunos?

a) ( ) Sim

b) ( ) Não

c) ( ) Não sei responder

2 Houve articulação entre a teoria e prática para o desenvolvimento do processo de aprendizagem?

a) ( ) Sim

b) ( ) Não

c) ( ) Não sei responder

3 Os objetivos da disciplina foram (estão sendo) alcançados?

a) ( ) Sim

b) ( ) Não

c) ( ) Não sei responder

4 A disciplina de estatística incentivou a participação (questionamento e discussão) dos alunos em sala de aula?

a) ( ) Sim

b) ( ) Não

c) ( ) Não sei responder

5 Você acha que a estrutura física (salas de aula, laboratórios, etc.) interfere no processo ensino-aprendizagem?

a) ( ) Sim

b) ( ) Não

c) ( ) Não sei responder

6 Houve diversificação dos tipos de avaliação de modo a promover a aprendizagem?

a) ( ) Sim

b) ( ) Não

c) ( ) Não sei responder

7 Você era assíduo às aulas de Estatística?

a) ( ) Sim

b) ( ) Não

c) ( ) Não sei responder

8 O professor discutia os resultados das avaliações?

a) ( ) Sim

b) ( ) Não

c) ( ) Não sei responder

## II. NÍVEL DE PERCEPÇÃO E DESEMPENHO PARA ESENVOLVIMENTO DE ATIVIDADES ESTATÍSTICAS

9 O professor promovia o uso de equipamentos tecnológico?

a) ( ) Muito

b) ( ) Às vezes

c) ( ) Raramente

d) ( ) Nunca

10 O professor era assíduo e pontual?

a) ( ) Sim

b) ( ) Não

c) ( ) Nem sempre

d) ( ) Não sei responder

11 Quando não entendia um conteúdo, você:

1. ( ) Questionava o professor no momento
2. ( ) Questionava o professor fora do horário da aula
3. ( ) Tirava as dúvidas com os colegas
4. ( ) Não procurava sanar as dúvidas desse conteúdo

12 O curso oferece uma formação humanista além dos conteúdos técnico/científico?

a) Não

b) Pouco

c) Razoável

d) Muito

13 Eu aprendo Estatística:

a) ( ) Fácil e rapidamente, sem nenhum esforço.

b) ( ) Facilmente, despendendo algum tempo e esforço.

c) ( ) Dificilmente, despendendo tempo e esforço.

d) ( )Não consigo aprender Estatística.

14 Eu compreendo as explicações do professor:

a) ( ) Na primeira vez que ouço.

b) ( ) Após resolver alguns exercícios sobre o assunto.

c) ( ) Após resolver muitos exercícios sobre o assunto.

d) ( ) Não compreendo em hipótese alguma.

15 Diante de problemas Estatísticos novos para mim:

a) ( ) Resolvo-os rapidamente.

b) ( ) Tenho dificuldades apenas até ser capaz de dominar o método de solução.

c) ( ) Tenho dificuldades em resolvê-los.

d) ( ) Não sou capaz de resolver nenhum problema estatístico.

16 Ao solucionar problemas Estatísticos:

a) ( ) Faço todos os cálculos mentalmente, sem recorrer a registro em papel ou uso de calculadora.

b) ( ) Faço todos os cálculos no papel.

c) ( ) Utilizo a calculadora em todos os cálculos.

d) ( ) Não chego a fazer cálculos.

e) ( ) Papel e Calculadora

17 Durante as aulas de exercícios de Estatística:

a) ( ) Sinto-me cansado pois a aula é monótona e os exercícios repetitivos.

b) ( ) Considero-a tão cansativa quanto qualquer outra aula de Estatística.

c) ( ) Sinto-me motivado a resolver todos os exercícios, para entender melhor.

d) ( ) Procuro copiar os exercícios de alguém que saiba mais que eu.

18 Durante as aulas de exercícios de Estatística, quando tem dificuldades de resolver algum exercício, você:

a) ( ) Pára, e passa para outro

b) ( ) Tenta resolvê-lo sozinho(a)

c) ( ) Tenta resolvê-lo com a ajuda de colegas

d) ( ) Tenta resolvê-lo com o auxílio do professor

19 Qual o motivo maior que dificultou o aprendizado em estatística?

a) ( ) A base fraca em matemática

b) ( ) A metodologia adotada pelo professor

c) ( ) A disciplina é muito difícil

d) ( ) A didática do professor

20 Você acha que uma boa relação professor-aluno:

a) ( ) Ajuda no processo ensino-aprendizagem

b) ( ) Não interfere no processo ensino-aprendizagem

c) ( ) Estimula os alunos a estudarem mais

d) ( ) Faz com que os alunos procurem mais o professor para sanar as dúvidas

21 O professor apresentou ou fez uso de algum software estatístico?

a) ( ) Sim

b) ( ) Não

c) ( ) Não sei responder

**III – ESCALA DE LIKERT – ATITUDES COM RELAÇÃO À ESTATÍSTICA**

22 Instrução: Compare os itens abaixo assinalando de acordo com sua percepção aquele que expresse sua opinião. (Adaptado e validado por Brito, 1998).

Cada uma das frases abaixo expressa o sentimento que pessoas apresentam com relação à Estatística. Você deve comparar o seu sentimento pessoal com aquele expresso em cada frase, assinalando um dentre os quatro pontos colocados abaixo de cada uma delas, de modo a indicar com a maior exatidão possível, o sentimento que você experimenta com relação à Estatística.

**ATITUDE NEGATIVA**

01 – Eu fico sempre sob uma terrível tensão na aula de Estatística.

( )Discordo totalmente ( )Discordo ( )Concordo ( )Concordo totalmente

02 – Eu não gosto de Estatística e me assusta ter que fazer essa matéria.

( )Discordo totalmente ( )Discordo ( )Concordo ( )Concordo totalmente

03 – Eu acho a Estatística muito interessante e gosto das aulas de Estatística.

( )Discordo totalmente ( )Discordo ( )Concordo ( )Concordo totalmente

04 – A Estatística é fascinante e divertida.

( )Discordo totalmente ( )Discordo ( )Concordo ( )Concordo totalmente

05 – A Estatística me faz sentir seguro (a) e é, ao mesmo tempo, estimulante.

( )Discordo totalmente ( )Discordo ( )Concordo ( )Concordo totalmente

06 – “Dá um branco” na minha cabeça e não consigo pensar claramente quando estuda Estatística.

( )Discordo totalmente ( )Discordo ( )Concordo ( )Concordo totalmente

07 – Eu tenho a sensação de insegurança quando me esforço em Estatística

( )Discordo totalmente ( )Discordo ( )Concordo ( )Concordo totalmente

08 – A Estatística me deixa inquieto (a), descontente, irritado (a) e impaciente.

( )Discordo totalmente ( )Discordo ( )Concordo ( )Concordo totalmente

09 – O sentimento que tenho com relação à Estatística é bom.

( )Discordo totalmente ( )Discordo ( )Concordo ( )Concordo totalmente

10 – A Estatística me faz sentir como se estivesse perdido (a) em selva de números e sem encontrar a saída.

( )Discordo totalmente ( )Discordo ( )Concordo ( )Concordo totalmente

**ATITUDE POSITIVA**

11 – A Estatística é algo que eu aprecio grandemente.

( )Discordo totalmente ( )Discordo ( )Concordo ( )Concordo totalmente

12 – Quando eu ouço a palavra Estatística, eu tenho um sentimento de aversão.

( )Discordo totalmente ( )Discordo ( )Concordo ( )Concordo totalmente

13 – Eu encaro a Estatística com um sentimento de indecisão, que é resultado do medo de não ser capaz em estatística.

( )Discordo totalmente ( )Discordo ( )Concordo ( )Concordo totalmente

14 – Eu gosto realmente de Estatística.

( )Discordo totalmente ( )Discordo ( )Concordo ( )Concordo totalmente

15 – A Estatística é uma das matérias que eu realmente gosto de estudar na universidade.

( )Discordo totalmente ( )Discordo ( )Concordo ( )Concordo totalmente

16 – Pensar sobre a obrigação de resolver um problema estatístico me deixa nervoso (a).

(  )Discordo totalmente (  )Discordo   (  )Concordo   (  )Concordo totalmente

17 – Eu nunca gostei de Estatística e é a matéria que me dá mais medo.

(  )Discordo totalmente (  )Discordo   (  )Concordo   (  )Concordo totalmente

18 – Eu fico mais feliz na aula de Estatística que na aula de qualquer outra matéria.

(  )Discordo totalmente (  )Discordo   (  )Concordo   (  )Concordo totalmente

19 – Eu sinto tranqüilo (a) em Estatística e gosto muito dessa matéria.

(  )Discordo totalmente (  )Discordo   (  )Concordo   (  )Concordo totalmente

20 – Eu tenho uma reação definitivamente positiva com relação à Estatística. Eu gosto e aprecio essa matéria.

(  )Discordo totalmente (  )Discordo   (  )Concordo   (  )Concordo totalmente

## Apêndice B - Questionário dos professores que lecionam a disciplina de estatística

**Caro professor**

Este questionário faz parte de uma pesquisa conduzida pelo Prof. Manoel Campos, que resultará em uma tese de Doutorado intitulada: "Uma proposta metodológica para a aprendizagem: Reflexão sobre as práticas pedagógicas da Estatística ao elaborar os instrumentos de pesquisa sociais". Pedimos por gentileza responder as questões que expressam a sua realidade quanto à Disciplina Estatística ministrada por você. Agradecemos à atenção e disponibilidade.

**Sexo:**

1( ) Masculino 2( ) Feminino

**Idade:**

1( ) Até 25 anos 2( ) 26 a 35 anos 3( ) 36 a 45 anos 4( ) Mais de 45 anos

**Tempo que ministra a disciplina Estatística:**

1( ) Menos de 2 anos 2( ) 2 a 7 anos 3( ) 8 a 13 anos 4( ) 14 a 19 anos 5 ( ) Mais de 19 anos

**Tipo de Vínculo:**

1( ) Efetivo 2( ) Substituto

**Tipo de Habilitação:**

1( ) Bacharelado 2( ) Licenciatura

**Titularidade:**

1( ) Graduado 2( ) Especialista 3( ) Mestre 4( ) Doutor

**Ciclo(s) que ministra a disciplina Estatística:**

1( ) 1º ano 2( ) 2º ano 3( ) 3º ano 4( ) 4º ano

1) Você acha que a disciplina Estatística é compatível com sua formação

( ) Sim, em todo ( ) Sim, em apenas alguns ( ) Não

( ) Sim, na maioria ( ) Não sei responder

2) Você se sente seguro e preparado/capacitado para trabalhar os conteúdos da disciplina Estatística?

( ) Sim, em todos ( ) Sim, em apenas alguns ( ) Não

( ) Sim, na maioria ( ) Não sei responder

3) Você acha que a ementa é adequada para alcançar os objetivos da disciplina?

( ) Sim, totalmente ( ) Não

( ) Sim, parcialmente ( ) Não sei responder

4) Você acha que a carga horária é adequada para alcançar os objetivos da disciplina?

( ) Sim, totalmente ( ) Não

( ) Sim, parcialmente ( ) Não sei responder

5) Você consegue fazer a articulação entre a teoria e a prática?

( ) Sim, em todos os conteúdos ( ) Não

( ) Sim, na maioria dos conteúdos ( ) Não sei responder

( ) Sim, em alguns conteúdos apenas

6) Existe articulação com outras disciplinas do(s) curso(s) em que trabalha?

( ) Sim ( ) Não

7) Você procura introduzir no conteúdo técnico/científico de suas disciplinas uma visão humanista quanto à sua aplicação?

( ) Sim ( ) Não

8) Você consegue diversificar as atividades, visando a melhoria do processo ensino-aprendizagem?

( ) Sempre ( ) Raramente ( ) Não sei responder

( ) Freqüentemente ( ) Não

9) Como você avalia os recursos audiovisuais (retroprojetor, datashow, computador, etc.) disponíveis para uso na disciplina?

( ) Ótimos ( ) Ruins ( ) Não sei responder

( ) Bons ( ) Péssimos

( ) Regulares ( ) Não utilizo

10) Você já utilizou algum software estatístico em suas aulas?

( ) Sim ()Não

11) Você dialoga com os alunos sobre os critérios de avaliação?

( ) Sim ( ) Nem sempre ( ) Não

12) Você discute e analisa os resultados das avaliações com os alunos?

( ) Sim ( ) Nem sempre ( ) Não

13) Qual é o nível de exigência na disciplina?

( ) Exijo muito
( ) Não sou muito exigente
( ) Exijo na medida
( ) Não sei responder

14) Você procura se manter atualizado na área de Estatística?

( ) Sempre
( ) Estatística não é minha prioridade
( ) Não

15) Você procura sanar as dúvidas dos alunos fora do horário de aula?

( ) Sim
( ) Não
( ) Nem Sempre

16) Você no primeiro dia de aula apresenta e discute o plano de ensino com os alunos?

( ) Sim
( ) Nem Sempre
( ) Não
( ) Não sei responder

17) Como você avalia o uso de equipamentos tecnológico? ( ) Muito

( ) Às vezes
( ) Nunca
( ) Raramente

18) Você acha importante trabalhar a disciplina de Estatística em Laboratório?

( ) Sim
( ) Nem Sempre
( ) Não

19) Como você avalia os alunos quanto à participação da discussão dos resultados das avaliações?

( ) Bom
( ) Regular

## Apêndice C - Prova de estatística para avaliar o desempenho na solução de problemas.

Resolva os seguintes problemas de 1 a 5.

Caso necessite fazer algum cálculo, não o apague.

1) Em um grupo de 600 hóspedes do hotel Mary Posa e Cia.ltda., têm-se os seguintes valores com relação ao tempo de permanência no hotel:

Médio ---------------------------- 9 dias;

1º Quartil ------------------------ 5 dias;

2º Quartil ------------------------15 dias;

Coeficiente de variação -------- 20 %

Pede-se:

a) Quantos hóspedes permanecem mais de 15 dias;

b) Quantos hóspedes permanecem entre 5 e 15 dias;

c) O desvio-padrão para o tempo de permanência;

d) Supondo que todos os hóspedes permaneçam mais dois dias, calcular a nova média, o desvio-padrão e o coeficiente de variação.

2) Uma pesquisa sobre a renda anual familiar realizada com uma amostra de 1000 pessoas na cidade de Tangará da Serra resultou na seguinte distribuição de freqüências:

| Salário Anual (em U$$ 1000) | Número de Funcionários |
|---|---|
| 0,00 ⟶ 10,00 | 250 |
| 10,00 ⟶ 20,00 | 300 |
| 20,00 ⟶ 30,00 | 200 |
| 30,00 ⟶ 40,00 | 120 |
| 40,00 ⟶ 50,00 | 60 |
| 50,00 ⟶ 60,00 | 40 |
| 60,00 ⟶ 70,00 | 20 |
| 70,00 ⟶ 80,00 | 10 |
| TOTAL | 1.000 |

Pede-se determinar a média, a moda, os quartis e o coeficiente de variação dos salários.

3) Considere a distribuição a seguir relativa a notas de dois alunos de informática durante determinado semestre:

| | | | | | | | | |
|---|---|---|---|---|---|---|---|---|
| Aluno A | 9,5 | 9,0 | 2,0 | 6,0 | 6,5 | 3,0 | 7,0 | 2,0 |
| Aluno B | 5,0 | 5,5 | 4,5 | 6,0 | 5,5 | 5,0 | 4,5 | 4,0 |

d) Calcule as notas médias de cada aluno.

e) Qual aluno apresentou resultado mais homogêneo? Justifique.

4) Considere as seguintes informações:

| **Característica** | **Estatística** | |
|---|---|---|
| | ***Média*** | ***Desvio-padrão*** |
| Salário em U$$ | 500 | 50 |
| Anos de Trabalho | 100 | 20 |

Pergunta-se: Quais das duas características variaram mais?

5) Os salários-hora de cinco funcionários de uma companhia são:

R$ 75,00; R$ 90,00; R$ 83,00; R$ 142,00; R$ 88,00

Determine:

e) A média dos salários-hora;

f) O salário-hora mediana.

## Apêndice D – Simulação do teste de Kruskal-Wallis

**Procedimento:**

**a)** Dispor, em ordem crescente, as observações de todos os K grupos, atribuindo-lhes postos de 1 a n. Caso haja empates, atribuir o posto médio.

**b)** Determinar o valor da soma dos postos para cada um dos K grupos: $R_i \quad i = 1, 2, \ldots, K$

**c)** Realizar o teste:

a) $H_o$ : As médias são iguais.

$H_1$ : Há pelo menos um par diferente.

2. Fixar α. Escolher uma variável qui-quadrado com φ = K-1.

3. Com auxílio da Tabela qui-quadrado determinam-se RA e RC.

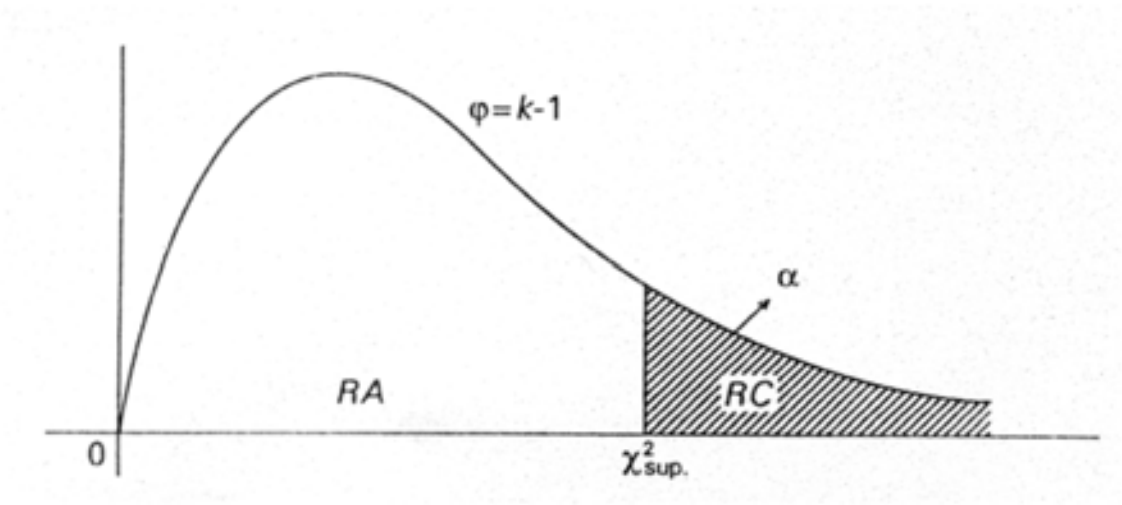

4. Calcula-se a estatística:

$$H = \frac{12}{n(n+1)} \sum_{i=1}^{K} \frac{(R_i)^2}{n_i} - 3(n+1)$$

5. Conclusão:

Se $H \leq \chi^2_{sup.}$, não se pode rejeitar $H_0$

Se $H > \chi^2_{sup.}$, rejeita-se $H_0$ concluindo-se com o risco α que há diferença entre médias dos *K* grupos.

## Apêndice E – Roteiro de observação

| ASPECTOS A CONSIDERAR | ELEMENTO A OBSERVAR |
|---|---|
| Espaço de apoio | Ambiente possui espaço adequado para recepção dos alunos. |
| Ocupação | Estimular a prática da pesquisa em sala de aula. |
| Método e técnica de ensino | Processos de ensino-aprendizagem da Matemática (modelagem, resolução de problemas, equipamento tecnológico, construção do conhecimento etc.) |
| Atividades de educação Matemática | Avaliação do processo de ensino-aprendizagem e avaliação sistêmica. |
| Atividades diferenciada de aprendizagem | Construir metodologias alternativas para situações específicas envolvendo dificuldades de aprendizagem. |

## ANEXO

## Anexo A – Tabela para cálculo do tamanho da amostra

Determinação do tamanho da amostra a partir do tamanho da população

| N* | A* | N | A | N | A |
|---|---|---|---|---|---|
| 10 | 10 | 220 | 140 | 1200 | 291 |
| 15 | 14 | 230 | 144 | 1300 | 297 |
| 20 | 19 | 240 | 148 | 1400 | 302 |
| 25 | 24 | 250 | 152 | 1500 | 306 |
| 30 | 28 | 260 | 155 | 1600 | 310 |
| 35 | 32 | 270 | 159 | 1700 | 313 |
| 40 | 36 | 280 | 162 | 1800 | 317 |
| 45 | 40 | 290 | 165 | 1900 | 320 |
| 50 | 44 | 300* | 169* | 2000 | 322 |
| 55 | 48 | 320 | 175 | 2200 | 327 |
| 60 | 52 | 340 | 181 | 2400 | 331 |
| 65 | 56 | 360 | 186 | 2600 | 335 |
| 70 | 59 | 380 | 191 | 2800 | 338 |
| 75 | 63 | 400 | 196 | 3000 | 341 |
| 80 | 66 | 420 | 201 | 3500 | 346 |
| 85 | 70 | 440 | 205 | 4000 | 351 |
| 90 | 73 | 460 | 210 | 4500 | 354 |
| 95 | 76 | 480 | 214 | 5000 | 357 |
| 100 | 80 | 500 | 217 | 6000 | 361 |
| 110 | 86 | 550 | 226 | 7000 | 364 |
| 120 | 92 | 600 | 234 | 8000 | 367 |
| 130 | 97 | 650 | 242 | 9000 | 368 |
| 140 | 103 | 700 | 248 | 10000 | 370 |
| 150 | 108 | 750 | 254 | 15000 | 375 |
| 160 | 113 | 800 | 260 | 20000 | 377 |
| 170 | 118 | 850 | 265 | 30000 | 379 |
| 180 | 123 | 900 | 269 | 40000 | 380 |
| 190 | 127 | 950 | 274 | 50000 | 381 |
| 200 | 132 | 1000 | 278 | 75000 | 382 |
| 210 | 136 | 1100 | 285 | 100000 | 384 |

---

*N = Tamanho da População

*A = Tamanho da Amostra

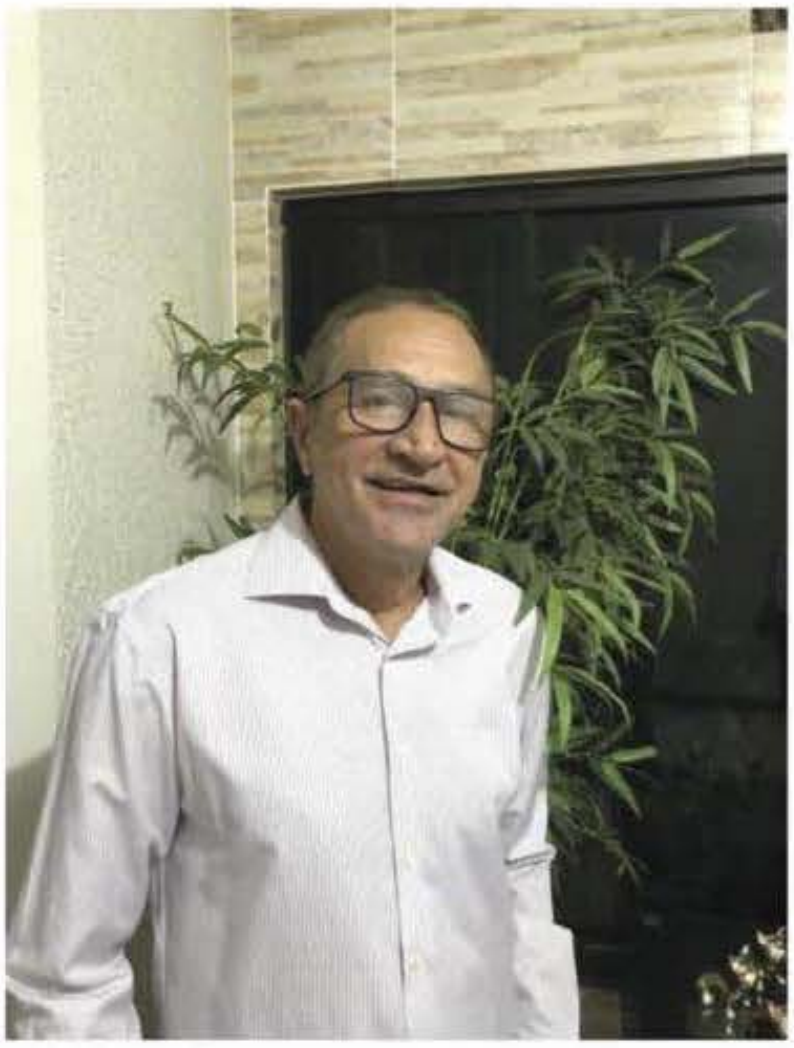

# MINHA HISTÓRIA

A minha primeira experiência com o ensino ocorreu em 1988, numa turma de Secretariado, na Escola Estadual de 1º e 2º Graus Raimundo Pinheiro em Cuiabá-MT. Fiz a graduação em Licenciatura Plena Em Matemática pela Universidade Federal de Mato Grosso. Em seguida concluí a Especialização em Formação Docente à Nível Superior com carga horária de 375/horas pela Universidade Federal de Mato Grosso, bem como, a Especialização em Estatística com carga horária de 360/horas pela Universidade de Marília, UNIMAR. Concluí o Mestrado em Geografia pelo Campus Universitário de Rondonópolis da Universidade Federal de Mato Grosso. O Doutorado em Ciências da Educação foi concluído na Universidad Tecnologica Intercontinental, UTIC, Paraguai e Convalidação pela Universidade Federal do Rio de Janeiro.

Atualmente sou professor Associado I do Departamento de Matemática do Instituto de Ciências Exatas e Naturais da Universidade Federal de Rondonópolis/MT. Tenho experiência na área de Matemática e Estatística, com ênfase em Ciências Exatas e da Terra, atuando principalmente com os seguintes temas: modelagem matemática, inclusão digital, processo de ensino-aprendizagem, software educativo, software livre e informática. No Departamento de Matemática, decidiu-se que todo docente vinculado a ele pode ministrar aula em qualquer curso daquelas disciplinas que está sob sua responsabilidade. Assim, além de responsáveis pela maior parte do curso de Matemática, nas modalidades Bacharelado e Licenciatura. Atuamos também nos cursos que têm em suas estruturas curriculares as disciplinas de Matemática e Estatística, a saber: Ciências Contábeis, Biologia, Engenharia Agrícola e Ambiental, Engenharia Mecânica, Pedagogia, Biologia, Sistema de Informação, Biblioteconomia, Psicologia e Enfermagem.

Ocupei os cargos de Pró-Reitor do Campus Universitário de Rondonópolis da Universidade Federal de Mato Grosso; Direção do Instituto de Ciências Exatas e Naturais do Campus Universitário de Rondonópolis da Universidade Federal de Mato Grosso, assumi também a Chefia do Departamento de Matemática e a Coordenações do Curso de Matemática e outras atividades administrativas pertinentes a Instituição. Destacamos que sempre procurei transformar minhas aulas num ambiente agradável e propício para a aprendizagem. Mesmo assim, tenho clareza do quanto ainda devo acrescentar à minha caminhada como Professor.